\documentclass[aps,prb,reprint,superscriptaddress]{revtex4-2}
\usepackage{graphicx}
\usepackage[english]{babel}
\DeclareUnicodeCharacter{2212}{-}

\usepackage{bm} 
\usepackage{color} 
\usepackage{amsmath} 
\usepackage{natbib}
\usepackage{SIunits} 
\usepackage{physics}

\usepackage{multirow}
\usepackage{tabularx}
\usepackage{booktabs}

\usepackage{xcolor}
\usepackage{soul}

\begin{document}
\title{Automated Workflow for Accurate High-Throughput GW Calculations}
\author{Lorenzo Varrassi}
\affiliation{Department of Physics and Astronomy, University of Bologna, 40127 Bologna, Italy}\email{lorenzo.varrassi3@unibo.it}
\author{Florian Ellinger}
\affiliation{University of Vienna, Faculty of Physics and Center for Computational Materials Science, Kolingasse 14-16, A-1090, Vienna, Austria.}
\author{Espen Flage-Larsen}
\affiliation{SINTEF Industry, Materials Physics, Oslo, Norway}
\affiliation{Department of Physics, University of Oslo, Oslo, Norway}
\author{Michael Wolloch}
\affiliation{University of Vienna, Faculty of Physics and Center for Computational Materials Science, Kolingasse 14-16, A-1090, Vienna, Austria.}
\affiliation{Vasp Software GmbH, Berggasse 21/14, 1090, Vienna, Austria}
\author{Georg Kresse}
\affiliation{University of Vienna, Faculty of Physics and Center for Computational Materials Science, Kolingasse 14-16, A-1090, Vienna, Austria.}
\affiliation{Vasp Software GmbH, Berggasse 21/14, 1090, Vienna, Austria}
\author{Nicola Marzari}
\affiliation{Theory and Simulations of Materials (THEOS) and National Centre for Computational Design and Discovery of Novel Materials (MARVEL), École Polytechnique Fédérale de Lausanne, CH-1015 Lausanne, Switzerland}
\author{Cesare Franchini}
\affiliation{Department of Physics and Astronomy, University of Bologna, 40127 Bologna, Italy}
\affiliation{University of Vienna, Faculty of Physics and Center for Computational Materials Science, Kolingasse 14-16, A-1090, Vienna, Austria.}

\date{\today}

\begin{abstract}
    The GW approximation represents the state-of-the-art ab-initio method for computing excited-state properties. 
    Its execution requires control over a larger number of (often interdependent) parameters, and therefore its application in high-throughput studies is hindered by the intricate and time-consuming convergence process across a multi-dimensional parameter space.
    To address these challenges, here we develop a fully-automated open-source workflow for G$_0$W$_0$ calculations within the AiiDA-VASP plugin architecture. 
    The workflow is based on an efficient estimation of the errors on the quasi-particle (QP) energies due to basis-set truncation and the pseudo-potential norm violation, which allows a reduction of the dimensionality of the parameter space and avoids the need for multi-dimensional convergence searches.
    Protocol validation is conducted through a systematic comparison against established experimental and state-of-the-art GW data.
    To demonstrate the effectiveness of the approach, we construct a database of QP energies for a diverse dataset of over 320 bulk structures.
   The openly accessible workflow and resulting dataset can serve as a valuable resource and reference for conducting accurate data-driven research.
\end{abstract}

\maketitle
\section*{INTRODUCTION}
    The rapid development of high-throughput (HT) approaches during the last decade has represented a significant advancement in the field of materials science.~\cite{Curtarolo2013,Thygesen_Science2016, Roadmap2019,Schaarschmidt2022} 
     The fast advancement  in computational power in combination with the reliability and efficiency of ab-initio codes and workflow engines~\cite{Fireworks2015,AiiDA2016,MyQueue2020,AFLOW2012,Atomate2018} has enabled researchers to conduct HT
    screening across increasingly larger chemical spaces. In turn, HT protocols have enabled the creation of large electronic-structure databases that have
    significantly contributed to the discovery and design of novel materials, providing a basis for future research efforts.~\cite{Kirklin2015,Jain2013,Draxl2019,Mounet2018,Talirz2020} 
    Furthermore, materials databases are an essential key for data-driven machine learning approaches,~\cite{Schmidt2019,Wang2020} enabling efficient data-analysis~\cite{Ashton2017,Cheon2017} and accelerating electronic structure simulations.
    Despite significant advancements several challenges persist, in particular regarding  limited verification standards and validation procedures, which are subject of ongoing efforts.~\cite{AiiDA2021_CommonWorkflow,DatabaseValidation_1,DatabaseValidation_2,Roadmap2019}
    \newline The vast majority of current HT schemes and materials databases employ density functional theory (DFT), due to its efficiency and reliability in predicting   structural,~\cite{Rasmussen2015,Ashton2017,Cheon2017,Choudhary2017,Mounet2018} thermodynamic~\cite{Kirklin2013,Bhattacharya2015} and ground-state electronic properties.~\cite{Hautier2013,Chen2016,Marrazzo2019,Zhang2019,Kahle2020}
    The extension of these HT approaches to the investigation of excited-state properties remains limited due to well-documented shortcomings of DFT~\cite{Godby1986}, such as the band-gap problem and the inability to accurately describe excitonic effects.~\cite{Golze2019,Onida_2002}
    A reliable description of these properties is crucial for predicting material's optical and transport behaviors – and in turn to design and discover novel materials for electronic, optoelectronic and photovoltaic applications.~\cite{Hachmann2011,He2019,Lee2018}
     In this respect, several studies have utilized ab-initio schemes based on extensions to conventional DFT,  for example by including on-site repulsion~\cite{DFTU}(DFT$+U$) or adding a portion of exact exchange~\cite{HSE06}(hybrids).
    These improved schemes have been employed to generate material databases or conduct HT screening of potential candidates for photocatalysis and photovoltaic applications.~\cite{Yan2017,Xiong2021,Kim2020,Hinuma2017,Liu2024,Castelli2012,Kuhar2018,Rasmussen2015}
    \newline In recent times, there has been a surge in efforts dedicated to advancing workflows and databases beyond local and semi-local functionals using the GW approximation.~\cite{VanSetten2007, Ergorenc2018,Haastrup2018,Rasmussen2021,Lee2016, VanSetten2017,Bonacci2023,Biswas2023,Großmann2024}
    The GW method~\cite{Hedin1965, Strinati1982, Onida_2002}, which relies on a direct approximated calculation of the electron self-energy, is widely recognized as the state of the art ab-initio method for calculating excited-state properties.~\cite{Reining2018,Golze2019}
    Within the GW approach, the energies levels can be identified as quasi-particle (QP) excitation energies and provide an improved account of bandgaps and dispersions relations.~\cite{Shishkin2007,Schilfgaarde_2006,vanSetten2015,VanSetten2017,Knight2016}
    Unfortunately, the execution of GW calculations poses technical and numerical challenges that can impact the accuracy of the results.
    The self-energy term displays a slow convergence with respect to the basis-set,~\cite{Shih2010,Friedrich2011,Klimes2014} which can results in under-converged QP gaps and severely increase the computational requirements.
    Moreover, standard implementations exhibit an interdependence between multiple numerical parameters, such as the plane-wave energy cutoff, number of k-points and basis-set dimension. ~\cite{Shih2010,falseConv_2,Klimes2014,Gao2016,ZnOwz_1}
    Therefore, typical convergence procedures must explore a multi-dimensional parameter space, which in turn increases the procedure complexity and the number of preliminary calculations required. In the worst case, not taking into account these dependencies properly may cause false convergence behaviors,~\cite{Shih2010,falseConv_2} which can compromise the accuracy of the QP energies.   
    \newline\newline In this paper, we develop an efficient high-throughput approach for computing accurate QP-energies based on the G$_0$W$_0$ scheme that specifically addresses these challenging aspects. 
    The proposed procedure offers two key advantages: first, it reduces the computational cost of multidimensional convergence procedures by significantly limiting the number of preliminary calculations needed. Second, it aims to achieve high accuracy QP energies by building upon the finite-basis-set correction concept,~\cite{Klimes2014} which identifies specific analytical constraints to correctly account for parameter interdependence.
    The workflow implementation relies on the Vienna Ab Initio Simulation Package (VASP)~\cite{Kresse1996,Kresse1996_2} within the projector augmented wave method (PAW),~\cite{Blochl1994}  integrated in the AiiDA framework~\cite{AiiDA2016,AiiDA2021,AiiDA2021_CommonWorkflow}  through a suitable extension of the AiiDA-VASP plugin.
    The open-source AiiDA platform enables the automation of multi-step procedures, including error handling, with minimal user intervention. Additionally, it has the capability to store the calculations' provenance to ensure reproducibility.~\cite{Mounet2018,Prandini2018,Mercado2018,Atambo2019,Vitale2020,Kahle2020} 
    We note, that the designed procedure is not specific to VASP and can be adapted to other ab-initio codes.
    The capability and efficiency of our HT protocol is demonstrated by the construction of an accurate G$_0$W$_0$ database containing QP gaps of more than 320 materials, easily extendable to contain additional materials. 
    Our database serves as a benchmark for validating the accuracy of the procedure, and the adopted standardized protocol ensures internal consistency among the parameters and pseudopotentials selection. This standardization not only improves reproducibility, but makes the database suitable as a platform for machine-learning purposes and established the dataset as high-quality reference for the QP-energies of the included compounds. 
    \newline\newline The paper is organized as follows. 
    We start by summarizing the basic theoretical aspects of the  GW method and basis-set extrapolation. In the results and discussion section we first report the technical aspects of the implemented workflow and then we discuss the construction of the database and assess the accuracy and the efficiency of the workflow procedure. Technical details of the VASP setup are collected at the end.

\section*{THEORETICAL ASPECTS}\label{sec:TheoreticalAspects}
\subsection*{The GW approximation}\label{section:GW}
   The quasi-particle energies presented in this paper are calculated using perturbative single-shot GW calculations (G$_0$W$_0$),\cite{Aryasetiawan1998, Golze2019,Shishkin2006,Shishkin2007} starting from Kohn-Sham single particle energies $E_{n \textbf{k}}^{DFT}$ and orbitals $\psi_{n \textbf{k}}^{DFT}$
   \begin{equation}
            E_{n \textbf{k}}^{QP} = E_{n \textbf{k}}^{DFT} + Z_{n \textbf{k}} \ev{\Sigma(E_{n \textbf{k}}^{DFT}) - V_{xc}}{\psi_{n \textbf{k}}^{DFT}}
   \end{equation}
   where $Z_{n \textbf{k}}$ is the renormalization factor and $V_{xc}$ the DFT exchange-correlation potential; $n$ and $\textbf{k}$ are the band and k-point indices. The orbitals $\psi$ are expanded on a plane-wave basis-set, associated to a cutoff parameter $G_{cut}^{pw}$. The diagonal elements of the self-energy $\Sigma_{n \textbf{k}}$ are calculated as the sum  of the exact Fock exchange $\Sigma^x_{n \textbf{k}}$ and the correlation term $\Sigma^c_{n \textbf{k}}$
   \begin{align}\label{eq_extrapolation}
        \Sigma^c_{n \textbf{k}}&(\omega)=
        \frac{1}{\Omega}
        \sum_{\textbf{q}}\sum_{m}^{N_{pw}}\sum_{\textbf{G}\textbf{G}'}^{\textbf{G}_{cut}^\chi}
        \frac{i}{2\pi}\int_0^\infty d \omega'
        {W}_{\textbf{q}}(\textbf{G},\textbf{G}',\omega')\nonumber\\
        &\times\rho_{nm}(\textbf{k},\textbf{q},\textbf{G})\rho_{nm}^*(\textbf{k},\textbf{q},\textbf{G}') \nonumber\\
        &\times\left[ 
        \frac{f_{m,\textbf{k}-\textbf{q}}}{\omega-\omega' - E^{DFT}_{m,\textbf{k}-\textbf{q}}-i\eta}+
        \frac{1-f_{m,\textbf{k}-\textbf{q}}}{\omega-\omega' - E^{DFT}_{m,\textbf{k}-\textbf{q}}+i\eta}     
        \right]
   \end{align}
   where ${W}_{\textbf{q}}(\textbf{G},\textbf{G}',\omega')$ is the screened Coulomb interaction, $\rho$ the overlap density $\rho_{nm}(\textbf{k},\textbf{q},\textbf{G}) = \mel{\psi_{n \textbf{k}}}{e^{i(\textbf{q}+\textbf{G})\textbf{r}}}{\psi_{m \textbf{k}-\textbf{q}}}$, $f$ the occupation function, $\Omega$ the cell volume and $\eta$ a positive infinitesimal. $N^{pw}$ defines the number of unoccupied bands included in the sum-over-bands in $\Sigma^c$ and $W$ expression;  $G_{cut}^\chi$ represents the energy cutoff on the response function and screened Coulomb potential elements.
   \begin{figure}[b!]
        \includegraphics[width=\columnwidth]{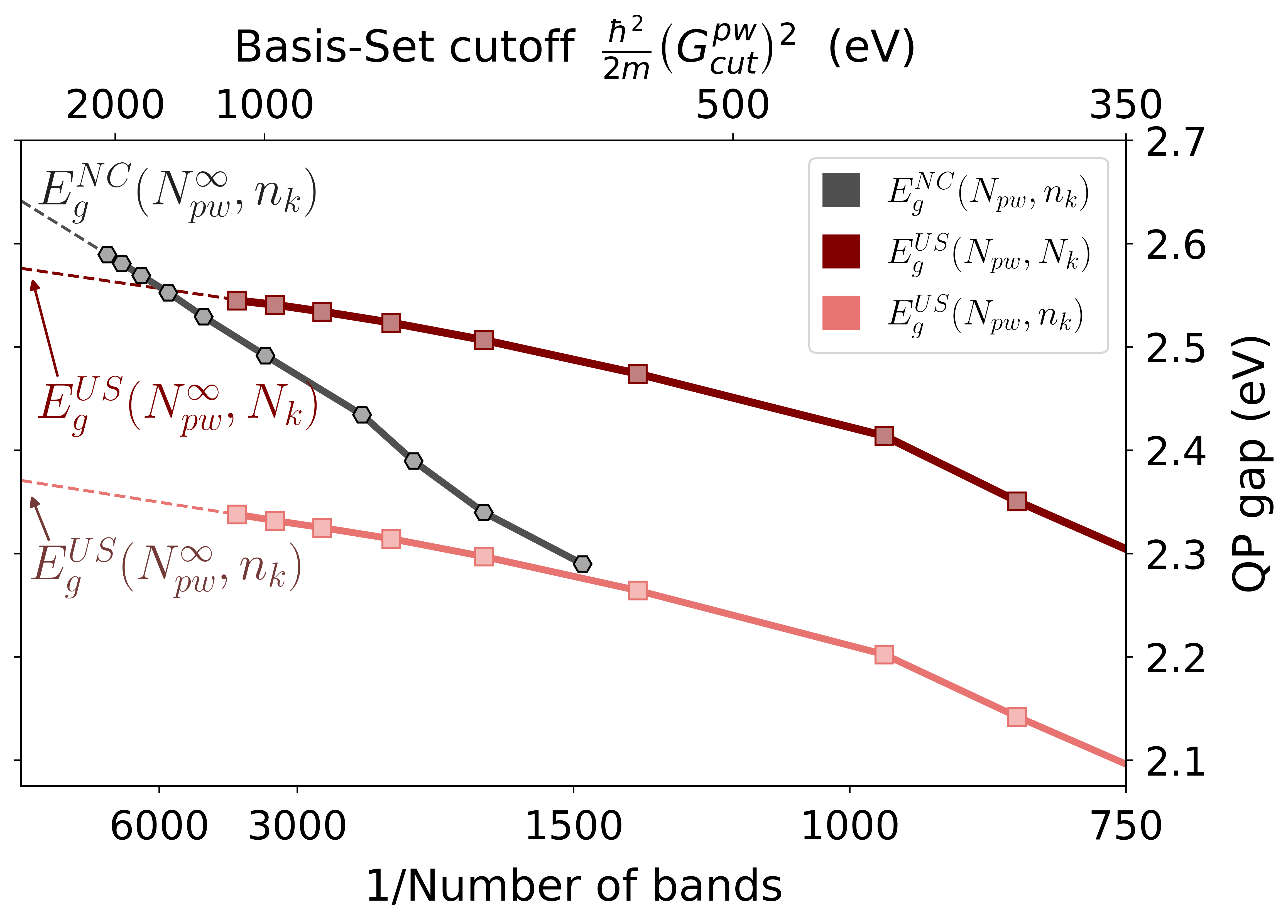}
        \caption{\label{fig:Img_AsymptoticCurves_ZnO} \textbf{ ZnO QP gap as a function of the number of bands $N_{pw}$.} The gap is shown are displayed for NC ($E_g^{NC}$) and ST-PAWs ($E_g^{ST}$), computed on the reduced $n_k \equiv 3\times3\times3$ k-point mesh. For comparison, the curve associated to $E_g^{ST}$ determined on dense $N_k \equiv 6\times6\times6$ k-meshes is also shown.  The corresponding cutoff energies, as defined by the full basis-set constraint, are shown on the upper axis.
        }
    \end{figure} 
\subsection*{Basis-set Extrapolation}\label{section:BSTheory}
    Several works~\cite{Klimes2014,AsymConvergProof1,AsymConvergProof2,AsymConvergProof3,AsymConvergProof4,AsymConvergProof5} proved analytically that  basis-set incompleteness error on the QP energies follow a linear dependence on the inverse number of plane waves
    \begin{equation}
        \Delta E_m \propto \frac{1}{(G^{pw}_{cut})^3 } \sum_{\textbf{g}} \rho_m(\textbf{g}) \rho(-\textbf{g}) \propto \frac{1}{N_{pw}} \sum_{\textbf{g}} \rho_m(\textbf{g}) \rho(-\textbf{g})
        \label{Eq:asymptoticLimit}
    \end{equation}
    where $\rho_m(\textbf{g})$ is the density of the $m$ band in reciprocal space and $\rho(\textbf{g})$ is the total density;
    $\textbf{g}$ is defined as  $\textbf{g} = \textbf{G} - \textbf{G}'$ with $\textbf{G}$ and $\textbf{G}'$  basis vectors of the reciprocal unit cell.
    The expression has been derived under the hypothesis of a complete set of unoccupied orbitals compatible with a given cutoff $G^{pw}_{cut}$ (all orbitals spanned by the finite plane-wave basis-set are included).
    \newline The inclusion of the full finite basis-set for the given $G^{pw}_{cut}$ (hereafter named full basis-set constraint) is crucial in order to avoid false convergence behaviors.
    In fact, the $1 /  N_{pw}$ asymptotic limit can be formally justified \textit{only if} the number of unoccupied bands $N^{pw}$ and both the cutoffs $G_{cut}^{pw}$  and $G_{cut}^{\chi}$ are increased \textit{simultaneously} and with a similar rate.~\cite{Klimes2014}
    This result clarifies why extrapolating with fixed parameters can result in false convergences.
    Crucially, this requirement is ensured by the full basis-set constraint and, moreover, the constraint implies that $N_{pw}$ is controlled by the orbital basis cutoff $G_{cut}^{pw}$.
    \newline\newline A second important factor that can affect the precision of the QP energies is the norm violation of the PAW pseudo-waves,  as pointed out in Refs.~\cite{Klimes2014,VanSetten2017,Golze2019}    
    { The PAW framework has been widely adopted in several popular GW implementations,~\cite{Abinit_Gonze2020,GPAW} thanks to its transferability properties and computational efficiency.~\cite{Golze2019}}
    { In particular, the completeness of the PAW partial-waves is another key assumption underlying} Eq.~(\ref{Eq:asymptoticLimit}).{While the assumption is likely satisfied for low-lying unoccupied states, standard (\textit{non} norm-conserving) PAW} potentials (ST-PAWs) can violate this constraint for the high-energy states included in the $\Sigma$ band summation.~\cite{Klimes2014} This violation  implies that their contributions to the $\rho(\textbf{g})$ density in Eq.~(\ref{Eq:asymptoticLimit}) are not properly described: the consequence is that, while the $1/N_{pw}$ asymptotic behavior still holds, the QP energies converge to an incorrect asymptotic limit.
    {A possible solution to this issue is to employ \textit{norm-conserving} PAW potentials (NC-PAWs), as enforcing norm conservation on the PAW partial-waves strongly mitigates the problem.
    However, this approach comes with a prominent drawback: NC-PAWs require significantly higher plane-wave cutoffs compared to ST-PAWs (up to $40-50\%$).}
    Fig.~\ref{fig:Img_AsymptoticCurves_ZnO} demonstrates this behaviour for the well-studied semiconductor ZnO. The figure displays the dependence of the QP gap versus $1/N_{pw}$ for NC and ST PAWs. While the band-gaps computed on a reduced k-point mesh $n_k \equiv 3\times3\times3$ with NC and ST-PAWs display similar values for $\sim1500$ bands, beyond that threshold they converge towards significantly different limits. Increasing the k-meshes to $N_k \equiv 6\times6\times6$ improves the ST prediction.

\section*{RESULTS AND DISCUSSION}\label{sec:Results}
    Our GW workflow achieves converged QP energies through the inclusion of two correction terms. These terms account for (i) the error committed by truncating the band summations in the self-energy and (ii) for the error related to the norm violation of the ST-PAW. The implementation details and their advantages are discussed in the next two subsections.

\subsection*{Basis-set Incompleteness correction}\label{Section:BSError}
    The protocol described in this section aims to estimate the basis incompleteness error associated to the QP-energies  $E_{QP}(N_{pw}^{(1)},N_k)$ determined for a given number of bands $N_{pw}^{(1)}$ and on a k-point mesh $N_k$.
    To estimate the error, { several (up to a maximum of 4 in our implementation)} G$_0$W$_0$ calculations are executed, and the QP energies are extrapolated with respect to the basis-set size by fitting Eq.~(\ref{eq_extrapolation}).
    The extrapolation is performed under two conditions: first, the computational parameters are increased simultaneously at the same rate between G$_0$W$_0$ runs, as discussed in the previous section. The response function cutoff $G_{cut}^\chi$ is determined as $G_{cut}^\chi = \frac{2}{3} G_{cut}^{pw}$, while the number of bands $N_{pw}$ is constrained by assuming a full basis for a given cutoff $G_{cut}^{pw}$. 
    Therefore, only $G_{cut}^{pw}$ is an independent parameter, while $N_{pw}$ and $G_{cut}^\chi$ are defined by the orbital cutoff: this crucially reduces the dimensionality of the parameter space that the workflow must explore while ensuring an accurate extrapolation. We emphasize that this represents a first important advantage with respect to conventional convergence procedures, which must sample multidimensional parameter spaces:~\cite{Bonacci2023,Biswas2023,Großmann2024}{ By minimizing the number of independent parameters, this approach allows for a efficient extrapolation strategy which avoids extensive parameter sweeps.}
    \newline Secondly, the convergence behavior has been proved to be insensitive~\cite{Klimes2014,Ergorenc2018,VanSetten2017} to the k-point density used: the extrapolation to the asymptotic limit is therefore performed on a reduced k-point grid $n_k$ and the errors due to truncation at $N_{pw}^{(1)}$ bands are estimated.
    Finally, the QP energies on the denser k-mesh $E_{QP}(N_{pw}^{(1)},N_k)$ can be corrected for the basis-set completeness error (labeled $\Delta BS$)~\cite{SOC_Gruneis,Klimes2014}
        \begin{multline}\label{eq:bsExtraKlimes}
        		E_{QP\text{-}\infty}(N_{pw}^{\infty},N_k) \approx E_{QP}(N_{pw}^{(1)},N_k) \\ + \underbrace{[E_{QP\text{-}\infty}(N_{pw}^{\infty},n_k) - E_{QP}(N_{pw}^{(1)},n_k)]}_\text{$\Delta BS$} 
        \end{multline}
        where $E_{\infty}(N_{pw}^{\infty},N_k)$ corresponds to the QP energies for infinite bands $N_{pw}^{\infty}$ and high density k-point mesh $N_k$. $N_{pw}^{(1)}$ and $n_k$ indicate respectively the sparse k-mesh and finite number of bands; $\Delta BS$ represents the estimate of the basis-set incompleteness error.
    \newline  Lastly, we note that the approach offers an additional computational advantage: the G$_0$W$_0$ calculation on the dense $N_k$ k-mesh can be executed with a reduced basis set for the $N_{pw}$ parameter which, depending on the system, can correspond to a significantly lower number of bands  than that obtained through  conventional convergence schemes.  In such cases, the $\Delta BS$ correction will result in correspondingly larger values to account for the discrepancy. 

    \begin{figure*}[th]
        \includegraphics[width=0.95\linewidth]{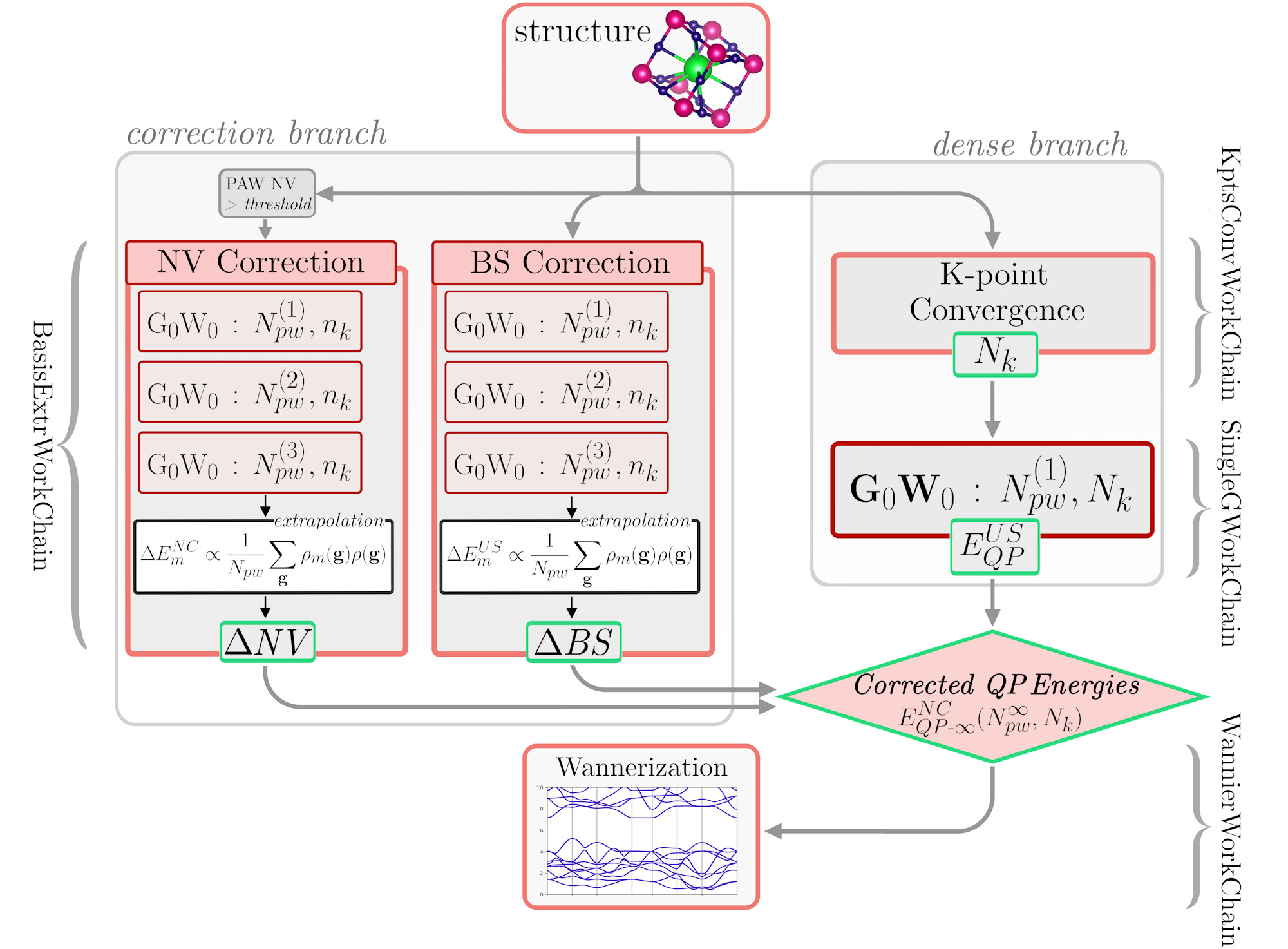}
        \caption{\textbf{Flowchart of the AiiDA \texttt{VaspGWorkChain} workflow. } 
        The structural data represent the only required input. For improved clarity, the workflow is organized into two branches, which are run in parallel. The main output of each step is shown in the green boxes, while the corresponding workchains are listed on the sides.
        In the dense branch, a k-point mesh convergence is first performed by \texttt{KptsConvWorkChain}, which returns the converged k-mesh $N_k$.
        Subsequently, an instance of \texttt{SingleGWorkChain}  is invoked to compute the QP energies on the k-mesh $E_{QP}(N_{pw}^{(1)},N_k)$.
        For the correction branch, the \texttt{BasisExtrWorkChain} is called to estimate the basis-set incompleteness error $\Delta BS$ on $E_{QP}(N_{pw},N_k)$. If the ST-PAW of the included elements possess a norm violation beyond a given threshold, a second \texttt{BasisExtrWorkChain} is called in parallel in order to estimate the Norm-Violation Error $\Delta NV$.  The resulting QP energies are then corrected and stored in the database. A Wannerization procedure is finally performed to interpolate the band-structure.}
        \label{fig:workflowProcedure}
    \end{figure*}

\subsection*{Norm-Violation correction}
    It has been noted\cite{Klimes2014,Ergorenc2018} that the basis-set extrapolation with ST-PAW potential often converges to a wrong limit, due to the incompleteness of the partial-waves.  Norm-conserving PAWs provide the most accurate extrapolations achievable within the PAW framework, albeit at the expense of increased computational resource cost. Therefore, employing exclusively NC-PAWs throughout the entire procedure could effectively resolve the issue.   However, computing the QP energies  on the dense $N_k$ k-mesh with the significantly harder cutoffs imposed by NC-PAW potentials can represent an additional notable computational bottleneck.
    \newline The Norm-Violation (NV) correction aims to restore the accuracy while limiting the additional computational cost by computing $E_{QP}(N_{pw}^{(1)},N_k)$ with the ST-PAWs and introducing an additional corrective term to restore the convergence of the basis-set correction to the precise NC value. This term represents the error of the basis-set extrapolated QP energy with ST-PAWs $ E^{ST}_{QP\text{-}\infty}(N_{pw}^{\infty},n_k)$ with respect to the reference NC ones $E^{NC}_{QP\text{-}\infty}(N_{pw}^{\infty},n_k)$ and is determined on the sparse k-mesh, i.e: 
    \begin{equation*}
        \Delta NV = E^{NC}_{QP\text{-}\infty}(N_{pw}^{\infty},n_k) - E^{ST}_{QP\text{-}\infty}(N_{pw}^{\infty},n_k)
    \end{equation*}
    This additional correction is then incorporated into the extrapolated QP energies on the converged k-mesh to compensate for the error:
    \begin{equation}
        E^{NC}_{QP\text{-}\infty}(N_{pw}^{\infty},N_k) \approx E^{ST}_{QP\text{-}\infty}(N_{pw}^{\infty},N_k) + \Delta NV
    \end{equation}

\subsection*{The AiiDA-VASP G$_0$W$_0$ workflows}
    The automation of the procedure described has been achieved through the development of a \texttt{VaspGWorkChain} workflow based on the AiiDA framework~\cite{AiiDA2016, AiiDA2021} and on the AiiDA-VASP plugin.~\cite{AiiDAVASP} The plugin provides the interface with the VASP software.~\cite{Kresse1996,Kresse1996_2} The plugin supports DFT and post-DFT ab-initio calculations (DFT+U, Hybrids and G$_0$W$_0$), spin-orbit and structural relaxation calculations, as well as optical routines (within the independent particle approximation). Furthermore, it includes error-handling routines for the most common errors.
    A general overview of the workflow flowchart and layout is illustrated in Figure~\ref{fig:workflowProcedure}; additional details regarding its main components will be described below.     
    \newline The workflow requires only the material's structure as mandatory input from the user; the PAWs and the k-point mesh are automatically selected. The preparation of the ab-initio DFT and GW inputs, submission to high-performance computing clusters, results parsing and storage are  handled internally by the software.
    The workflow proceeding is structured for clarity into two different branches. In the so-called dense branch (see Fig.~\ref{fig:workflowProcedure}) first the k-point convergence and then \textit{single} calculation on the dense k-point mesh $N_k$ are performed.
    Concurrently, in the correction branch both the basis-set incompleteness errors associated with this G$_0$W$_0$ run and the norm-violation errors (if needed) are estimated.
    \newline The main outputs of the workflow are the QP energies $E_{n \textbf{k}}^{QP}$ on the dense k-mesh $N_k$, including corrections to account     the basis-set incompleteness and, if required, norm violation errors.
    Additionally, the workflow can perform an automatic Wannierization of the QP band-structure; in this case, the Wannier-interpolated band-structure is additionally returned by the \texttt{VaspGWorkChain}. All outputs, together with their detailed provenance, are automatically stored in the AiiDA database. 
    \begin{figure}[b!]
        \includegraphics[width=\columnwidth]{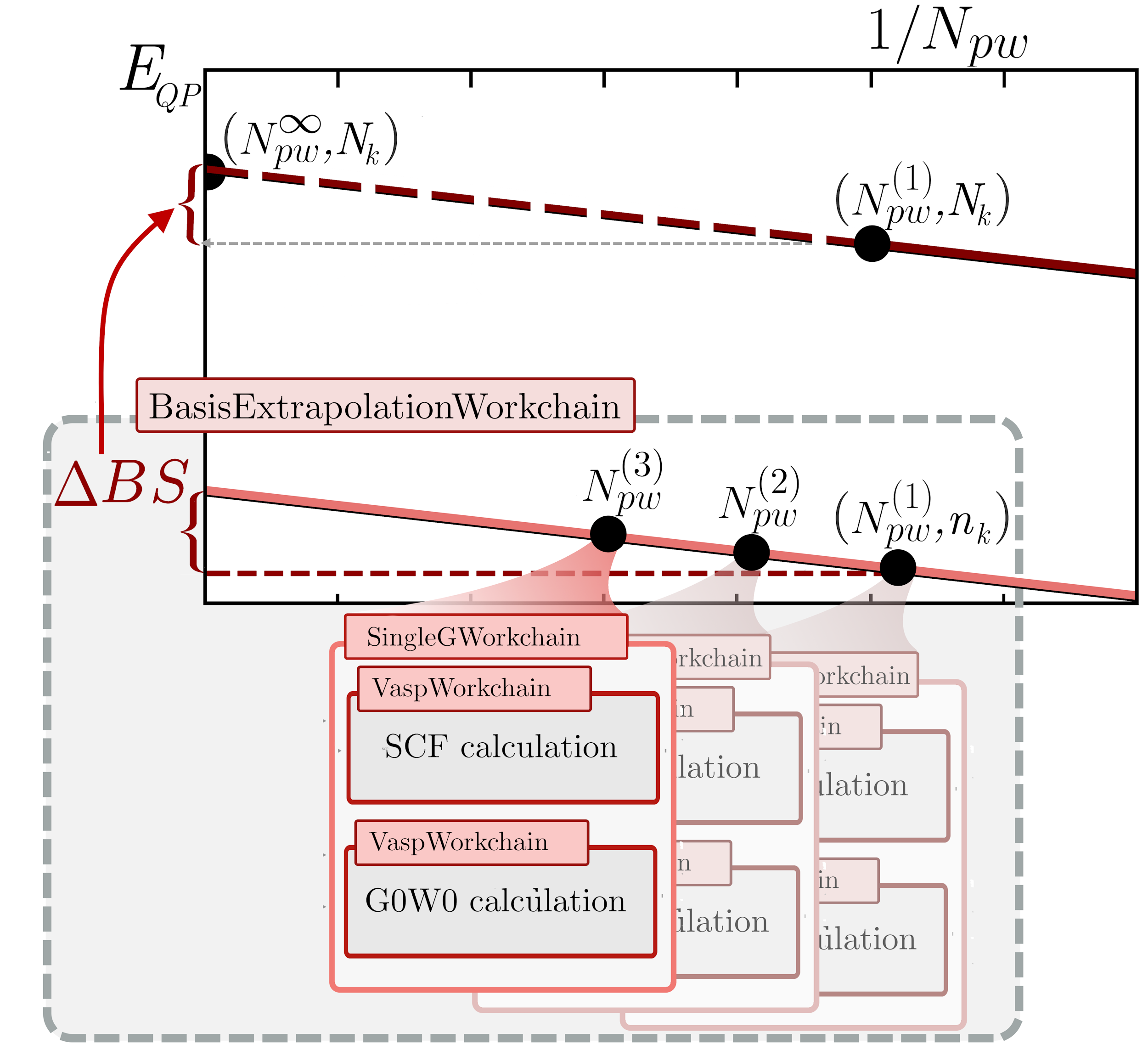}
        \caption{\label{fig:Img_basisSetExtrapolation} \textbf{Workflow scheme of the Basis Extrapolation workchain and its sub-workflows.} The higher-level \texttt{BasisExtrWorkChain} launches three instances of \texttt{SingleGWorkChain} to compute the QP energies on the low k-point density ($n_k$) for the parameters $N_{pw}^{(1)} , N_{pw}^{(2)}$ and  $N_{pw}^{(3)}$. The asymptotic limit is extrapolated and the correction $\Delta BS$, which is the main output of the \texttt{BasisExtrWorkChain} is determined and returned. Finally, the QP-energies on the dense k-point grid $N_k$ are corrected on dense k-point grid ($N_k$). }
   \end{figure} 
   \newline\newline The main workflow  does not launch directly any ab-initio calculation, but prepares the inputs and calls the different sub-workflows, each dedicated to a specific purpose:
   \begin{itemize}   
        \item K-point Convergence: \textbf{\texttt{KptsConvWorkChain}}. It automatizes the convergence of the k-point mesh with respect to the direct and indirect QP  gaps and returns the dense k-point mesh $N_k$.
        \item $\Delta BS$, $\Delta NV$  corrections: \textbf{\texttt{BasisExtrWorkChain}}. It automatizes the computation of $E^{ST}_{QP\text{-}\infty}(N_{pw}^{\infty},n_k)$ and of the corresponding BS corrections $\Delta BS$. The workchain provides a higher level interface to the upper level logic, requiring as main inputs only the structure and the PAW choice. A schematic flowchart of the workflow is outline in Fig.~\ref{fig:Img_basisSetExtrapolation}.
        \newline The main \texttt{VaspGWorkChain} may call a second \texttt{BasisExtrWorkChain} instance and pass the NC-PAWs as inputs in order to compute $E^{NC}_{QP\text{-}\infty}(N_{pw}^{\infty},n_k)$ and the corresponding NV corrections. By default, this correction is skipped unless the structure contains elements whose ST-PAWs exhibit non-negligible norm violations (the threshold is defined at $20\%$ for the \textit{d} partial-waves). The choice is performed automatically at runtime.
        \item QP eigenvalues on the converged (dense) k-point mesh $E_{n \textbf{k}}(N_{pw}^{(1)},N_k)$: \textbf{\texttt{SingleGWorkChain}}. This sub-workflow is an abstraction layer representing a complete G$_0$W$_0$ run.
        The workflow defines the logic needed to compute the QP energies for  a specific structure, k-point mesh, and a specific number of bands and cutoff. It executes internally two different ab-initio simulations, the actual G$_0$W$_0$ calculation and the starting point DFT simulation, and returns the corresponding QP eigenvalues. A single \texttt{SingleGWorkChain} is launched to determine the QP energies $E^{ST}_{QP}(N_{pw},N_k)$, on the dense k-mesh with ST-PAW potentials. The $N_{pw}$, $G_{cut}^{pw}$  parameters are selected as the less computationally expensive pair employed within the \texttt{BasisExtrWorkChain}.
        \item Wannier-interpolation:  \textbf{\texttt{WannierWorkChain}}. As a last step, this workchain can be called to interpolate the obtained QP energies through a Wannierization procedure to generate a QP band-structure along high-symmetry k-point directions.
    \end{itemize}
    We note that the \texttt{BasisExtrWorkChain} and \texttt{KptsConvWorkChain} both internally call \texttt{SingleGWorkChain} to perform the G$_0$W$_0$ runs. 
    Furthermore, the lower level workflow is represented by \texttt{VaspWorkchain}, a core part of the AiiDA-VASP plugin which handles the actual ab-initio simulations on the remote high-performance clusters. It serves as a wrapper of a single generic VASP simulation and manages directly the construction of the VASP-specific files (INCAR, POSCAR, etc.), the submission and the output parsing.
   \newline\newline Beside the structural data, which represents the only mandatory inputs, the workchain accepts several other optional input parameters, which allow the user to override the default behavior of the workchain.
    \begin{itemize}
        \item \textit{Deactivate\_NVcorrection}: Disable the NV correction.
        \item \textit{Selected\_Kpoint\_mesh}: If this input is provided, the k-points convergence procedure is bypassed, and the provided k-mesh is utilized as the dense $N_k$ mesh.
        \item \textit{Selected\_Mode}: Enforce a specific protocol for determining the corrections.
        \item \textit{Kpts\_convergence\_threshold}:  threshold value used to determine k-point convergence; the default value is set to 50 meV.
        \item \textit{Deactivate\_Wannierization}: Skip the last Wannier interpolation step, activated by default.
        \item \textit{Kpts\_Wannierization\_spacing}: upper limit on the k-point spacing of a uniform grid employed for the Wannierization procedure; following Vitale \textit{et al}, the default is set to $\rho_{\textbf{k}} = 0.2 \text{\AA}^{-1}$.
    \end{itemize}
    In the next sections, the main sub-workflows and their algorithms will be discussed in detail, starting from the workflow computing the Basis-Set Incompleteness error.

\subsubsection{Basis Extrapolation workchain}
    The algorithm encoded in the workchain is based on Eq.~(\ref{eq:bsExtraKlimes}), and aims to extrapolate $E_{QP\text{-}\infty}(N_{pw}^{\infty},n_k)$ and the correction $\Delta BS = E_{QP\text{-}\infty}(N_{pw}^{\infty},n_k) - E_{QP}(N_{pw}^{(1)},n_k)$. We describe below its architecture:
    \begin{itemize}
        \item Three G$_0$W$_0$ calculations are performed in parallel with different plane-wave cutoffs and band numbers, denoted respectively as ${G_{cut}^{pw}}^{(1)},{G_{cut}^{pw}}^{(2)},{G_{cut}^{pw}}^{(3)}$
        and $N_{pw}^{(1)}, N_{pw}^{(2)}$ and $N_{pw}^{(3)}$.
       \item The cutoff of the first G$_0$W$_0$ data point is determined as the maximum energy cutoff (\texttt{ENMAX} tag) given in the pseudo-potentials, labeled $G_{pw}^{PAW}$, i.e. 
       $G_{pw}^{(1)} = G_{pw}^{PAW}$. $N_{pw}^{(1)}$ is determined by the full basis-set constraint.
       \item The parameters of the subsequent two G$_0$W$_0$ data points are chosen in order to progressively increase the number of bands in steps of $0.20 \times N_{pw}^{(1)}$, i.e.  $N_{pw}^{(2)}=1.2\times N_{pw}^{(1)}$ and $N_{pw}^{(3)}=1.4\times N_{pw}^{(1)}$. The corresponding ${G_{cut}^{pw}}^{(2)},{G_{cut}^{pw}}^{(3)}$ are defined by the full basis-set constraint.
       \item The workflow performs a first extrapolation to compute $E^{ST}_{QP\text{-}\infty}(N_{pw}^{\infty},n_k)$ (or $E^{NC}_{QP\text{-}\infty}(N_{pw}^{\infty},n_k)$, depending on the PAWs used). If the $R^2$ determination coefficients of the linear fits exceed a predefined threshold (with a default 0.85, adjustable by the user), the extrapolations are deemed accurate and the extrapolated $E_{QP\text{-}\infty}$ values are returned. In cases where the condition is not satisfied, an additional fourth G$_0$W$_0$ calculation is performed with $N_{pw}^{(4)} = 1.6\times N_{pw}^{(1)}$ and the  fits are updated.
    \end{itemize}
    This protocol represents a computationally efficient alternative to the one introduced by Ref.~\cite{Klimes2014}, where the G$_0$W$_0$ calculations were performed employing  a considerably larger number of bands, up to $N_{pw}^{(3)} \sim\hspace{-0.09cm}2.0\times N_{pw}^{PAW}$.  
    This choice is adopted as the default scheme. However, the application of this protocol to larger cells or supercells can be computationally demanding, as large cells are typically associated to a denser bandstructure which can potentially greatly increase the number of bands that need to be considered for the same $G_{cut}$. Therefore, we have introduced an alternative memory-conserving variant, which is suited for rapid screenings or for materials with large cells. 
    \newline In the \textit{memory-conserving} scheme three separate calculations with ${G_{cut}^{pw}}^{(1)},{G_{cut}^{pw}}^{(2)},{G_{cut}^{pw}}^{(3)}$ defined by $ 0.75 \times G_{pw}^{PAW}  , 1.00 \times G_{pw}^{PAW}, 1.25 \times G_{pw}^{PAW}$ are used for the extrapolation. The memory variant is active (default choice) for volume larger than $150 \text{\AA}^3$.

\subsubsection{K-point convergence workchain}
    The \texttt{KptsConvWorkChain} finds the minimally dense k-point mesh which achieves convergence of the QP direct and indirect gaps within a given convergence tolerance. The search is restricted to uniform k-point meshes which include the high-symmetry k-points of the irreducible Brillouin Zone.
    To maintain computational efficiency, the workflow leverages the decoupling between the convergences of basis-set dimension and the k-point mesh density: the G$_0$W$_0$ calculations used to determine the converged k-point mesh are performed with a fixed and reduced basis-set dimension defined by $G_{cut}^{pw – kptsConv} = 0.70 \times G_{pw}^{PAW}$; the corresponding $N_{pw}$ is determined by the full basis-set constraint.
    \newline When the Wannierization procedure is enabled, an additional constraint must be taken into account: as outlined by Vitale \textit{et al},~\cite{Vitale2020} the automated Wannierization procedure requires a single input parameter, namely the k-point spacing of the uniform grid employed for the Wannierization procedure. 
    Their work demonstrated how this parameter significantly impacts the accuracy of the resulting Wannier-interpolated band-structure, and noted how a spacing of $\rho_{\textbf{k}\text{-min}} = 0.2 \text{\AA}^{-1}$ is sufficient for achieving interpolations with errors less than 20 meV.
    This consideration is incorporated as an additional constraint on the convergence procedure. The minimum k-point spacing for the Wannier interpolation is taken by the workchain as an optional argument; the k-point meshes considered during the search are restricted to grids with a spacing equal or higher than $\rho_{\textbf{k}\text{-min}}$.

\subsubsection{Wannierization workchain}
\begin{figure}[bt!]
        \includegraphics[width=\columnwidth]{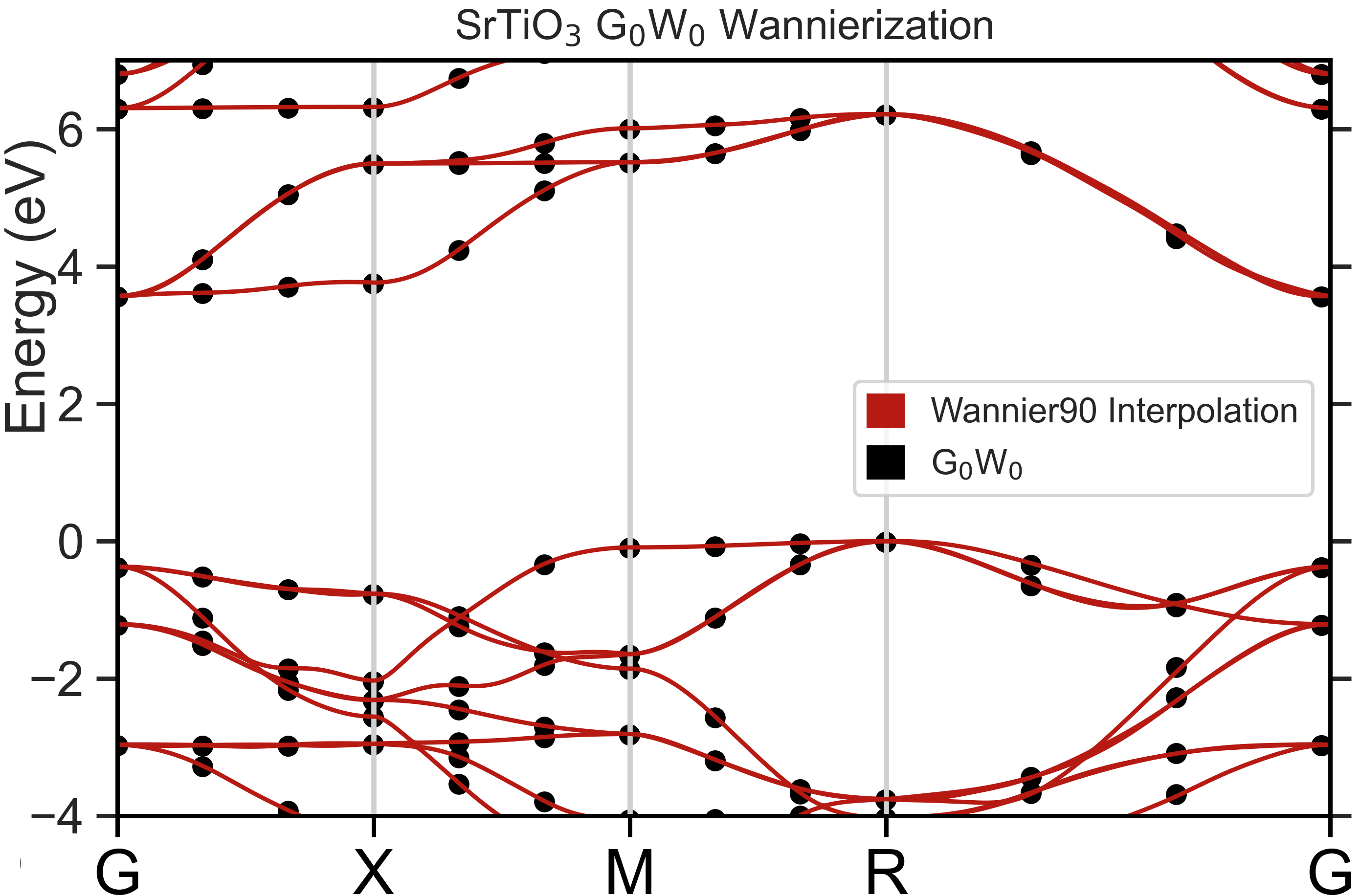}
        \caption{\label{fig:Img_Wannier} 
        \textbf{Automatic Wannier interpolation of the G$_0$W$_0$ band-structure.}
        The G$_0$W$_0$ band-structure of cubic SrTiO$_3$ is
        automatically wannierized through the SCDM scheme and 
        plotted alongside the G$_0$W$_0$ eigenvalues.}
    \end{figure}

    The interpolation of the GW band-structure through an automatic Wannierization is performed by the \texttt{WannierWorkChain} sub-workflow.
    The starting projections are automatically determined using the selected columns of the density matrix (SCDM).~\cite{SCDM_1,SCDM_2}
    \newline This sub-workchain executes a VASP simulation (via the \texttt{VaspWorkchain}) with Wannier90~\cite{Wannier_Pizzi_2020} and its VASP interface~\cite{Franchini2012} starting from the wavefunctions on the dense k-point mesh $N_k$ determined in the previous steps. An example of Wannierization obtained from this workflow is displayed in Fig.~\ref{fig:Img_Wannier}.
    \begin{table*}[t]
    \newcolumntype{Y}{>{\centering\arraybackslash}X}   
    \newcolumntype{s}{>{\hsize=.025\hsize}X}    
    \begin{tabularx}{0.95\textwidth}{llsYYsYYsYY}
    \cline{1-11}
      &&
      &   \multicolumn{2}{c}{$E_{QP\text{-}\infty}^{NC} - E_{QP}^{ST}(N_{pw}^{(1)})$} 
      &&  \multicolumn{2}{c}{$E_{QP\text{-}\infty}^{NC} - [E_{QP}^{ST}(N_{pw}^{(1)})\text{+}\Delta BS]$} 
      &&  \multicolumn{2}{c}{$E_{QP\text{-}\infty}^{NC} - [E_{QP}^{ST}(N_{pw}^{(1)})\text{+}\Delta BS\text{+}\Delta NV]$} 
      \\
    Systems       &$N_{pw}^{(1)}$ && VBM (meV)   & CBM (meV)    && VBM (meV)    & CBM (meV)   && VBM (meV)     & CBM (meV)    \\ 
    \cline{2-2}\cline{4-5}\cline{7-8}\cline{10-11}
    AlAs           & 910           && -164         & -173         && -119          & -115         && 1             & -2         \\
    AlP            & 820           && -77          & -88          && -18           & -19          && -14           & -7         \\
    AlSb           & 1110          && -123         & -128         && -64           & -57          && 1             &  -2        \\
    CdO            & 580           && -179         & -28          && -113          & -22          && -6            & 2          \\
    CdS            & 860           && -95          & -97          && 20            & -47          && -3            & 1          \\
    CdSe           & 1300          && -325         & -189         && -258          & -176         && -43           & -21        \\
    CdTe           & 1150          && -231         & -135         && -105          & -66          && 5             & 8          \\
    GaAs           & 910           && -204         & -221         && -163          & -171         && 2             & 5          \\
    GaP            & 760           && -103         & -134         && -36           & -58          && 35            & 3          \\
    GaN            & 960           && -212         & -199         && -26           & -80          && 50            & -20        \\
    InP            & 860           && -138         & -118         && -37           & -102         && -4            & -2         \\
    InSb           & 1150          && -166         & -100         && -78           & -47          && 1             & 6          \\
    MgO            & 430           && -361         & -274         && -169          & -165         && -26           & -16        \\
    SiC            & 600           && -96          & -86          && -12           & -32          && -9            & -11        \\
    ZnS            & 760           && -269         & -97          && -125          & -36          && 4             & -1         \\
    ZnSe           & 1100          && -328         & -180         && -268          & -174         && 5             & 7          \\
    ZnTe           & 1060          && -302         & -93          && -191          & -23          && 21            & 9          \\ \cline{1-11}
    \textbf{MAE}   &               && \textbf{198} & \textbf{138} && \textbf{106}  & \textbf{82}  && \textbf{14}   & \textbf{7}  \\
    \textbf{$\sigma$}&             && \textbf{91}  & \textbf{61}  && \textbf{82}   & \textbf{58}  && \textbf{16}   & \textbf{6}  \\ \botrule
    \end{tabularx}
    \label{tab:ValidationAgainstNC}
    \caption{\textbf{Energy differences from the reference basis-set extrapolated QP energies $E_{QP\text{-}\infty}^{NC}$ and impact of the corrections.}
    The table illustrates how the $\Delta BS$ and  $\Delta NV$ corrections effectively approximate the reference quasiparticle energies for a group of 19 typical semiconductors and insulators.
    The third and fourth columns list the deviations from the reference for $E_{QP}^{ST}(N_{pw}^{(1)},N_k)$ at the VBM and CBM $\Gamma$ k-point, as determined in the dense branch of the workflow in Fig.~\ref{fig:workflowProcedure}. The inclusion of $\Delta BS$ lowers the differences by $\sim 40\%$ on average, while the combined corrections reduce it by more than 
    $90\%$.}
\end{table*}

\subsection*{Validation against ab-initio references}\label{sec:VerificationOfTheProcedure}
    This section is dedicated to the validation of the protocol implemented the computational workflow: the main goal is to verify that the application of the NV and BS corrections can reliably reproduce the target quasi-particle energies $E_{QP-\infty}^{NC}(N_{pw}^\infty,N_k)$.
    For this purpose, a reference dataset comprising the basis-set extrapolated $E_{QP-\infty}^{NC}(N_{pw}^\infty,N_k)$ for 19 typical~\cite{Shishkin2007_Vertex, Schilfgaarde_2006, ElPhonRen_5,STO_2} group III–VI semiconductors and insulators are determined through a careful extrapolation performed directly on the dense $N_k$ grid (thus without the need of $\Delta BS$ and $\Delta NV$). 
    The results of the automated workflow for the same dataset are benchmarked against these reference $E_{QP-\infty}^{NC}$, and the differences (with and without the inclusion of the corrections) are compiled in Tab.~\ref{tab:ValidationAgainstNC} for valence band minimum (VBM) and conduction band maximum (CBM) at the $\Gamma$ k-point. The complete QP-energies for the considered set are compiled in Table S3 and S4 in the SM.
    The inclusion of both corrections inside the workflow achieves a remarkable agreement with the reference QP energies, exhibiting a mean absolute error (MAE) of $\sim\hspace{-0.09cm}15$ meV for the VBM and less than $10$ meV for the CBM; this represents a sizeable error reduction compared to the data without the corrections, associated to MAEs of $\sim\hspace{-0.09cm}200$ meV;
    The basis-set corrections alone accounts for the $45-50\%$ of the error reductions for the considered dataset. However, the QP energies on ST-PAW potentials $E_{QP}^{ST}(N_{pw}^{(1)})\text{+}\Delta BS$ still display a residual underestimation, in particular for group V and VI compounds (with ZnTe, MgO, InAs, GaAs and AlAs at $\sim\hspace{-0.09cm}−150$ meV average) and markedly high for the ZnSe  and CdSe VBMs.
    The inclusion of the NV correction successfully improves over the remaining errors, reducing the deviation from the reference $E_{QP-\infty}^{NC}$ to an average of  under 20 meV.
    The proposed protocol is therefore able to reproduce the highly precise reference results with a reduced computational cost and without need of direct user interventions. 
     
    \begin{figure*}[hbt]
        \includegraphics[width=\textwidth]{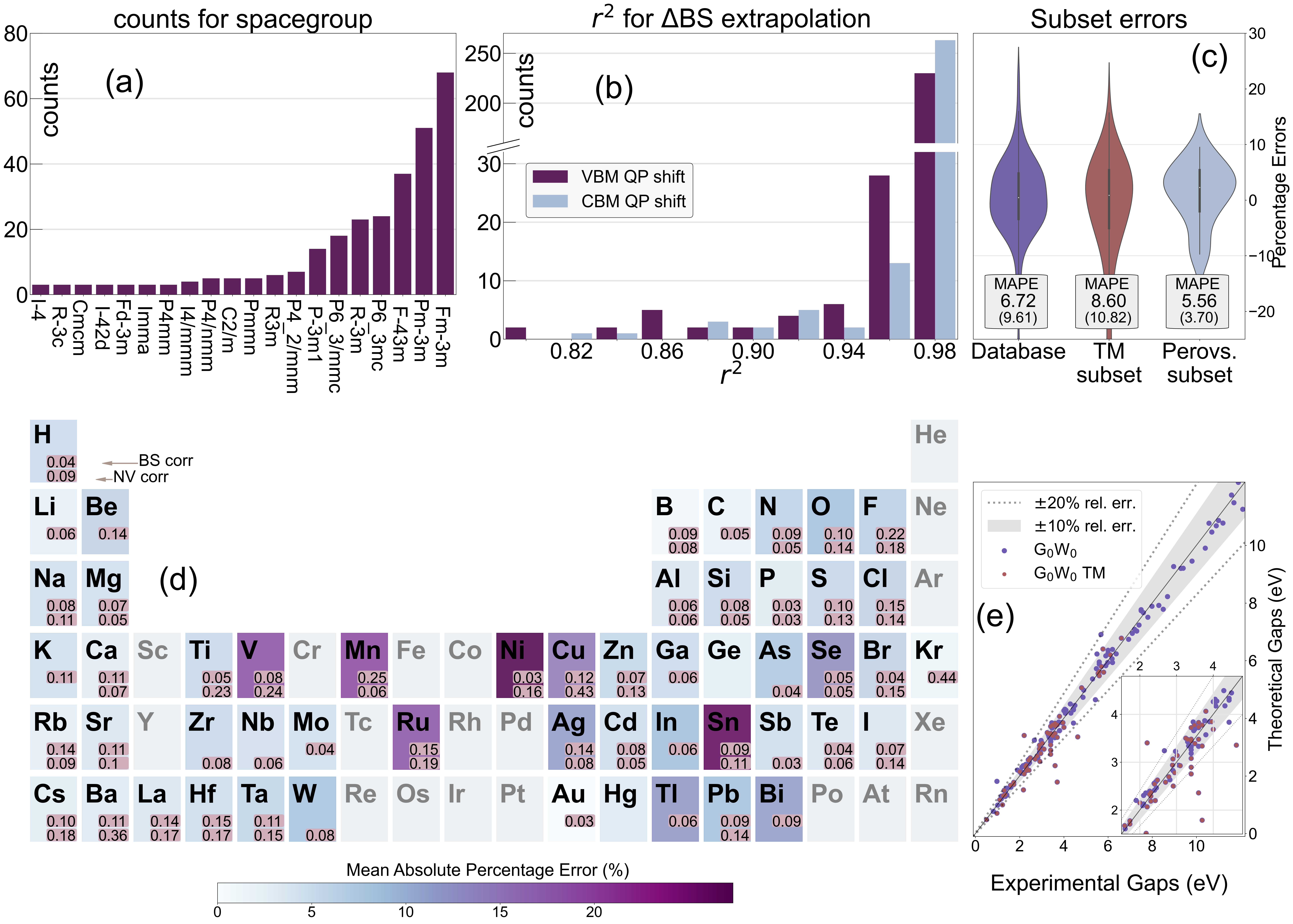}
        
        \caption{    \label{fig:datasetStatistics}
        \textbf{G$_0$W$_0$ Database statistics.}
        \textbf{(a)} Distribution of compounds across the twenty most populated space-groups in the database.
        \textbf{(b)} Histogram showing the distribution of $r^2$ determination coefficients for the extrapolations performed for the basis-set correction $\Delta BS$, for the valence band maximum and conduction band minimum.
        \textbf{(c)} Violin plots of the percentage errors with respect to experiments for the entire G$_0$W$_0$ database and for the two subsets comprising all TM compounds and perovskite systems.  The MAPEs and their corresponding standard deviations for each dataset are displayed within the boxes.
        \textbf{(d)} For each element the MAPE across all compounds containing that element (relative to experimental bandgaps) is displayed. The average Basis-Set (BS) and Norm-Violation (NV) corrections for each element are also included (shown only for values exceeding $0.03$ eV for clarity).
        \textbf{(e)} Comparison between experimental and extrapolated G$_0$W$_0$ bandgaps; the TM points are highlighted in a different color. The shaded area and the dashed lines identify regions where errors are below $10\%$ and $20\%$, respectively.
        }
    \end{figure*} 
\subsection*{The G$_0$W$_0$ Database}\label{sec:results_ComparisonKlimes}
\begin{figure}[b]
        \includegraphics[width=0.49\textwidth]{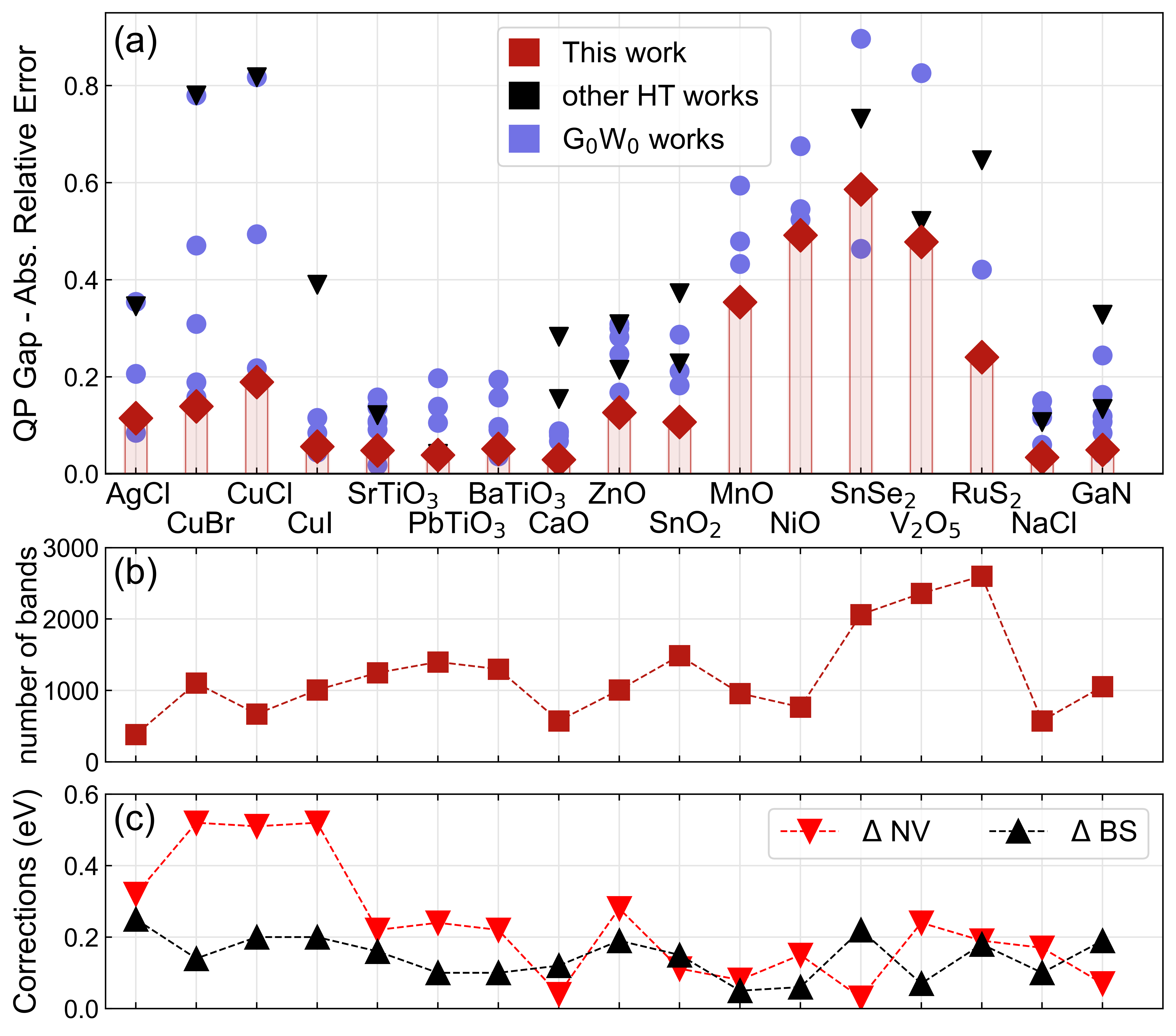}
        \caption{\textbf{(a)} Comparison of absolute relative error with respect to experimental bandgaps for the benchmark set.
        This subset includes compounds exhibiting substantial dependencies on the number of bands or large errors in previous HT studies.~\cite{VanSetten2017,Bonacci2023,HTReference_abedi2022benchmark}
        The G$_0$W$_0$ gaps determined via the BS and NV corrections are depicted in red, alongside results from other HT studies (black),~\cite{VanSetten2017,Bonacci2023,HTReference_abedi2022benchmark} and G$_0$W$_0$@PBE values from references~\cite{Gao2016, AgCl_1, AgCl_2, CuCl_1, CuBr_1, CuI_1, CaO_1, CaO_2, Schilfgaarde_2006, CaO_3, Friedrich2011, falseConv_2, ZnO_GW_1, ZnO_GW_2, ZnO_GW_3, SnO2_1, SnO2_2, SnO2_3, NaCl_1, NaCl_2, NaCl_3, NaCl_4, RuS_1, V2O5_1, SnSe2_1, SnSe2_2, GaN_1, STO_AbInitio_test_1, GaN_2, GaN_4, GaN_5, STO_BTO_PhysRevB.103.035128, BTO_1, BTO_2, SeMa_STO_1, PbTiO3_1, PbTiO3_2, PbTiO_3, Ergorenc2018, STO_1, STO_2, STO_3, STO_4, MnO_1,MnO_2,MnO_3,NiO_1} (blue).
        \textbf{(b)} Number of bands $N_{pw}^{(1)}$ included in the calculation on the dense k-mesh. \textbf{(c)} values of the Basis-Set and Norm-Violation corrections.}
        \label{fig:nontrivialMat}
    \end{figure}

    The workflow has been used to generate a database of QP gaps and energies comprising 327 distinct bulk structures. This database stands out, {to the best of our knowledge, as one of the largest G$_0$W$_0$ datasets for bulk compounds available.}
    It encompasses $\sim 220$ binary and $\sim 100$ ternary compounds, covering a gap range from 0.7 eV  to 14 eV and containing $\sim 40$ diverse space groups (see Fig.~\ref{fig:datasetStatistics} a) for a distribution of the most represented spacegroup in the database). The full list of materials, including the predicted G$_0$W$_0$ gaps, the $\Delta BS$ and $\Delta NV$ corrections, is given in Table S5 in the SM. 
    {This expanded dataset enables a comprehensive evaluation of the accuracy of the predicted results, specifically assessing the performance of NV and BS corrections against experimental references (which were identified for 163 systems) and comparing them with prior G$_0$W$_0$ literature data.}    
    \newline  In order to ensure consistency with previous works,~\cite{Klimes2014,Ergorenc2018} we adopt a $\Gamma$-centered $6\times6\times6$ grid as the dense k-point mesh $N_k$ for the database.
    This grid choice has been established as converged up to around $\sim\hspace{-0.09cm}50$ meV for insulators~\cite{VanSetten2017,Shishkin2006,Shishkin2007_Vertex,Gant2022} including transition metal oxides,~\cite{Klimes2014,Shishkin2007,Ergorenc2018} or halides.~\cite{Cu_Gao} For materials displaying slow k-point convergence, like the TM oxide ZnO, the selected grid results in errors on the order of 100 meV.~\cite{falseConv_2,Klimes2014} 
    \newline Two additional subsets have been integrated into the database to serve as additional test-cases for further assessing the workflow’s efficacy.
    The first comprises 36 TM oxide perovskites, a class of materials which has often been used as proving grounds to propose or compare different computational schemes.~\cite{Varrassi2021,Varrassi_2024,STO_AbInitio_test_1,STO_AbInitio_test_2,STO_AbInitio_test_3,SeMa_STO_1}
    Furthermore, the presence of stronger degrees of electronic correlation establishes these systems as challenging benchmarks for electronic structure schemes, as exemplified by the substantial discrepancies and variability observed among theoretical predictions (covering for example a range from $3.36$ eV~\cite{SeMa_STO_1} to $4.05$ eV~\cite{Ergorenc2018} for SrTiO$_3$ and from $3.18$ eV~\cite{SeMa_STO_1} to $3.7$ eV~\cite{BaTiO3_1} for BaTiO$_3$).
    The systems in the second subset were selected from compounds known in the literature to present significant challenges to GW methods, in terms of severe dependence on the number of orbitals or because they yielded inconsistent and contrasting results in previous G$_0$W$_0$ studies.
    The former category includes ZnO and TM halides, which are noted as extreme cases due to the exceptionally high number of states required (estimated at more than $4000$ for ZnO~\cite{ZnOwz_1,ZnOwz_2,ZnOwz_3} and $\sim\hspace{-0.09cm}8000$ and $\sim\hspace{-0.09cm}4000$ for CuCl and AgCl respectively~\cite{Cu_Gao}). 
    Similar computational demands are additionally recognized for the perovskites SrTiO$_3$ and BaTiO$_3$.
    The latter category encompasses TM oxides like MnO and NiO, along with several other compounds that demonstrated significant errors in a previous G$_0$W$_0$ HT study~\cite{VanSetten2017} (SnO$_2$, SnSe$_2$, RuS$_2$, V$_2$O$_5$ and CaO). The comparisons for this dataset are summarized in Fig.~\ref{fig:nontrivialMat}.
    \newline\newline Before discussing the workflow's accuracy, we briefly examine the robustness of the extrapolation.
    The protocol relies on the assumption that the QP energies used to evaluate the corrections follow the 1 over $N_{pw}$ limit; to assess the validity of this hypothesis the workflow automatically computes the  $r^2$ determination coefficients.
    The final $r^2$ distributions (considering all materials in the database) for the Valence Band Maximum (VBM) and Conduction Band Minimum (CBM) are characterized by an average of respectively $\sim\hspace{-0.09cm}97\%$ and $\sim\hspace{-0.09cm}96\%$, with a standard deviation of $\sim\hspace{-0.09cm}10\%$ and $\sim\hspace{-0.09cm}12\%$ (see Fig~\ref{fig:datasetStatistics} b) for the histograms of the distributions). These results confirm the extrapolations' reliability.

    \begin{table}[]
        \begin{tabular}{lccccc}
        \hline
         & \multirow{2}{*}{\begin{tabular}[c]{@{}l@{}}N° of\\ structures\end{tabular}} & \multicolumn{2}{c}{Error (eV)}\hspace{0.10cm} & \multicolumn{2}{c}{Perc. Error (\%)} \\
        Set                                                           &         & MAE    & SD     & MAPE    & SD      \\ \cline{3-6} 
        \begin{tabular}[c]{@{}l@{}}G$_0$W$_0$\\ Database\end{tabular} & 163  & 0.24  & 0.27\hspace{0.10cm} & 6.72  & 9.61   \\
        TM subset                                                     & 98   & 0.26  & 0.34\hspace{0.10cm} & 8.60  & 10.82  \\
        \begin{tabular}[c]{@{}l@{}}Perovskite \\ subset\end{tabular}  & 36   & 0.21  & 0.12\hspace{0.10cm} & 5.56  & 3.70   \\ \hline
        \end{tabular}
        
        \caption{Statistical analysis of errors on QP gaps relative to experimental bandgaps.
        The mean absolute errors (MAE), mean absolute percentage errors (MAPE) and relative standard deviations (SD) on the QP Gaps for the G$_0$W$_0$ Database computed with the workflow are tabulated.
        The statistics of transitions metals (TM) compounds, and more specifically of TM oxide perovskites are also highlighted separately.}
        \label{tab:GapStatistics}
    \end{table}
     A graphical summary of the errors with respect to the experimental references over the entire dataset is presented in Figure~\ref{fig:datasetStatistics}:
    Fig.~\ref{fig:datasetStatistics} c) represents a violin plot of the percentage errors for the entire database and for the TM and perovskite, subsets, Fig.~\ref{fig:datasetStatistics} e) shows a scatter plot of the extrapolated G$_0$W$_0$ gaps against the experimental ones, while the distributions of the mean absolute percentage errors over the materials containing the given element is shown in Figure~\ref{fig:datasetStatistics} d), along with the average BS and NV corrections.
    The workflow is able to achieve a mean absolute percentage error (MAPE) of $\sim\hspace{-0.09cm}6.7\%$ (see Tab.~\ref{tab:GapStatistics}), corresponding to state-of-the-art accuracy.
    This value is reached including all 163 systems with experimental references, representing a robust and comprehensive evaluation of the workflow performance.
    Only a minority fraction, comprising 22 materials, show errors exceeding 10$\%$:
    the outliers include chalcogenides (BeS, SnSe$_2$, WSe$_2$ and TlSe), as well as cuprates (Cu$_2$O and CuCl) and heavy 4-\textit{d} and 5-\textit{d} compounds (Bi$_2$Te$_3$, BiOCl, PbF$_2$ and HfO$_2$).
    The larger discrepancies identified above for the heavy 4-\textit{d} and 5-\textit{d} compounds~\cite{SOC_WSe2,SOC_BiOCl,SOC_HfO2} can be explained by the omission of the spin-orbit coupling (SOC), which is known to result in significant errors, up to $0.3-0.4$ eV.~\cite{SOC_Kuhn2015,SOC_Scherpelz2016,SOC_Gruneis} Additionally, electron-phonon coupling can also affect predicted band gaps. Several studies demonstrated how zero-point renormalization (ZPR) can lead to reduction of the gap by $\sim\hspace{-0.09cm}0.15-0.20$ eV for several systems present in the dataset, including BaSnO$_3$, SnO$_2$ or ZnO.~\cite{BaSnO3_Aggoune2022,ElPhonRen_4,ElPhonRen_5}
    We must also note that the heterogeneity of temperatures, quality of the samples (e.g. degree of crystallinity, defects, etc.) and experimental techniques employed in the reference measurements introduces non-controllable uncertainties in the comparison with our data.
    \newline\newline 
    For systems without transition metals, the workflow achieves a MAPE around $5.5\%$. 
    Conversely, transition metal compounds are recognized as some of the more critical and demanding cases for G$_0$W$_0$@DFT, as evidenced by higher deviations from the experimental data (with a MAPE of $8.6\%$).
    The description of the localized partially filled \textit{d} states presents several difficulties, beginning from using DFT as a starting point.~\cite{Golze2019,SRPdO3_Franchini,MnO_1} In particular, the requirements on the number of states for the self-energy and the norm violations associated with TM PAWs are typically exacerbated,  potentially evolving into critical issues for such materials.~\cite{Golze2019,Klimes2014,Ergorenc2018} For example, the 3-\textit{d} ST-PAWs exhibit the largest norm violations among the entire dataset, ranging from $\sim\hspace{-0.09cm}25\%$ for the $d$ partial-waves of Ti and V up to $\sim\hspace{-0.09cm}57\%$ for Cu.
    These two factors explain the higher on average NV and BS corrections for TM systems observed in Fig.~\ref{fig:datasetStatistics} d), which reach the maximum for the strongly localized 3-\textit{d} states or 2-\textit{p} states~\cite{Maggio2017} (as in the case of fluorine). In fact, the higher $\Delta BS$ and $\Delta NV$ correct the larger basis-set truncation errors on the energies $E_{QP}^{ST}$ involving localized states.
    \newline The inclusion of $\Delta NV$ and $\Delta BS$ proves notably accurate for all transition metal compounds in the second subset: as shown in Fig.~\ref{fig:nontrivialMat}, the protocol consistently reproduces or slight improves over the most accurate G$_0$W$_0$@DFT results reported in the literature for nearly all these materials. Importantly, the workflow maintains an efficient computational setup:  for almost all the TM systems in Fig.~\ref{fig:nontrivialMat} the workflow requires at most $\sim\hspace{-0.09cm}1000$ bands and cutoffs $G^{pw}_{cut}$ up to approximately $\sim\hspace{-0.09cm}450$ eV to compute the QP energies on the dense k-mesh. In particular, copper halides have been noted to exhibit the largest errors in a previous HT G$_0$W$_0$ study,~\cite{VanSetten2017} and demands exceptionally high band requirements.
    Nevertheless, the proposed protocol allows to reduce the computational cost needed for calculating the QP energies (requiring for $E_{QP}^{ST}(N_{pw}^{(1)},N_k)$ at most a quarter of the number of bands referenced in literature) without compromising accuracy. We note lastly that compounds containing Cu are associated with the highest values of $\Delta NV$ and $\Delta BS$ among the database, with the halides displaying $\Delta BS \sim\hspace{-0.09cm}0.20$ eV and $\Delta NV \sim\hspace{-0.09cm}0.50$ eV, respectively.
    \newline The workflow proves effective also for the perovskite subset, achieving a MAPE of $\sim\hspace{-0.09cm}5.6\%$ associated to a similar MAE of $0.20$ eV. 
    The previously cited perovskites SrTiO$_3$ and BaTiO$_3$ exhibit similar convergence criticalities, with different studies (employing conventional convergence procedures) identifying convergence at between 2000 and 5000 bands;~\cite{Ergorenc2018,STO_BTO_PhysRevB.103.035128} our workflow in turn achieves state of the art accuracy with $\sim\hspace{-0.09cm}1000$ bands (and $\sim\hspace{-0.09cm}440$ eV of $G^{pw}_{cut}$ cutoff), owing to $\Delta BS$ and $\Delta NV$ around $\sim\hspace{-0.09cm}0.20$ eV.
    Lastly, we note that the protocol requires a larger number of bands for layered oxide V$_2$O$_5$ ($\sim\hspace{-0.09cm}2300$), due to the larger volume of the unit cell. Nonetheless, our setup employed a cutoff $\sim\hspace{-0.09cm}440$ eV to determine QP energies on the dense $N_k$ mesh for V$_2$O$_5$, a comparably lower value with respect to the $\sim\hspace{-0.09cm} 1100-1400$ eV demanded by recent GW and QSGW studies.~\cite{V2O5_1,V2O5_2,V2O5_3}
    \newline\newline In conclusion, we have presented the development and validation of a  high-throughput automatized approach for computing G$_0$W$_0$ quasi-particle energies using the AiiDA-VASP framework.
    \newline The approach is based on the estimation and correction of errors related to the basis-set truncation and PAW norm violation. To showcase its effectiveness and for benchmark purposes, a comprehensive database encompassing 327 materials was constructed using the proposed workflow.
    From a theoretical point of view, the correction scheme respects the full basis-set constraint, which formally ensures the correct asymptotic limit. An extensive validation, performed involving more than 160 different systems, shows that the automated procedure is able to achieve state-of-the-art accuracy while requiring minimal user intervention.
    The workflow's computational efficiency represents a second important advantage of the protocol. The scheme does not need to sample multidimensional parameter spaces, strongly limiting the total number of calculations required. 
    Further developments of the workflow, aimed at improving accuracy for critical cases, could involve integrating ZPR as well as the additional contributions due to vertex corrections and SOC.
    The presented results illustrate how the proposed workflow, which requires only the structural data, can represent a powerful resource for the material science community for high-throughput excited-state studies with high accuracy.
   Finally, the complete database collected in the supplementary data offers a valuable reference for future studies, facilitating comparison and benchmarking across ab-initio codes.

\section*{METHODS}
All calculations are performed using \textit{Vienna Ab-initio Simulation Package} (VASP).~\cite{Kresse1996,Kresse1996_2}
The GW versions of the ST-PAW pseudo-potentials,\cite{Perdew1996, KresseG;Joubert1999} with relativistic effects taken into account only at scalar level and semicore electrons included (where available), are selected for all elements. The complete list of the ST-PAWs and NC-PAWs are listed in SM (see Tables SM1 and SM2). The table also lists the maximum norm violation among the pseudo-waves for each PAW potential.
Unless explicitly states, the standard (non norm-conserving) version of the pseudo-potentials is chosen.
\newline All results discussed in the text are obtained using the quartic-scaling G$_0$W$_0$ scheme, using DFT (PBE) as a starting point. The Spin-Orbit effects are not taken into account. The full frequency dependent self-energy is evaluated through the Hilbert transform technique~\cite{Shishkin2006} including 200 frequency points. Furthermore, all G$_0$W$_0$ results presented include the settings \texttt{PREC = Accurate}, which forces VASP to employ a denser than standard FFT grid. The \texttt{NMAXFOCKAE} flag, which controls the cutoff used for the reconstruction for the overlap densities in the PAW scheme, is set equal to \texttt{NMAXFOCKAE = }2. The Wannierization is performed with the Wannier90 code ~\cite{Wannier_Pizzi_2020}m version 3.1,  which is called inside the VASP calculations in library mode.

\section*{CODE AVAILABILITY}
The source code is available open-source on GitHub (https://github.com/lorenzovarro/GW-VASP-workflow) and will be part of the next release of the AiiDA-VASP Plugin.

\section*{DATA AVAILABILITY}
The dataset composing the G$_0$W$_0$ database will be made publicly available and accessible; the NC-PAW potentials will be released in the next version of the VASP software.

\section*{ACKNOWLEDGMENTS}
L.V. thanks Jusong Yu, Giovanni Pizzi, Domenico di Sante for fruitful discussions.
This work is partly funded by the European Union – Next Generation EU - “PNRR - M4C2, investimento 1.1 - Fondo PRIN 2022” - “Superlattices of relativistic oxides” (ID 2022L28H97, CUP D53D23002260006).
The authors acknowledge the CINECA award under the ISCRA initiative, for the availability of high-performance computing resources and support, as well as computing time granted by the Vienna Scientific Cluster.

\section*{AUTHOR CONTRIBUTIONS}
C.F. conceived the project. L. V. has written the GW extension of the AiiDA-VASP plugin with the help of E.F.-L and M.W. and executed all GW calculations with the support of F.E. 
L.V. wrote the manuscript assisted by C.F. with input from all the authors. G.K. contributed to the revision of the manuscript, developed the GW code and all used PAW potentials.

\section*{COMPETING INTERESTS}
G.K. is shareholder of the VASP Software GmbH, and M.W. is an employee of the VASP Software GmbH.
The remaining authors declare no competing interests.


\bibliography{main}

\begin{thebibliography}{157}
\expandafter\ifx\csname natexlab\endcsname\relax\def\natexlab#1{#1}\fi
\expandafter\ifx\csname bibnamefont\endcsname\relax
  \def\bibnamefont#1{#1}\fi
\expandafter\ifx\csname bibfnamefont\endcsname\relax
  \def\bibfnamefont#1{#1}\fi
\expandafter\ifx\csname citenamefont\endcsname\relax
  \def\citenamefont#1{#1}\fi
\expandafter\ifx\csname url\endcsname\relax
  \def\url#1{\texttt{#1}}\fi
\expandafter\ifx\csname urlprefix\endcsname\relax\def\urlprefix{URL }\fi
\providecommand{\bibinfo}[2]{#2}
\providecommand{\eprint}[2][]{\url{#2}}

\bibitem[{\citenamefont{Curtarolo et~al.}(2013)\citenamefont{Curtarolo, Hart,
  Mingo, Sanvito, and Levy}}]{Curtarolo2013}
\bibinfo{author}{\bibfnamefont{S.}~\bibnamefont{Curtarolo}},
  \bibinfo{author}{\bibfnamefont{M.~B.} \bibnamefont{Hart}, \bibfnamefont{Gus
  L. W.and~Nardelli}}, \bibinfo{author}{\bibfnamefont{N.}~\bibnamefont{Mingo}},
  \bibinfo{author}{\bibfnamefont{S.}~\bibnamefont{Sanvito}}, \bibnamefont{and}
  \bibinfo{author}{\bibfnamefont{O.}~\bibnamefont{Levy}},
  \bibinfo{journal}{Nature Materials} \textbf{\bibinfo{volume}{12}},
  \bibinfo{pages}{191} (\bibinfo{year}{2013}), ISSN \bibinfo{issn}{1476-4660},
  \urlprefix\url{https://doi.org/10.1038/nmat3568}.

\bibitem[{\citenamefont{Thygesen and Jacobsen}(2016)}]{Thygesen_Science2016}
\bibinfo{author}{\bibfnamefont{K.~S.} \bibnamefont{Thygesen}} \bibnamefont{and}
  \bibinfo{author}{\bibfnamefont{K.~W.} \bibnamefont{Jacobsen}},
  \bibinfo{journal}{Science} \textbf{\bibinfo{volume}{354}},
  \bibinfo{pages}{180} (\bibinfo{year}{2016}),
  \eprint{https://www.science.org/doi/pdf/10.1126/science.aah4776},
  \urlprefix\url{https://www.science.org/doi/abs/10.1126/science.aah4776}.

\bibitem[{\citenamefont{Alberi et~al.}(2018)\citenamefont{Alberi, Nardelli,
  Zakutayev, Mitas, Curtarolo, Jain, Fornari, Marzari, Takeuchi, Green
  et~al.}}]{Roadmap2019}
\bibinfo{author}{\bibfnamefont{K.}~\bibnamefont{Alberi}},
  \bibinfo{author}{\bibfnamefont{M.~B.} \bibnamefont{Nardelli}},
  \bibinfo{author}{\bibfnamefont{A.}~\bibnamefont{Zakutayev}},
  \bibinfo{author}{\bibfnamefont{L.}~\bibnamefont{Mitas}},
  \bibinfo{author}{\bibfnamefont{S.}~\bibnamefont{Curtarolo}},
  \bibinfo{author}{\bibfnamefont{A.}~\bibnamefont{Jain}},
  \bibinfo{author}{\bibfnamefont{M.}~\bibnamefont{Fornari}},
  \bibinfo{author}{\bibfnamefont{N.}~\bibnamefont{Marzari}},
  \bibinfo{author}{\bibfnamefont{I.}~\bibnamefont{Takeuchi}},
  \bibinfo{author}{\bibfnamefont{M.~L.} \bibnamefont{Green}},
  \bibnamefont{et~al.}, \bibinfo{journal}{Journal of Physics D: Applied
  Physics} \textbf{\bibinfo{volume}{52}}, \bibinfo{pages}{013001}
  (\bibinfo{year}{2018}),
  \urlprefix\url{https://dx.doi.org/10.1088/1361-6463/aad926}.

\bibitem[{\citenamefont{Schaarschmidt et~al.}(2022)\citenamefont{Schaarschmidt,
  Yuan, Strunk, Kondov, Huber, Pizzi, Kahle, Bölle, Castelli, Vegge
  et~al.}}]{Schaarschmidt2022}
\bibinfo{author}{\bibfnamefont{J.}~\bibnamefont{Schaarschmidt}},
  \bibinfo{author}{\bibfnamefont{J.}~\bibnamefont{Yuan}},
  \bibinfo{author}{\bibfnamefont{T.}~\bibnamefont{Strunk}},
  \bibinfo{author}{\bibfnamefont{I.}~\bibnamefont{Kondov}},
  \bibinfo{author}{\bibfnamefont{S.~P.} \bibnamefont{Huber}},
  \bibinfo{author}{\bibfnamefont{G.}~\bibnamefont{Pizzi}},
  \bibinfo{author}{\bibfnamefont{L.}~\bibnamefont{Kahle}},
  \bibinfo{author}{\bibfnamefont{F.~T.} \bibnamefont{Bölle}},
  \bibinfo{author}{\bibfnamefont{I.~E.} \bibnamefont{Castelli}},
  \bibinfo{author}{\bibfnamefont{T.}~\bibnamefont{Vegge}},
  \bibnamefont{et~al.}, \bibinfo{journal}{Advanced Energy Materials}
  \textbf{\bibinfo{volume}{12}}, \bibinfo{pages}{2102638}
  (\bibinfo{year}{2022}),
  \eprint{https://onlinelibrary.wiley.com/doi/pdf/10.1002/aenm.202102638},
  \urlprefix\url{https://onlinelibrary.wiley.com/doi/abs/10.1002/aenm.202102638}.

\bibitem[{\citenamefont{Jain et~al.}(2015)\citenamefont{Jain, Ong, Chen,
  Medasani, Qu, Kocher, Brafman, Petretto, Rignanese, Hautier
  et~al.}}]{Fireworks2015}
\bibinfo{author}{\bibfnamefont{A.}~\bibnamefont{Jain}},
  \bibinfo{author}{\bibfnamefont{S.~P.} \bibnamefont{Ong}},
  \bibinfo{author}{\bibfnamefont{W.}~\bibnamefont{Chen}},
  \bibinfo{author}{\bibfnamefont{B.}~\bibnamefont{Medasani}},
  \bibinfo{author}{\bibfnamefont{X.}~\bibnamefont{Qu}},
  \bibinfo{author}{\bibfnamefont{M.}~\bibnamefont{Kocher}},
  \bibinfo{author}{\bibfnamefont{M.}~\bibnamefont{Brafman}},
  \bibinfo{author}{\bibfnamefont{G.}~\bibnamefont{Petretto}},
  \bibinfo{author}{\bibfnamefont{G.-M.} \bibnamefont{Rignanese}},
  \bibinfo{author}{\bibfnamefont{G.}~\bibnamefont{Hautier}},
  \bibnamefont{et~al.}, \bibinfo{journal}{Concurrency and Computation: Practice
  and Experience} \textbf{\bibinfo{volume}{27}}, \bibinfo{pages}{5037}
  (\bibinfo{year}{2015}),
  \eprint{https://onlinelibrary.wiley.com/doi/pdf/10.1002/cpe.3505},
  \urlprefix\url{https://onlinelibrary.wiley.com/doi/abs/10.1002/cpe.3505}.

\bibitem[{\citenamefont{Pizzi et~al.}(2016)\citenamefont{Pizzi, Cepellotti,
  Sabatini, Marzari, and Kozinsky}}]{AiiDA2016}
\bibinfo{author}{\bibfnamefont{G.}~\bibnamefont{Pizzi}},
  \bibinfo{author}{\bibfnamefont{A.}~\bibnamefont{Cepellotti}},
  \bibinfo{author}{\bibfnamefont{R.}~\bibnamefont{Sabatini}},
  \bibinfo{author}{\bibfnamefont{N.}~\bibnamefont{Marzari}}, \bibnamefont{and}
  \bibinfo{author}{\bibfnamefont{B.}~\bibnamefont{Kozinsky}},
  \bibinfo{journal}{Computational Materials Science}
  \textbf{\bibinfo{volume}{111}}, \bibinfo{pages}{218} (\bibinfo{year}{2016}),
  ISSN \bibinfo{issn}{0927-0256},
  \urlprefix\url{https://www.sciencedirect.com/science/article/pii/S0927025615005820}.

\bibitem[{\citenamefont{Mortensen et~al.}(2020)\citenamefont{Mortensen,
  Gjerding, and Thygesen}}]{MyQueue2020}
\bibinfo{author}{\bibfnamefont{J.~J.} \bibnamefont{Mortensen}},
  \bibinfo{author}{\bibfnamefont{M.}~\bibnamefont{Gjerding}}, \bibnamefont{and}
  \bibinfo{author}{\bibfnamefont{K.~S.} \bibnamefont{Thygesen}},
  \bibinfo{journal}{Journal of Open Source Software}
  \textbf{\bibinfo{volume}{5}}, \bibinfo{pages}{1844} (\bibinfo{year}{2020}),
  \urlprefix\url{https://doi.org/10.21105/joss.01844}.

\bibitem[{\citenamefont{Curtarolo et~al.}(2012)\citenamefont{Curtarolo,
  Setyawan, Hart, Jahnatek, Chepulskii, Taylor, Wang, Xue, Yang, Levy
  et~al.}}]{AFLOW2012}
\bibinfo{author}{\bibfnamefont{S.}~\bibnamefont{Curtarolo}},
  \bibinfo{author}{\bibfnamefont{W.}~\bibnamefont{Setyawan}},
  \bibinfo{author}{\bibfnamefont{G.~L.} \bibnamefont{Hart}},
  \bibinfo{author}{\bibfnamefont{M.}~\bibnamefont{Jahnatek}},
  \bibinfo{author}{\bibfnamefont{R.~V.} \bibnamefont{Chepulskii}},
  \bibinfo{author}{\bibfnamefont{R.~H.} \bibnamefont{Taylor}},
  \bibinfo{author}{\bibfnamefont{S.}~\bibnamefont{Wang}},
  \bibinfo{author}{\bibfnamefont{J.}~\bibnamefont{Xue}},
  \bibinfo{author}{\bibfnamefont{K.}~\bibnamefont{Yang}},
  \bibinfo{author}{\bibfnamefont{O.}~\bibnamefont{Levy}}, \bibnamefont{et~al.},
  \bibinfo{journal}{Computational Materials Science}
  \textbf{\bibinfo{volume}{58}}, \bibinfo{pages}{218} (\bibinfo{year}{2012}),
  ISSN \bibinfo{issn}{0927-0256},
  \urlprefix\url{https://www.sciencedirect.com/science/article/pii/S0927025612000717}.

\bibitem[{\citenamefont{Mathew et~al.}(2017)\citenamefont{Mathew, Montoya,
  Faghaninia, Dwarakanath, Aykol, Tang, heng Chu, Smidt, Bocklund, Horton
  et~al.}}]{Atomate2018}
\bibinfo{author}{\bibfnamefont{K.}~\bibnamefont{Mathew}},
  \bibinfo{author}{\bibfnamefont{J.~H.} \bibnamefont{Montoya}},
  \bibinfo{author}{\bibfnamefont{A.}~\bibnamefont{Faghaninia}},
  \bibinfo{author}{\bibfnamefont{S.}~\bibnamefont{Dwarakanath}},
  \bibinfo{author}{\bibfnamefont{M.}~\bibnamefont{Aykol}},
  \bibinfo{author}{\bibfnamefont{H.}~\bibnamefont{Tang}},
  \bibinfo{author}{\bibfnamefont{I.}~\bibnamefont{heng Chu}},
  \bibinfo{author}{\bibfnamefont{T.}~\bibnamefont{Smidt}},
  \bibinfo{author}{\bibfnamefont{B.}~\bibnamefont{Bocklund}},
  \bibinfo{author}{\bibfnamefont{M.}~\bibnamefont{Horton}},
  \bibnamefont{et~al.}, \bibinfo{journal}{Computational Materials Science}
  \textbf{\bibinfo{volume}{139}}, \bibinfo{pages}{140} (\bibinfo{year}{2017}),
  ISSN \bibinfo{issn}{0927-0256},
  \urlprefix\url{https://www.sciencedirect.com/science/article/pii/S0927025617303919}.

\bibitem[{\citenamefont{Kirklin et~al.}(2015)\citenamefont{Kirklin, Saal,
  Meredig, Thompson, Doak, Aykol, R{\"u}hl, and Wolverton}}]{Kirklin2015}
\bibinfo{author}{\bibfnamefont{S.}~\bibnamefont{Kirklin}},
  \bibinfo{author}{\bibfnamefont{J.~E.} \bibnamefont{Saal}},
  \bibinfo{author}{\bibfnamefont{B.}~\bibnamefont{Meredig}},
  \bibinfo{author}{\bibfnamefont{A.}~\bibnamefont{Thompson}},
  \bibinfo{author}{\bibfnamefont{J.~W.} \bibnamefont{Doak}},
  \bibinfo{author}{\bibfnamefont{M.}~\bibnamefont{Aykol}},
  \bibinfo{author}{\bibfnamefont{S.}~\bibnamefont{R{\"u}hl}}, \bibnamefont{and}
  \bibinfo{author}{\bibfnamefont{C.}~\bibnamefont{Wolverton}},
  \bibinfo{journal}{npj Computational Materials} \textbf{\bibinfo{volume}{1}},
  \bibinfo{pages}{15010} (\bibinfo{year}{2015}), ISSN
  \bibinfo{issn}{2057-3960},
  \urlprefix\url{https://doi.org/10.1038/npjcompumats.2015.10}.

\bibitem[{\citenamefont{Jain et~al.}(2013)\citenamefont{Jain, Ong, Hautier,
  Chen, Richards, Dacek, Cholia, Gunter, Skinner, Ceder et~al.}}]{Jain2013}
\bibinfo{author}{\bibfnamefont{A.}~\bibnamefont{Jain}},
  \bibinfo{author}{\bibfnamefont{S.~P.} \bibnamefont{Ong}},
  \bibinfo{author}{\bibfnamefont{G.}~\bibnamefont{Hautier}},
  \bibinfo{author}{\bibfnamefont{W.}~\bibnamefont{Chen}},
  \bibinfo{author}{\bibfnamefont{W.~D.} \bibnamefont{Richards}},
  \bibinfo{author}{\bibfnamefont{S.}~\bibnamefont{Dacek}},
  \bibinfo{author}{\bibfnamefont{S.}~\bibnamefont{Cholia}},
  \bibinfo{author}{\bibfnamefont{D.}~\bibnamefont{Gunter}},
  \bibinfo{author}{\bibfnamefont{D.}~\bibnamefont{Skinner}},
  \bibinfo{author}{\bibfnamefont{G.}~\bibnamefont{Ceder}},
  \bibnamefont{et~al.}, \bibinfo{journal}{APL Materials}
  \textbf{\bibinfo{volume}{1}} (\bibinfo{year}{2013}), ISSN
  \bibinfo{issn}{2166-532X}, \bibinfo{note}{011002},
  \urlprefix\url{https://doi.org/10.1063/1.4812323}.

\bibitem[{\citenamefont{Draxl and Scheffler}(2019)}]{Draxl2019}
\bibinfo{author}{\bibfnamefont{C.}~\bibnamefont{Draxl}} \bibnamefont{and}
  \bibinfo{author}{\bibfnamefont{M.}~\bibnamefont{Scheffler}},
  \bibinfo{journal}{Journal of Physics: Materials}
  \textbf{\bibinfo{volume}{2}}, \bibinfo{pages}{036001} (\bibinfo{year}{2019}),
  \urlprefix\url{https://dx.doi.org/10.1088/2515-7639/ab13bb}.

\bibitem[{\citenamefont{Mounet et~al.}(2018)\citenamefont{Mounet, Gibertini,
  Schwaller, Campi, Merkys, Marrazzo, Sohier, Castelli, Cepellotti, Pizzi
  et~al.}}]{Mounet2018}
\bibinfo{author}{\bibfnamefont{N.}~\bibnamefont{Mounet}},
  \bibinfo{author}{\bibfnamefont{M.}~\bibnamefont{Gibertini}},
  \bibinfo{author}{\bibfnamefont{P.}~\bibnamefont{Schwaller}},
  \bibinfo{author}{\bibfnamefont{D.}~\bibnamefont{Campi}},
  \bibinfo{author}{\bibfnamefont{A.}~\bibnamefont{Merkys}},
  \bibinfo{author}{\bibfnamefont{A.}~\bibnamefont{Marrazzo}},
  \bibinfo{author}{\bibfnamefont{T.}~\bibnamefont{Sohier}},
  \bibinfo{author}{\bibfnamefont{I.~E.} \bibnamefont{Castelli}},
  \bibinfo{author}{\bibfnamefont{A.}~\bibnamefont{Cepellotti}},
  \bibinfo{author}{\bibfnamefont{G.}~\bibnamefont{Pizzi}},
  \bibnamefont{et~al.}, \bibinfo{journal}{Nature Nanotechnology}
  \textbf{\bibinfo{volume}{13}}, \bibinfo{pages}{246} (\bibinfo{year}{2018}),
  ISSN \bibinfo{issn}{1748-3395},
  \urlprefix\url{https://doi.org/10.1038/s41565-017-0035-5}.

\bibitem[{\citenamefont{Talirz et~al.}(2020)\citenamefont{Talirz, Kumbhar,
  Passaro, Yakutovich, Granata, Gargiulo, Borelli, Uhrin, Huber, Zoupanos
  et~al.}}]{Talirz2020}
\bibinfo{author}{\bibfnamefont{L.}~\bibnamefont{Talirz}},
  \bibinfo{author}{\bibfnamefont{S.}~\bibnamefont{Kumbhar}},
  \bibinfo{author}{\bibfnamefont{E.}~\bibnamefont{Passaro}},
  \bibinfo{author}{\bibfnamefont{A.~V.} \bibnamefont{Yakutovich}},
  \bibinfo{author}{\bibfnamefont{V.}~\bibnamefont{Granata}},
  \bibinfo{author}{\bibfnamefont{F.}~\bibnamefont{Gargiulo}},
  \bibinfo{author}{\bibfnamefont{M.}~\bibnamefont{Borelli}},
  \bibinfo{author}{\bibfnamefont{M.}~\bibnamefont{Uhrin}},
  \bibinfo{author}{\bibfnamefont{S.~P.} \bibnamefont{Huber}},
  \bibinfo{author}{\bibfnamefont{S.}~\bibnamefont{Zoupanos}},
  \bibnamefont{et~al.}, \bibinfo{journal}{Scientific Data}
  \textbf{\bibinfo{volume}{7}}, \bibinfo{pages}{299} (\bibinfo{year}{2020}),
  ISSN \bibinfo{issn}{2052-4463},
  \urlprefix\url{https://doi.org/10.1038/s41597-020-00637-5}.

\bibitem[{\citenamefont{Schmidt et~al.}(2019)\citenamefont{Schmidt, Marques,
  Botti, and Marques}}]{Schmidt2019}
\bibinfo{author}{\bibfnamefont{J.}~\bibnamefont{Schmidt}},
  \bibinfo{author}{\bibfnamefont{M.~R.~G.} \bibnamefont{Marques}},
  \bibinfo{author}{\bibfnamefont{S.}~\bibnamefont{Botti}}, \bibnamefont{and}
  \bibinfo{author}{\bibfnamefont{M.~A.~L.} \bibnamefont{Marques}},
  \bibinfo{journal}{npj Computational Materials} \textbf{\bibinfo{volume}{5}},
  \bibinfo{pages}{83} (\bibinfo{year}{2019}), ISSN \bibinfo{issn}{2057-3960},
  \urlprefix\url{https://doi.org/10.1038/s41524-019-0221-0}.

\bibitem[{\citenamefont{Wang et~al.}(2020)\citenamefont{Wang, Murdock, Kauwe,
  Oliynyk, Gurlo, Brgoch, Persson, and Sparks}}]{Wang2020}
\bibinfo{author}{\bibfnamefont{A.~Y.-T.} \bibnamefont{Wang}},
  \bibinfo{author}{\bibfnamefont{R.~J.} \bibnamefont{Murdock}},
  \bibinfo{author}{\bibfnamefont{S.~K.} \bibnamefont{Kauwe}},
  \bibinfo{author}{\bibfnamefont{A.~O.} \bibnamefont{Oliynyk}},
  \bibinfo{author}{\bibfnamefont{A.}~\bibnamefont{Gurlo}},
  \bibinfo{author}{\bibfnamefont{J.}~\bibnamefont{Brgoch}},
  \bibinfo{author}{\bibfnamefont{K.~A.} \bibnamefont{Persson}},
  \bibnamefont{and} \bibinfo{author}{\bibfnamefont{T.~D.}
  \bibnamefont{Sparks}}, \bibinfo{journal}{Chemistry of Materials}
  \textbf{\bibinfo{volume}{32}}, \bibinfo{pages}{4954} (\bibinfo{year}{2020}),
  ISSN \bibinfo{issn}{0897-4756},
  \urlprefix\url{https://doi.org/10.1021/acs.chemmater.0c01907}.

\bibitem[{\citenamefont{Ashton et~al.}(2017)\citenamefont{Ashton, Paul,
  Sinnott, and Hennig}}]{Ashton2017}
\bibinfo{author}{\bibfnamefont{M.}~\bibnamefont{Ashton}},
  \bibinfo{author}{\bibfnamefont{J.}~\bibnamefont{Paul}},
  \bibinfo{author}{\bibfnamefont{S.~B.} \bibnamefont{Sinnott}},
  \bibnamefont{and} \bibinfo{author}{\bibfnamefont{R.~G.}
  \bibnamefont{Hennig}}, \bibinfo{journal}{Phys. Rev. Lett.}
  \textbf{\bibinfo{volume}{118}}, \bibinfo{pages}{106101}
  (\bibinfo{year}{2017}),
  \urlprefix\url{https://link.aps.org/doi/10.1103/PhysRevLett.118.106101}.

\bibitem[{\citenamefont{Cheon et~al.}(2017)\citenamefont{Cheon, Duerloo,
  Sendek, Porter, Chen, and Reed}}]{Cheon2017}
\bibinfo{author}{\bibfnamefont{G.}~\bibnamefont{Cheon}},
  \bibinfo{author}{\bibfnamefont{K.-A.~N.} \bibnamefont{Duerloo}},
  \bibinfo{author}{\bibfnamefont{A.~D.} \bibnamefont{Sendek}},
  \bibinfo{author}{\bibfnamefont{C.}~\bibnamefont{Porter}},
  \bibinfo{author}{\bibfnamefont{Y.}~\bibnamefont{Chen}}, \bibnamefont{and}
  \bibinfo{author}{\bibfnamefont{E.~J.} \bibnamefont{Reed}},
  \bibinfo{journal}{Nano Letters} \textbf{\bibinfo{volume}{17}},
  \bibinfo{pages}{1915} (\bibinfo{year}{2017}), ISSN \bibinfo{issn}{1530-6984},
  \urlprefix\url{https://doi.org/10.1021/acs.nanolett.6b05229}.

\bibitem[{\citenamefont{Huber et~al.}(2021)\citenamefont{Huber, Bosoni, Bercx,
  Br{\"o}der, Degomme, Dikan, Eimre, Flage-Larsen, Garcia, Genovese
  et~al.}}]{AiiDA2021_CommonWorkflow}
\bibinfo{author}{\bibfnamefont{S.~P.} \bibnamefont{Huber}},
  \bibinfo{author}{\bibfnamefont{E.}~\bibnamefont{Bosoni}},
  \bibinfo{author}{\bibfnamefont{M.}~\bibnamefont{Bercx}},
  \bibinfo{author}{\bibfnamefont{J.}~\bibnamefont{Br{\"o}der}},
  \bibinfo{author}{\bibfnamefont{A.}~\bibnamefont{Degomme}},
  \bibinfo{author}{\bibfnamefont{V.}~\bibnamefont{Dikan}},
  \bibinfo{author}{\bibfnamefont{K.}~\bibnamefont{Eimre}},
  \bibinfo{author}{\bibfnamefont{E.}~\bibnamefont{Flage-Larsen}},
  \bibinfo{author}{\bibfnamefont{A.}~\bibnamefont{Garcia}},
  \bibinfo{author}{\bibfnamefont{L.}~\bibnamefont{Genovese}},
  \bibnamefont{et~al.}, \bibinfo{journal}{npj Computational Materials}
  \textbf{\bibinfo{volume}{7}}, \bibinfo{pages}{136} (\bibinfo{year}{2021}),
  ISSN \bibinfo{issn}{2057-3960},
  \urlprefix\url{https://doi.org/10.1038/s41524-021-00594-6}.

\bibitem[{\citenamefont{Bosoni et~al.}(2024)\citenamefont{Bosoni, Beal, Bercx,
  Blaha, Bl{\"u}gel, Br{\"o}der, Callsen, Cottenier, Degomme, Dikan
  et~al.}}]{DatabaseValidation_1}
\bibinfo{author}{\bibfnamefont{E.}~\bibnamefont{Bosoni}},
  \bibinfo{author}{\bibfnamefont{L.}~\bibnamefont{Beal}},
  \bibinfo{author}{\bibfnamefont{M.}~\bibnamefont{Bercx}},
  \bibinfo{author}{\bibfnamefont{P.}~\bibnamefont{Blaha}},
  \bibinfo{author}{\bibfnamefont{S.}~\bibnamefont{Bl{\"u}gel}},
  \bibinfo{author}{\bibfnamefont{J.}~\bibnamefont{Br{\"o}der}},
  \bibinfo{author}{\bibfnamefont{M.}~\bibnamefont{Callsen}},
  \bibinfo{author}{\bibfnamefont{S.}~\bibnamefont{Cottenier}},
  \bibinfo{author}{\bibfnamefont{A.}~\bibnamefont{Degomme}},
  \bibinfo{author}{\bibfnamefont{V.}~\bibnamefont{Dikan}},
  \bibnamefont{et~al.}, \bibinfo{journal}{Nature Reviews Physics}
  \textbf{\bibinfo{volume}{6}}, \bibinfo{pages}{45} (\bibinfo{year}{2024}),
  ISSN \bibinfo{issn}{2522-5820},
  \urlprefix\url{https://doi.org/10.1038/s42254-023-00655-3}.

\bibitem[{\citenamefont{Carbogno et~al.}(2022)\citenamefont{Carbogno, Thygesen,
  Bieniek, Draxl, Ghiringhelli, Gulans, Hofmann, Jacobsen, Lubeck, Mortensen
  et~al.}}]{DatabaseValidation_2}
\bibinfo{author}{\bibfnamefont{C.}~\bibnamefont{Carbogno}},
  \bibinfo{author}{\bibfnamefont{K.~S.} \bibnamefont{Thygesen}},
  \bibinfo{author}{\bibfnamefont{B.}~\bibnamefont{Bieniek}},
  \bibinfo{author}{\bibfnamefont{C.}~\bibnamefont{Draxl}},
  \bibinfo{author}{\bibfnamefont{L.~M.} \bibnamefont{Ghiringhelli}},
  \bibinfo{author}{\bibfnamefont{A.}~\bibnamefont{Gulans}},
  \bibinfo{author}{\bibfnamefont{O.~T.} \bibnamefont{Hofmann}},
  \bibinfo{author}{\bibfnamefont{K.~W.} \bibnamefont{Jacobsen}},
  \bibinfo{author}{\bibfnamefont{S.}~\bibnamefont{Lubeck}},
  \bibinfo{author}{\bibfnamefont{J.~J.} \bibnamefont{Mortensen}},
  \bibnamefont{et~al.}, \bibinfo{journal}{npj Computational Materials}
  \textbf{\bibinfo{volume}{8}}, \bibinfo{pages}{69} (\bibinfo{year}{2022}),
  ISSN \bibinfo{issn}{2057-3960},
  \urlprefix\url{https://doi.org/10.1038/s41524-022-00744-4}.

\bibitem[{\citenamefont{Rasmussen and
  Thygesen}(2015{\natexlab{a}})}]{Rasmussen2015}
\bibinfo{author}{\bibfnamefont{F.~A.} \bibnamefont{Rasmussen}}
  \bibnamefont{and} \bibinfo{author}{\bibfnamefont{K.~S.}
  \bibnamefont{Thygesen}}, \bibinfo{journal}{The Journal of Physical Chemistry
  C} \textbf{\bibinfo{volume}{119}}, \bibinfo{pages}{13169}
  (\bibinfo{year}{2015}{\natexlab{a}}), ISSN \bibinfo{issn}{1932-7447},
  \urlprefix\url{https://doi.org/10.1021/acs.jpcc.5b02950}.

\bibitem[{\citenamefont{Choudhary et~al.}(2017)\citenamefont{Choudhary, Kalish,
  Beams, and Tavazza}}]{Choudhary2017}
\bibinfo{author}{\bibfnamefont{K.}~\bibnamefont{Choudhary}},
  \bibinfo{author}{\bibfnamefont{I.}~\bibnamefont{Kalish}},
  \bibinfo{author}{\bibfnamefont{R.}~\bibnamefont{Beams}}, \bibnamefont{and}
  \bibinfo{author}{\bibfnamefont{F.}~\bibnamefont{Tavazza}},
  \bibinfo{journal}{Scientific Reports} \textbf{\bibinfo{volume}{7}},
  \bibinfo{pages}{5179} (\bibinfo{year}{2017}), ISSN \bibinfo{issn}{2045-2322},
  \urlprefix\url{https://doi.org/10.1038/s41598-017-05402-0}.

\bibitem[{\citenamefont{Kirklin et~al.}(2013)\citenamefont{Kirklin, Meredig,
  and Wolverton}}]{Kirklin2013}
\bibinfo{author}{\bibfnamefont{S.}~\bibnamefont{Kirklin}},
  \bibinfo{author}{\bibfnamefont{B.}~\bibnamefont{Meredig}}, \bibnamefont{and}
  \bibinfo{author}{\bibfnamefont{C.}~\bibnamefont{Wolverton}},
  \bibinfo{journal}{Advanced Energy Materials} \textbf{\bibinfo{volume}{3}},
  \bibinfo{pages}{252} (\bibinfo{year}{2013}),
  \eprint{https://onlinelibrary.wiley.com/doi/pdf/10.1002/aenm.201200593},
  \urlprefix\url{https://onlinelibrary.wiley.com/doi/abs/10.1002/aenm.201200593}.

\bibitem[{\citenamefont{Bhattacharya and Madsen}(2015)}]{Bhattacharya2015}
\bibinfo{author}{\bibfnamefont{S.}~\bibnamefont{Bhattacharya}}
  \bibnamefont{and} \bibinfo{author}{\bibfnamefont{G.~K.~H.}
  \bibnamefont{Madsen}}, \bibinfo{journal}{Phys. Rev. B}
  \textbf{\bibinfo{volume}{92}}, \bibinfo{pages}{085205}
  (\bibinfo{year}{2015}),
  \urlprefix\url{https://link.aps.org/doi/10.1103/PhysRevB.92.085205}.

\bibitem[{\citenamefont{Hautier et~al.}(2013)\citenamefont{Hautier, Miglio,
  Ceder, Rignanese, and Gonze}}]{Hautier2013}
\bibinfo{author}{\bibfnamefont{G.}~\bibnamefont{Hautier}},
  \bibinfo{author}{\bibfnamefont{A.}~\bibnamefont{Miglio}},
  \bibinfo{author}{\bibfnamefont{G.}~\bibnamefont{Ceder}},
  \bibinfo{author}{\bibfnamefont{G.-M.} \bibnamefont{Rignanese}},
  \bibnamefont{and} \bibinfo{author}{\bibfnamefont{X.}~\bibnamefont{Gonze}},
  \bibinfo{journal}{Nature Communications} \textbf{\bibinfo{volume}{4}},
  \bibinfo{pages}{2292} (\bibinfo{year}{2013}), ISSN \bibinfo{issn}{2041-1723},
  \urlprefix\url{https://doi.org/10.1038/ncomms3292}.

\bibitem[{\citenamefont{Chen et~al.}(2016)\citenamefont{Chen, Pöhls, Hautier,
  Broberg, Bajaj, Aydemir, Gibbs, Zhu, Asta, Snyder et~al.}}]{Chen2016}
\bibinfo{author}{\bibfnamefont{W.}~\bibnamefont{Chen}},
  \bibinfo{author}{\bibfnamefont{J.-H.} \bibnamefont{Pöhls}},
  \bibinfo{author}{\bibfnamefont{G.}~\bibnamefont{Hautier}},
  \bibinfo{author}{\bibfnamefont{D.}~\bibnamefont{Broberg}},
  \bibinfo{author}{\bibfnamefont{S.}~\bibnamefont{Bajaj}},
  \bibinfo{author}{\bibfnamefont{U.}~\bibnamefont{Aydemir}},
  \bibinfo{author}{\bibfnamefont{Z.~M.} \bibnamefont{Gibbs}},
  \bibinfo{author}{\bibfnamefont{H.}~\bibnamefont{Zhu}},
  \bibinfo{author}{\bibfnamefont{M.}~\bibnamefont{Asta}},
  \bibinfo{author}{\bibfnamefont{G.~J.} \bibnamefont{Snyder}},
  \bibnamefont{et~al.}, \bibinfo{journal}{J. Mater. Chem. C}
  \textbf{\bibinfo{volume}{4}}, \bibinfo{pages}{4414} (\bibinfo{year}{2016}),
  \urlprefix\url{http://dx.doi.org/10.1039/C5TC04339E}.

\bibitem[{\citenamefont{Marrazzo et~al.}(2019)\citenamefont{Marrazzo,
  Gibertini, Campi, Mounet, and Marzari}}]{Marrazzo2019}
\bibinfo{author}{\bibfnamefont{A.}~\bibnamefont{Marrazzo}},
  \bibinfo{author}{\bibfnamefont{M.}~\bibnamefont{Gibertini}},
  \bibinfo{author}{\bibfnamefont{D.}~\bibnamefont{Campi}},
  \bibinfo{author}{\bibfnamefont{N.}~\bibnamefont{Mounet}}, \bibnamefont{and}
  \bibinfo{author}{\bibfnamefont{N.}~\bibnamefont{Marzari}},
  \bibinfo{journal}{Nano Letters} \textbf{\bibinfo{volume}{19}},
  \bibinfo{pages}{8431} (\bibinfo{year}{2019}), ISSN \bibinfo{issn}{1530-6984},
  \urlprefix\url{https://doi.org/10.1021/acs.nanolett.9b02689}.

\bibitem[{\citenamefont{Zhang et~al.}(2019)\citenamefont{Zhang, Zhang, Zhao,
  Yao, Chen, and Zhou}}]{Zhang2019}
\bibinfo{author}{\bibfnamefont{Z.}~\bibnamefont{Zhang}},
  \bibinfo{author}{\bibfnamefont{X.}~\bibnamefont{Zhang}},
  \bibinfo{author}{\bibfnamefont{X.}~\bibnamefont{Zhao}},
  \bibinfo{author}{\bibfnamefont{S.}~\bibnamefont{Yao}},
  \bibinfo{author}{\bibfnamefont{A.}~\bibnamefont{Chen}}, \bibnamefont{and}
  \bibinfo{author}{\bibfnamefont{Z.}~\bibnamefont{Zhou}}, \bibinfo{journal}{ACS
  Omega} \textbf{\bibinfo{volume}{4}}, \bibinfo{pages}{7822}
  (\bibinfo{year}{2019}),
  \urlprefix\url{https://doi.org/10.1021/acsomega.9b00482}.

\bibitem[{\citenamefont{Kahle et~al.}(2020)\citenamefont{Kahle, Marcolongo, and
  Marzari}}]{Kahle2020}
\bibinfo{author}{\bibfnamefont{L.}~\bibnamefont{Kahle}},
  \bibinfo{author}{\bibfnamefont{A.}~\bibnamefont{Marcolongo}},
  \bibnamefont{and} \bibinfo{author}{\bibfnamefont{N.}~\bibnamefont{Marzari}},
  \bibinfo{journal}{Energy Environ. Sci.} \textbf{\bibinfo{volume}{13}},
  \bibinfo{pages}{928} (\bibinfo{year}{2020}),
  \urlprefix\url{http://dx.doi.org/10.1039/C9EE02457C}.

\bibitem[{\citenamefont{Godby et~al.}(1986)\citenamefont{Godby, Schl\"uter, and
  Sham}}]{Godby1986}
\bibinfo{author}{\bibfnamefont{R.~W.} \bibnamefont{Godby}},
  \bibinfo{author}{\bibfnamefont{M.}~\bibnamefont{Schl\"uter}},
  \bibnamefont{and} \bibinfo{author}{\bibfnamefont{L.~J.} \bibnamefont{Sham}},
  \bibinfo{journal}{Phys. Rev. Lett.} \textbf{\bibinfo{volume}{56}},
  \bibinfo{pages}{2415} (\bibinfo{year}{1986}),
  \urlprefix\url{https://link.aps.org/doi/10.1103/PhysRevLett.56.2415}.

\bibitem[{\citenamefont{Golze et~al.}(2019)\citenamefont{Golze, Dvorak, and
  Rinke}}]{Golze2019}
\bibinfo{author}{\bibfnamefont{D.}~\bibnamefont{Golze}},
  \bibinfo{author}{\bibfnamefont{M.}~\bibnamefont{Dvorak}}, \bibnamefont{and}
  \bibinfo{author}{\bibfnamefont{P.}~\bibnamefont{Rinke}},
  \bibinfo{journal}{Front. Chem.} \textbf{\bibinfo{volume}{7}}
  (\bibinfo{year}{2019}).

\bibitem[{\citenamefont{Onida et~al.}(2002)\citenamefont{Onida, Reining, and
  Rubio}}]{Onida_2002}
\bibinfo{author}{\bibfnamefont{G.}~\bibnamefont{Onida}},
  \bibinfo{author}{\bibfnamefont{L.}~\bibnamefont{Reining}}, \bibnamefont{and}
  \bibinfo{author}{\bibfnamefont{A.}~\bibnamefont{Rubio}},
  \bibinfo{journal}{Rev. Mod. Phys.} \textbf{\bibinfo{volume}{74}},
  \bibinfo{pages}{601} (\bibinfo{year}{2002}),
  \urlprefix\url{https://link.aps.org/doi/10.1103/RevModPhys.74.601}.

\bibitem[{\citenamefont{Hachmann et~al.}(2011)\citenamefont{Hachmann,
  Olivares-Amaya, Atahan-Evrenk, Amador-Bedolla, S{\'a}nchez-Carrera,
  Gold-Parker, Vogt, Brockway, and Aspuru-Guzik}}]{Hachmann2011}
\bibinfo{author}{\bibfnamefont{J.}~\bibnamefont{Hachmann}},
  \bibinfo{author}{\bibfnamefont{R.}~\bibnamefont{Olivares-Amaya}},
  \bibinfo{author}{\bibfnamefont{S.}~\bibnamefont{Atahan-Evrenk}},
  \bibinfo{author}{\bibfnamefont{C.}~\bibnamefont{Amador-Bedolla}},
  \bibinfo{author}{\bibfnamefont{R.~S.} \bibnamefont{S{\'a}nchez-Carrera}},
  \bibinfo{author}{\bibfnamefont{A.}~\bibnamefont{Gold-Parker}},
  \bibinfo{author}{\bibfnamefont{L.}~\bibnamefont{Vogt}},
  \bibinfo{author}{\bibfnamefont{A.~M.} \bibnamefont{Brockway}},
  \bibnamefont{and}
  \bibinfo{author}{\bibfnamefont{A.}~\bibnamefont{Aspuru-Guzik}},
  \bibinfo{journal}{The Journal of Physical Chemistry Letters}
  \textbf{\bibinfo{volume}{2}}, \bibinfo{pages}{2241} (\bibinfo{year}{2011}),
  \urlprefix\url{https://doi.org/10.1021/jz200866s}.

\bibitem[{\citenamefont{He et~al.}(2019)\citenamefont{He, Yu, Li, and
  Zhao}}]{He2019}
\bibinfo{author}{\bibfnamefont{Q.}~\bibnamefont{He}},
  \bibinfo{author}{\bibfnamefont{B.}~\bibnamefont{Yu}},
  \bibinfo{author}{\bibfnamefont{Z.}~\bibnamefont{Li}}, \bibnamefont{and}
  \bibinfo{author}{\bibfnamefont{Y.}~\bibnamefont{Zhao}},
  \bibinfo{journal}{ENERGY \& ENVIRONMENTAL MATERIALS}
  \textbf{\bibinfo{volume}{2}}, \bibinfo{pages}{264} (\bibinfo{year}{2019}),
  \eprint{https://onlinelibrary.wiley.com/doi/pdf/10.1002/eem2.12056},
  \urlprefix\url{https://onlinelibrary.wiley.com/doi/abs/10.1002/eem2.12056}.

\bibitem[{\citenamefont{Lee et~al.}(2018)\citenamefont{Lee, Youn, Yim, and
  Han}}]{Lee2018}
\bibinfo{author}{\bibfnamefont{M.}~\bibnamefont{Lee}},
  \bibinfo{author}{\bibfnamefont{Y.}~\bibnamefont{Youn}},
  \bibinfo{author}{\bibfnamefont{K.}~\bibnamefont{Yim}}, \bibnamefont{and}
  \bibinfo{author}{\bibfnamefont{S.}~\bibnamefont{Han}},
  \bibinfo{journal}{Scientific Reports} \textbf{\bibinfo{volume}{8}},
  \bibinfo{pages}{14794} (\bibinfo{year}{2018}), ISSN
  \bibinfo{issn}{2045-2322},
  \urlprefix\url{https://doi.org/10.1038/s41598-018-33095-6}.

\bibitem[{\citenamefont{Dudarev et~al.}(1998)\citenamefont{Dudarev, Botton,
  Savrasov, Humphreys, and Sutton}}]{DFTU}
\bibinfo{author}{\bibfnamefont{S.~L.} \bibnamefont{Dudarev}},
  \bibinfo{author}{\bibfnamefont{G.~A.} \bibnamefont{Botton}},
  \bibinfo{author}{\bibfnamefont{S.~Y.} \bibnamefont{Savrasov}},
  \bibinfo{author}{\bibfnamefont{C.~J.} \bibnamefont{Humphreys}},
  \bibnamefont{and} \bibinfo{author}{\bibfnamefont{A.~P.}
  \bibnamefont{Sutton}}, \bibinfo{journal}{Phys. Rev. B}
  \textbf{\bibinfo{volume}{57}}, \bibinfo{pages}{1505} (\bibinfo{year}{1998}),
  \urlprefix\url{https://link.aps.org/doi/10.1103/PhysRevB.57.1505}.

\bibitem[{\citenamefont{Krukau et~al.}(2006)\citenamefont{Krukau, Vydrov,
  Izmaylov, and Scuseria}}]{HSE06}
\bibinfo{author}{\bibfnamefont{A.~V.} \bibnamefont{Krukau}},
  \bibinfo{author}{\bibfnamefont{O.~A.} \bibnamefont{Vydrov}},
  \bibinfo{author}{\bibfnamefont{A.~F.} \bibnamefont{Izmaylov}},
  \bibnamefont{and} \bibinfo{author}{\bibfnamefont{G.~E.}
  \bibnamefont{Scuseria}}, \bibinfo{journal}{The Journal of Chemical Physics}
  \textbf{\bibinfo{volume}{125}}, \bibinfo{pages}{224106}
  (\bibinfo{year}{2006}), ISSN \bibinfo{issn}{0021-9606},
  \eprint{https://pubs.aip.org/aip/jcp/article-pdf/doi/10.1063/1.2404663/13263224/224106\_1\_online.pdf},
  \urlprefix\url{https://doi.org/10.1063/1.2404663}.

\bibitem[{\citenamefont{{Yan} et~al.}(2017)\citenamefont{{Yan}, {Yu}, {Suram},
  {Zhou}, {Shinde}, {Newhouse}, {Chen}, {Li}, {Persson}, {Gregoire}
  et~al.}}]{Yan2017}
\bibinfo{author}{\bibfnamefont{Q.}~\bibnamefont{{Yan}}},
  \bibinfo{author}{\bibfnamefont{J.}~\bibnamefont{{Yu}}},
  \bibinfo{author}{\bibfnamefont{S.~K.} \bibnamefont{{Suram}}},
  \bibinfo{author}{\bibfnamefont{L.}~\bibnamefont{{Zhou}}},
  \bibinfo{author}{\bibfnamefont{A.}~\bibnamefont{{Shinde}}},
  \bibinfo{author}{\bibfnamefont{P.~F.} \bibnamefont{{Newhouse}}},
  \bibinfo{author}{\bibfnamefont{W.}~\bibnamefont{{Chen}}},
  \bibinfo{author}{\bibfnamefont{G.}~\bibnamefont{{Li}}},
  \bibinfo{author}{\bibfnamefont{K.~A.} \bibnamefont{{Persson}}},
  \bibinfo{author}{\bibfnamefont{J.~M.} \bibnamefont{{Gregoire}}},
  \bibnamefont{et~al.}, \bibinfo{journal}{Proceedings of the National Academy
  of Science} \textbf{\bibinfo{volume}{114}}, \bibinfo{pages}{3040}
  (\bibinfo{year}{2017}).

\bibitem[{\citenamefont{Xiong et~al.}(2021)\citenamefont{Xiong, Campbell,
  Fanghanel, Badding, Wang, Kirchner-Hall, Theibault, Timrov, Mondschein, Seth
  et~al.}}]{Xiong2021}
\bibinfo{author}{\bibfnamefont{Y.}~\bibnamefont{Xiong}},
  \bibinfo{author}{\bibfnamefont{Q.~T.} \bibnamefont{Campbell}},
  \bibinfo{author}{\bibfnamefont{J.}~\bibnamefont{Fanghanel}},
  \bibinfo{author}{\bibfnamefont{C.~K.} \bibnamefont{Badding}},
  \bibinfo{author}{\bibfnamefont{H.}~\bibnamefont{Wang}},
  \bibinfo{author}{\bibfnamefont{N.~E.} \bibnamefont{Kirchner-Hall}},
  \bibinfo{author}{\bibfnamefont{M.~J.} \bibnamefont{Theibault}},
  \bibinfo{author}{\bibfnamefont{I.}~\bibnamefont{Timrov}},
  \bibinfo{author}{\bibfnamefont{J.~S.} \bibnamefont{Mondschein}},
  \bibinfo{author}{\bibfnamefont{K.}~\bibnamefont{Seth}}, \bibnamefont{et~al.},
  \bibinfo{journal}{Energy Environ. Sci.} \textbf{\bibinfo{volume}{14}},
  \bibinfo{pages}{2335} (\bibinfo{year}{2021}),
  \urlprefix\url{http://dx.doi.org/10.1039/D0EE02984J}.

\bibitem[{\citenamefont{Kim et~al.}(2020)\citenamefont{Kim, Lee, Hong, Yoon,
  An, Lee, Jeong, Yoo, Kang, Youn et~al.}}]{Kim2020}
\bibinfo{author}{\bibfnamefont{S.}~\bibnamefont{Kim}},
  \bibinfo{author}{\bibfnamefont{M.}~\bibnamefont{Lee}},
  \bibinfo{author}{\bibfnamefont{C.}~\bibnamefont{Hong}},
  \bibinfo{author}{\bibfnamefont{Y.}~\bibnamefont{Yoon}},
  \bibinfo{author}{\bibfnamefont{H.}~\bibnamefont{An}},
  \bibinfo{author}{\bibfnamefont{D.}~\bibnamefont{Lee}},
  \bibinfo{author}{\bibfnamefont{W.}~\bibnamefont{Jeong}},
  \bibinfo{author}{\bibfnamefont{D.}~\bibnamefont{Yoo}},
  \bibinfo{author}{\bibfnamefont{Y.}~\bibnamefont{Kang}},
  \bibinfo{author}{\bibfnamefont{Y.}~\bibnamefont{Youn}}, \bibnamefont{et~al.},
  \bibinfo{journal}{Scientific Data} \textbf{\bibinfo{volume}{7}},
  \bibinfo{pages}{387} (\bibinfo{year}{2020}), ISSN \bibinfo{issn}{2052-4463},
  \urlprefix\url{https://doi.org/10.1038/s41597-020-00723-8}.

\bibitem[{\citenamefont{Hinuma et~al.}(2017)\citenamefont{Hinuma, Kumagai,
  Tanaka, and Oba}}]{Hinuma2017}
\bibinfo{author}{\bibfnamefont{Y.}~\bibnamefont{Hinuma}},
  \bibinfo{author}{\bibfnamefont{Y.}~\bibnamefont{Kumagai}},
  \bibinfo{author}{\bibfnamefont{I.}~\bibnamefont{Tanaka}}, \bibnamefont{and}
  \bibinfo{author}{\bibfnamefont{F.}~\bibnamefont{Oba}},
  \bibinfo{journal}{Phys. Rev. B} \textbf{\bibinfo{volume}{95}},
  \bibinfo{pages}{075302} (\bibinfo{year}{2017}),
  \urlprefix\url{https://link.aps.org/doi/10.1103/PhysRevB.95.075302}.

\bibitem[{\citenamefont{Liu et~al.}(2024)\citenamefont{Liu, Gopakumar, Hegde,
  He, and Wolverton}}]{Liu2024}
\bibinfo{author}{\bibfnamefont{M.}~\bibnamefont{Liu}},
  \bibinfo{author}{\bibfnamefont{A.}~\bibnamefont{Gopakumar}},
  \bibinfo{author}{\bibfnamefont{V.~I.} \bibnamefont{Hegde}},
  \bibinfo{author}{\bibfnamefont{J.}~\bibnamefont{He}}, \bibnamefont{and}
  \bibinfo{author}{\bibfnamefont{C.}~\bibnamefont{Wolverton}},
  \bibinfo{journal}{Phys. Rev. Materials}  (\bibinfo{year}{2024}),
  \bibinfo{note}{accepted}.

\bibitem[{\citenamefont{Castelli et~al.}(2012)\citenamefont{Castelli, Olsen,
  Datta, Landis, Dahl, Thygesen, and Jacobsen}}]{Castelli2012}
\bibinfo{author}{\bibfnamefont{I.~E.} \bibnamefont{Castelli}},
  \bibinfo{author}{\bibfnamefont{T.}~\bibnamefont{Olsen}},
  \bibinfo{author}{\bibfnamefont{S.}~\bibnamefont{Datta}},
  \bibinfo{author}{\bibfnamefont{D.~D.} \bibnamefont{Landis}},
  \bibinfo{author}{\bibfnamefont{S.}~\bibnamefont{Dahl}},
  \bibinfo{author}{\bibfnamefont{K.~S.} \bibnamefont{Thygesen}},
  \bibnamefont{and} \bibinfo{author}{\bibfnamefont{K.~W.}
  \bibnamefont{Jacobsen}}, \bibinfo{journal}{Energy Environ. Sci.}
  \textbf{\bibinfo{volume}{5}}, \bibinfo{pages}{5814} (\bibinfo{year}{2012}),
  \urlprefix\url{http://dx.doi.org/10.1039/C1EE02717D}.

\bibitem[{\citenamefont{Kuhar et~al.}(2018)\citenamefont{Kuhar, Pandey,
  Thygesen, and Jacobsen}}]{Kuhar2018}
\bibinfo{author}{\bibfnamefont{K.}~\bibnamefont{Kuhar}},
  \bibinfo{author}{\bibfnamefont{M.}~\bibnamefont{Pandey}},
  \bibinfo{author}{\bibfnamefont{K.~S.} \bibnamefont{Thygesen}},
  \bibnamefont{and} \bibinfo{author}{\bibfnamefont{K.~W.}
  \bibnamefont{Jacobsen}}, \bibinfo{journal}{ACS Energy Letters}
  \textbf{\bibinfo{volume}{3}}, \bibinfo{pages}{436} (\bibinfo{year}{2018}),
  \urlprefix\url{https://doi.org/10.1021/acsenergylett.7b01312}.

\bibitem[{\citenamefont{{Van Setten} et~al.}(2007)\citenamefont{{Van Setten},
  Popa, {De Wijs}, and Brocks}}]{VanSetten2007}
\bibinfo{author}{\bibfnamefont{M.~J.} \bibnamefont{{Van Setten}}},
  \bibinfo{author}{\bibfnamefont{V.~A.} \bibnamefont{Popa}},
  \bibinfo{author}{\bibfnamefont{G.~A.} \bibnamefont{{De Wijs}}},
  \bibnamefont{and} \bibinfo{author}{\bibfnamefont{G.}~\bibnamefont{Brocks}},
  \bibinfo{journal}{Phys. Rev. B - Condens. Matter Mater. Phys.}
  \textbf{\bibinfo{volume}{75}}, \bibinfo{pages}{035204}
  (\bibinfo{year}{2007}), ISSN \bibinfo{issn}{10980121}, \eprint{0609189},
  \urlprefix\url{https://journals.aps.org/prb/abstract/10.1103/PhysRevB.75.035204}.

\bibitem[{\citenamefont{Erg{\"{o}}nenc
  et~al.}(2018)\citenamefont{Erg{\"{o}}nenc, Kim, Liu, Kresse, and
  Franchini}}]{Ergorenc2018}
\bibinfo{author}{\bibfnamefont{Z.}~\bibnamefont{Erg{\"{o}}nenc}},
  \bibinfo{author}{\bibfnamefont{B.}~\bibnamefont{Kim}},
  \bibinfo{author}{\bibfnamefont{P.}~\bibnamefont{Liu}},
  \bibinfo{author}{\bibfnamefont{G.}~\bibnamefont{Kresse}}, \bibnamefont{and}
  \bibinfo{author}{\bibfnamefont{C.}~\bibnamefont{Franchini}},
  \bibinfo{journal}{Phys. Rev. Mater.} \textbf{\bibinfo{volume}{2}}
  (\bibinfo{year}{2018}).

\bibitem[{\citenamefont{Haastrup et~al.}(2018)\citenamefont{Haastrup, Strange,
  Pandey, Deilmann, Schmidt, Hinsche, Gjerding, Torelli, Larsen, Riis-Jensen
  et~al.}}]{Haastrup2018}
\bibinfo{author}{\bibfnamefont{S.}~\bibnamefont{Haastrup}},
  \bibinfo{author}{\bibfnamefont{M.}~\bibnamefont{Strange}},
  \bibinfo{author}{\bibfnamefont{M.}~\bibnamefont{Pandey}},
  \bibinfo{author}{\bibfnamefont{T.}~\bibnamefont{Deilmann}},
  \bibinfo{author}{\bibfnamefont{P.~S.} \bibnamefont{Schmidt}},
  \bibinfo{author}{\bibfnamefont{N.~F.} \bibnamefont{Hinsche}},
  \bibinfo{author}{\bibfnamefont{M.~N.} \bibnamefont{Gjerding}},
  \bibinfo{author}{\bibfnamefont{D.}~\bibnamefont{Torelli}},
  \bibinfo{author}{\bibfnamefont{P.~M.} \bibnamefont{Larsen}},
  \bibinfo{author}{\bibfnamefont{A.~C.} \bibnamefont{Riis-Jensen}},
  \bibnamefont{et~al.}, \bibinfo{journal}{2D Materials}
  \textbf{\bibinfo{volume}{5}}, \bibinfo{pages}{042002} (\bibinfo{year}{2018}),
  \urlprefix\url{https://dx.doi.org/10.1088/2053-1583/aacfc1}.

\bibitem[{\citenamefont{Rasmussen et~al.}(2021)\citenamefont{Rasmussen,
  Deilmann, and Thygesen}}]{Rasmussen2021}
\bibinfo{author}{\bibfnamefont{A.}~\bibnamefont{Rasmussen}},
  \bibinfo{author}{\bibfnamefont{T.}~\bibnamefont{Deilmann}}, \bibnamefont{and}
  \bibinfo{author}{\bibfnamefont{K.~S.} \bibnamefont{Thygesen}},
  \bibinfo{journal}{npj Computational Materials} \textbf{\bibinfo{volume}{7}},
  \bibinfo{pages}{22} (\bibinfo{year}{2021}), ISSN \bibinfo{issn}{2057-3960},
  \urlprefix\url{https://doi.org/10.1038/s41524-020-00480-7}.

\bibitem[{\citenamefont{Lee et~al.}(2016)\citenamefont{Lee, Seko, Shitara,
  Nakayama, and Tanaka}}]{Lee2016}
\bibinfo{author}{\bibfnamefont{J.}~\bibnamefont{Lee}},
  \bibinfo{author}{\bibfnamefont{A.}~\bibnamefont{Seko}},
  \bibinfo{author}{\bibfnamefont{K.}~\bibnamefont{Shitara}},
  \bibinfo{author}{\bibfnamefont{K.}~\bibnamefont{Nakayama}}, \bibnamefont{and}
  \bibinfo{author}{\bibfnamefont{I.}~\bibnamefont{Tanaka}},
  \bibinfo{journal}{Phys. Rev. B} \textbf{\bibinfo{volume}{93}},
  \bibinfo{pages}{115104} (\bibinfo{year}{2016}),
  \urlprefix\url{https://link.aps.org/doi/10.1103/PhysRevB.93.115104}.

\bibitem[{\citenamefont{van Setten et~al.}(2017)\citenamefont{van Setten,
  Giantomassi, Gonze, Rignanese, and Hautier}}]{VanSetten2017}
\bibinfo{author}{\bibfnamefont{M.~J.} \bibnamefont{van Setten}},
  \bibinfo{author}{\bibfnamefont{M.}~\bibnamefont{Giantomassi}},
  \bibinfo{author}{\bibfnamefont{X.}~\bibnamefont{Gonze}},
  \bibinfo{author}{\bibfnamefont{G.-M.} \bibnamefont{Rignanese}},
  \bibnamefont{and} \bibinfo{author}{\bibfnamefont{G.}~\bibnamefont{Hautier}},
  \bibinfo{journal}{Phys. Rev. B} \textbf{\bibinfo{volume}{96}},
  \bibinfo{pages}{155207} (\bibinfo{year}{2017}),
  \urlprefix\url{https://link.aps.org/doi/10.1103/PhysRevB.96.155207}.

\bibitem[{\citenamefont{Bonacci et~al.}(2023)\citenamefont{Bonacci, Qiao,
  Spallanzani, Marrazzo, Pizzi, Molinari, Varsano, Ferretti, and
  Prezzi}}]{Bonacci2023}
\bibinfo{author}{\bibfnamefont{M.}~\bibnamefont{Bonacci}},
  \bibinfo{author}{\bibfnamefont{J.}~\bibnamefont{Qiao}},
  \bibinfo{author}{\bibfnamefont{N.}~\bibnamefont{Spallanzani}},
  \bibinfo{author}{\bibfnamefont{A.}~\bibnamefont{Marrazzo}},
  \bibinfo{author}{\bibfnamefont{G.}~\bibnamefont{Pizzi}},
  \bibinfo{author}{\bibfnamefont{E.}~\bibnamefont{Molinari}},
  \bibinfo{author}{\bibfnamefont{D.}~\bibnamefont{Varsano}},
  \bibinfo{author}{\bibfnamefont{A.}~\bibnamefont{Ferretti}}, \bibnamefont{and}
  \bibinfo{author}{\bibfnamefont{D.}~\bibnamefont{Prezzi}},
  \emph{\bibinfo{title}{Towards high-throughput many-body perturbation theory:
  efficient algorithms and automated workflows}} (\bibinfo{year}{2023}),
  \eprint{2301.06407}.

\bibitem[{\citenamefont{Biswas and Singh}(2023)}]{Biswas2023}
\bibinfo{author}{\bibfnamefont{T.}~\bibnamefont{Biswas}} \bibnamefont{and}
  \bibinfo{author}{\bibfnamefont{A.~K.} \bibnamefont{Singh}},
  \bibinfo{journal}{npj Computational Materials} \textbf{\bibinfo{volume}{9}},
  \bibinfo{pages}{22} (\bibinfo{year}{2023}), ISSN \bibinfo{issn}{2057-3960},
  \urlprefix\url{https://doi.org/10.1038/s41524-023-00976-y}.

\bibitem[{\citenamefont{Gro{\ss}mann et~al.}(2024)\citenamefont{Gro{\ss}mann,
  Grunert, and Runge}}]{Großmann2024}
\bibinfo{author}{\bibfnamefont{M.}~\bibnamefont{Gro{\ss}mann}},
  \bibinfo{author}{\bibfnamefont{M.}~\bibnamefont{Grunert}}, \bibnamefont{and}
  \bibinfo{author}{\bibfnamefont{E.}~\bibnamefont{Runge}},
  \bibinfo{journal}{npj Computational Materials} \textbf{\bibinfo{volume}{10}},
  \bibinfo{pages}{135} (\bibinfo{year}{2024}), ISSN \bibinfo{issn}{2057-3960},
  \urlprefix\url{https://doi.org/10.1038/s41524-024-01311-9}.

\bibitem[{\citenamefont{Hedin}(1965)}]{Hedin1965}
\bibinfo{author}{\bibfnamefont{L.}~\bibnamefont{Hedin}},
  \bibinfo{journal}{Phys. Rev.} \textbf{\bibinfo{volume}{139}},
  \bibinfo{pages}{A796} (\bibinfo{year}{1965}),
  \urlprefix\url{https://link.aps.org/doi/10.1103/PhysRev.139.A796}.

\bibitem[{\citenamefont{Strinati et~al.}(1982)\citenamefont{Strinati,
  Mattausch, and Hanke}}]{Strinati1982}
\bibinfo{author}{\bibfnamefont{G.}~\bibnamefont{Strinati}},
  \bibinfo{author}{\bibfnamefont{H.~J.} \bibnamefont{Mattausch}},
  \bibnamefont{and} \bibinfo{author}{\bibfnamefont{W.}~\bibnamefont{Hanke}},
  \bibinfo{journal}{Phys. Rev. B} \textbf{\bibinfo{volume}{25}},
  \bibinfo{pages}{2867} (\bibinfo{year}{1982}),
  \urlprefix\url{https://link.aps.org/doi/10.1103/PhysRevB.25.2867}.

\bibitem[{\citenamefont{Reining}(2018)}]{Reining2018}
\bibinfo{author}{\bibfnamefont{L.}~\bibnamefont{Reining}},
  \bibinfo{journal}{WIREs Computational Molecular Science}
  \textbf{\bibinfo{volume}{8}}, \bibinfo{pages}{e1344} (\bibinfo{year}{2018}),
  \eprint{https://wires.onlinelibrary.wiley.com/doi/pdf/10.1002/wcms.1344},
  \urlprefix\url{https://wires.onlinelibrary.wiley.com/doi/abs/10.1002/wcms.1344}.

\bibitem[{\citenamefont{Shishkin and Kresse}(2007)}]{Shishkin2007}
\bibinfo{author}{\bibfnamefont{M.}~\bibnamefont{Shishkin}} \bibnamefont{and}
  \bibinfo{author}{\bibfnamefont{G.}~\bibnamefont{Kresse}},
  \bibinfo{journal}{Phys. Rev. B} \textbf{\bibinfo{volume}{75}},
  \bibinfo{pages}{235102} (\bibinfo{year}{2007}),
  \urlprefix\url{https://link.aps.org/doi/10.1103/PhysRevB.75.235102}.

\bibitem[{\citenamefont{van Schilfgaarde
  et~al.}(2006{\natexlab{a}})\citenamefont{van Schilfgaarde, Kotani, and
  Faleev}}]{Schilfgaarde_2006}
\bibinfo{author}{\bibfnamefont{M.}~\bibnamefont{van Schilfgaarde}},
  \bibinfo{author}{\bibfnamefont{T.}~\bibnamefont{Kotani}}, \bibnamefont{and}
  \bibinfo{author}{\bibfnamefont{S.}~\bibnamefont{Faleev}},
  \bibinfo{journal}{Phys. Rev. Lett.} \textbf{\bibinfo{volume}{96}},
  \bibinfo{pages}{226402} (\bibinfo{year}{2006}{\natexlab{a}}),
  \urlprefix\url{https://link.aps.org/doi/10.1103/PhysRevLett.96.226402}.

\bibitem[{\citenamefont{van Setten et~al.}(2015)\citenamefont{van Setten,
  Caruso, Sharifzadeh, Ren, Scheffler, Liu, Lischner, Lin, Deslippe, Louie
  et~al.}}]{vanSetten2015}
\bibinfo{author}{\bibfnamefont{M.~J.} \bibnamefont{van Setten}},
  \bibinfo{author}{\bibfnamefont{F.}~\bibnamefont{Caruso}},
  \bibinfo{author}{\bibfnamefont{S.}~\bibnamefont{Sharifzadeh}},
  \bibinfo{author}{\bibfnamefont{X.}~\bibnamefont{Ren}},
  \bibinfo{author}{\bibfnamefont{M.}~\bibnamefont{Scheffler}},
  \bibinfo{author}{\bibfnamefont{F.}~\bibnamefont{Liu}},
  \bibinfo{author}{\bibfnamefont{J.}~\bibnamefont{Lischner}},
  \bibinfo{author}{\bibfnamefont{L.}~\bibnamefont{Lin}},
  \bibinfo{author}{\bibfnamefont{J.~R.} \bibnamefont{Deslippe}},
  \bibinfo{author}{\bibfnamefont{S.~G.} \bibnamefont{Louie}},
  \bibnamefont{et~al.}, \bibinfo{journal}{Journal of Chemical Theory and
  Computation} \textbf{\bibinfo{volume}{11}}, \bibinfo{pages}{5665}
  (\bibinfo{year}{2015}), ISSN \bibinfo{issn}{1549-9618},
  \urlprefix\url{https://doi.org/10.1021/acs.jctc.5b00453}.

\bibitem[{\citenamefont{Knight et~al.}(2016)\citenamefont{Knight, Wang,
  Gallandi, Dolgounitcheva, Ren, Ortiz, Rinke, Körzdörfer, and
  Marom}}]{Knight2016}
\bibinfo{author}{\bibfnamefont{J.~W.} \bibnamefont{Knight}},
  \bibinfo{author}{\bibfnamefont{X.}~\bibnamefont{Wang}},
  \bibinfo{author}{\bibfnamefont{L.}~\bibnamefont{Gallandi}},
  \bibinfo{author}{\bibfnamefont{O.}~\bibnamefont{Dolgounitcheva}},
  \bibinfo{author}{\bibfnamefont{X.}~\bibnamefont{Ren}},
  \bibinfo{author}{\bibfnamefont{J.~V.} \bibnamefont{Ortiz}},
  \bibinfo{author}{\bibfnamefont{P.}~\bibnamefont{Rinke}},
  \bibinfo{author}{\bibfnamefont{T.}~\bibnamefont{Körzdörfer}},
  \bibnamefont{and} \bibinfo{author}{\bibfnamefont{N.}~\bibnamefont{Marom}},
  \bibinfo{journal}{Journal of Chemical Theory and Computation}
  \textbf{\bibinfo{volume}{12}}, \bibinfo{pages}{615} (\bibinfo{year}{2016}),
  \bibinfo{note}{pMID: 26731609},
  \urlprefix\url{https://doi.org/10.1021/acs.jctc.5b00871}.

\bibitem[{\citenamefont{Shih et~al.}(2010{\natexlab{a}})\citenamefont{Shih,
  Xue, Zhang, Cohen, and Louie}}]{Shih2010}
\bibinfo{author}{\bibfnamefont{B.-C.} \bibnamefont{Shih}},
  \bibinfo{author}{\bibfnamefont{Y.}~\bibnamefont{Xue}},
  \bibinfo{author}{\bibfnamefont{P.}~\bibnamefont{Zhang}},
  \bibinfo{author}{\bibfnamefont{M.~L.} \bibnamefont{Cohen}}, \bibnamefont{and}
  \bibinfo{author}{\bibfnamefont{S.~G.} \bibnamefont{Louie}},
  \bibinfo{journal}{Phys. Rev. Lett.} \textbf{\bibinfo{volume}{105}},
  \bibinfo{pages}{146401} (\bibinfo{year}{2010}{\natexlab{a}}),
  \urlprefix\url{https://link.aps.org/doi/10.1103/PhysRevLett.105.146401}.

\bibitem[{\citenamefont{Friedrich
  et~al.}(2011{\natexlab{a}})\citenamefont{Friedrich, M\"uller, and
  Bl\"ugel}}]{Friedrich2011}
\bibinfo{author}{\bibfnamefont{C.}~\bibnamefont{Friedrich}},
  \bibinfo{author}{\bibfnamefont{M.~C.} \bibnamefont{M\"uller}},
  \bibnamefont{and} \bibinfo{author}{\bibfnamefont{S.}~\bibnamefont{Bl\"ugel}},
  \bibinfo{journal}{Phys. Rev. B} \textbf{\bibinfo{volume}{83}},
  \bibinfo{pages}{081101} (\bibinfo{year}{2011}{\natexlab{a}}),
  \urlprefix\url{https://link.aps.org/doi/10.1103/PhysRevB.83.081101}.

\bibitem[{\citenamefont{Klime\ifmmode~\check{s}\else \v{s}\fi{}
  et~al.}(2014)\citenamefont{Klime\ifmmode~\check{s}\else \v{s}\fi{}, Kaltak,
  and Kresse}}]{Klimes2014}
\bibinfo{author}{\bibfnamefont{J.~c.~v.}
  \bibnamefont{Klime\ifmmode~\check{s}\else \v{s}\fi{}}},
  \bibinfo{author}{\bibfnamefont{M.}~\bibnamefont{Kaltak}}, \bibnamefont{and}
  \bibinfo{author}{\bibfnamefont{G.}~\bibnamefont{Kresse}},
  \bibinfo{journal}{Phys. Rev. B} \textbf{\bibinfo{volume}{90}},
  \bibinfo{pages}{075125} (\bibinfo{year}{2014}),
  \urlprefix\url{https://link.aps.org/doi/10.1103/PhysRevB.90.075125}.

\bibitem[{\citenamefont{Stankovski et~al.}(2011)\citenamefont{Stankovski,
  Antonius, Waroquiers, Miglio, Dixit, Sankaran, Giantomassi, Gonze, C\^ot\'e,
  and Rignanese}}]{falseConv_2}
\bibinfo{author}{\bibfnamefont{M.}~\bibnamefont{Stankovski}},
  \bibinfo{author}{\bibfnamefont{G.}~\bibnamefont{Antonius}},
  \bibinfo{author}{\bibfnamefont{D.}~\bibnamefont{Waroquiers}},
  \bibinfo{author}{\bibfnamefont{A.}~\bibnamefont{Miglio}},
  \bibinfo{author}{\bibfnamefont{H.}~\bibnamefont{Dixit}},
  \bibinfo{author}{\bibfnamefont{K.}~\bibnamefont{Sankaran}},
  \bibinfo{author}{\bibfnamefont{M.}~\bibnamefont{Giantomassi}},
  \bibinfo{author}{\bibfnamefont{X.}~\bibnamefont{Gonze}},
  \bibinfo{author}{\bibfnamefont{M.}~\bibnamefont{C\^ot\'e}}, \bibnamefont{and}
  \bibinfo{author}{\bibfnamefont{G.-M.} \bibnamefont{Rignanese}},
  \bibinfo{journal}{Phys. Rev. B} \textbf{\bibinfo{volume}{84}},
  \bibinfo{pages}{241201} (\bibinfo{year}{2011}),
  \urlprefix\url{https://link.aps.org/doi/10.1103/PhysRevB.84.241201}.

\bibitem[{\citenamefont{Gao et~al.}(2016)\citenamefont{Gao, Xia, Gao, and
  Zhang}}]{Gao2016}
\bibinfo{author}{\bibfnamefont{W.}~\bibnamefont{Gao}},
  \bibinfo{author}{\bibfnamefont{W.}~\bibnamefont{Xia}},
  \bibinfo{author}{\bibfnamefont{X.}~\bibnamefont{Gao}}, \bibnamefont{and}
  \bibinfo{author}{\bibfnamefont{P.}~\bibnamefont{Zhang}},
  \bibinfo{journal}{Scientific Reports} \textbf{\bibinfo{volume}{6}},
  \bibinfo{pages}{36849} (\bibinfo{year}{2016}), ISSN
  \bibinfo{issn}{2045-2322}, \urlprefix\url{https://doi.org/10.1038/srep36849}.

\bibitem[{\citenamefont{Rangel et~al.}(2020)\citenamefont{Rangel, {Del Ben},
  Varsano, Antonius, Bruneval, {da Jornada}, {van Setten}, Orhan, O’Regan,
  Canning et~al.}}]{ZnOwz_1}
\bibinfo{author}{\bibfnamefont{T.}~\bibnamefont{Rangel}},
  \bibinfo{author}{\bibfnamefont{M.}~\bibnamefont{{Del Ben}}},
  \bibinfo{author}{\bibfnamefont{D.}~\bibnamefont{Varsano}},
  \bibinfo{author}{\bibfnamefont{G.}~\bibnamefont{Antonius}},
  \bibinfo{author}{\bibfnamefont{F.}~\bibnamefont{Bruneval}},
  \bibinfo{author}{\bibfnamefont{F.~H.} \bibnamefont{{da Jornada}}},
  \bibinfo{author}{\bibfnamefont{M.~J.} \bibnamefont{{van Setten}}},
  \bibinfo{author}{\bibfnamefont{O.~K.} \bibnamefont{Orhan}},
  \bibinfo{author}{\bibfnamefont{D.~D.} \bibnamefont{O’Regan}},
  \bibinfo{author}{\bibfnamefont{A.}~\bibnamefont{Canning}},
  \bibnamefont{et~al.}, \bibinfo{journal}{Computer Physics Communications}
  \textbf{\bibinfo{volume}{255}}, \bibinfo{pages}{107242}
  (\bibinfo{year}{2020}), ISSN \bibinfo{issn}{0010-4655},
  \urlprefix\url{https://www.sciencedirect.com/science/article/pii/S0010465520300734}.

\bibitem[{\citenamefont{Kresse and Furthm{\"{u}}ller}(1996)}]{Kresse1996}
\bibinfo{author}{\bibfnamefont{G.}~\bibnamefont{Kresse}} \bibnamefont{and}
  \bibinfo{author}{\bibfnamefont{J.}~\bibnamefont{Furthm{\"{u}}ller}},
  \bibinfo{journal}{Phys. Rev. B - Condens. Matter Mater. Phys.}
  \textbf{\bibinfo{volume}{54}}, \bibinfo{pages}{11169} (\bibinfo{year}{1996}),
  ISSN \bibinfo{issn}{1550235X},
  \urlprefix\url{https://journals.aps.org/prb/abstract/10.1103/PhysRevB.54.11169}.

\bibitem[{\citenamefont{Kresse and Furthmüller}(1996)}]{Kresse1996_2}
\bibinfo{author}{\bibfnamefont{G.}~\bibnamefont{Kresse}} \bibnamefont{and}
  \bibinfo{author}{\bibfnamefont{J.}~\bibnamefont{Furthmüller}},
  \bibinfo{journal}{Computational Materials Science}
  \textbf{\bibinfo{volume}{6}}, \bibinfo{pages}{15} (\bibinfo{year}{1996}),
  ISSN \bibinfo{issn}{0927-0256},
  \urlprefix\url{https://www.sciencedirect.com/science/article/pii/0927025696000080}.

\bibitem[{\citenamefont{Bl\"ochl}(1994)}]{Blochl1994}
\bibinfo{author}{\bibfnamefont{P.~E.} \bibnamefont{Bl\"ochl}},
  \bibinfo{journal}{Phys. Rev. B} \textbf{\bibinfo{volume}{50}},
  \bibinfo{pages}{17953} (\bibinfo{year}{1994}),
  \urlprefix\url{https://link.aps.org/doi/10.1103/PhysRevB.50.17953}.

\bibitem[{\citenamefont{Uhrin et~al.}(2021)\citenamefont{Uhrin, Huber, Yu,
  Marzari, and Pizzi}}]{AiiDA2021}
\bibinfo{author}{\bibfnamefont{M.}~\bibnamefont{Uhrin}},
  \bibinfo{author}{\bibfnamefont{S.~P.} \bibnamefont{Huber}},
  \bibinfo{author}{\bibfnamefont{J.}~\bibnamefont{Yu}},
  \bibinfo{author}{\bibfnamefont{N.}~\bibnamefont{Marzari}}, \bibnamefont{and}
  \bibinfo{author}{\bibfnamefont{G.}~\bibnamefont{Pizzi}},
  \bibinfo{journal}{Computational Materials Science}
  \textbf{\bibinfo{volume}{187}}, \bibinfo{pages}{110086}
  (\bibinfo{year}{2021}), ISSN \bibinfo{issn}{0927-0256},
  \urlprefix\url{https://www.sciencedirect.com/science/article/pii/S0927025620305772}.

\bibitem[{\citenamefont{Prandini et~al.}(2018)\citenamefont{Prandini, Marrazzo,
  Castelli, Mounet, and Marzari}}]{Prandini2018}
\bibinfo{author}{\bibfnamefont{G.}~\bibnamefont{Prandini}},
  \bibinfo{author}{\bibfnamefont{A.}~\bibnamefont{Marrazzo}},
  \bibinfo{author}{\bibfnamefont{I.~E.} \bibnamefont{Castelli}},
  \bibinfo{author}{\bibfnamefont{N.}~\bibnamefont{Mounet}}, \bibnamefont{and}
  \bibinfo{author}{\bibfnamefont{N.}~\bibnamefont{Marzari}},
  \bibinfo{journal}{npj Computational Materials} \textbf{\bibinfo{volume}{4}},
  \bibinfo{pages}{72} (\bibinfo{year}{2018}), ISSN \bibinfo{issn}{2057-3960},
  \urlprefix\url{https://doi.org/10.1038/s41524-018-0127-2}.

\bibitem[{\citenamefont{Mercado et~al.}(2018)\citenamefont{Mercado, Fu,
  Yakutovich, Talirz, Haranczyk, and Smit}}]{Mercado2018}
\bibinfo{author}{\bibfnamefont{R.}~\bibnamefont{Mercado}},
  \bibinfo{author}{\bibfnamefont{R.-S.} \bibnamefont{Fu}},
  \bibinfo{author}{\bibfnamefont{A.~V.} \bibnamefont{Yakutovich}},
  \bibinfo{author}{\bibfnamefont{L.}~\bibnamefont{Talirz}},
  \bibinfo{author}{\bibfnamefont{M.}~\bibnamefont{Haranczyk}},
  \bibnamefont{and} \bibinfo{author}{\bibfnamefont{B.}~\bibnamefont{Smit}},
  \bibinfo{journal}{Chemistry of Materials} \textbf{\bibinfo{volume}{30}},
  \bibinfo{pages}{5069} (\bibinfo{year}{2018}), ISSN \bibinfo{issn}{0897-4756},
  \urlprefix\url{https://doi.org/10.1021/acs.chemmater.8b01425}.

\bibitem[{\citenamefont{Atambo et~al.}(2019)\citenamefont{Atambo, Varsano,
  Ferretti, Ataei, Caldas, Molinari, and Selloni}}]{Atambo2019}
\bibinfo{author}{\bibfnamefont{M.~O.} \bibnamefont{Atambo}},
  \bibinfo{author}{\bibfnamefont{D.}~\bibnamefont{Varsano}},
  \bibinfo{author}{\bibfnamefont{A.}~\bibnamefont{Ferretti}},
  \bibinfo{author}{\bibfnamefont{S.~S.} \bibnamefont{Ataei}},
  \bibinfo{author}{\bibfnamefont{M.~J.} \bibnamefont{Caldas}},
  \bibinfo{author}{\bibfnamefont{E.}~\bibnamefont{Molinari}}, \bibnamefont{and}
  \bibinfo{author}{\bibfnamefont{A.}~\bibnamefont{Selloni}},
  \bibinfo{journal}{Phys. Rev. Mater.} \textbf{\bibinfo{volume}{3}},
  \bibinfo{pages}{045401} (\bibinfo{year}{2019}),
  \urlprefix\url{https://link.aps.org/doi/10.1103/PhysRevMaterials.3.045401}.

\bibitem[{\citenamefont{Vitale et~al.}(2020)\citenamefont{Vitale, Pizzi,
  Marrazzo, Yates, Marzari, and Mostofi}}]{Vitale2020}
\bibinfo{author}{\bibfnamefont{V.}~\bibnamefont{Vitale}},
  \bibinfo{author}{\bibfnamefont{G.}~\bibnamefont{Pizzi}},
  \bibinfo{author}{\bibfnamefont{A.}~\bibnamefont{Marrazzo}},
  \bibinfo{author}{\bibfnamefont{J.~R.} \bibnamefont{Yates}},
  \bibinfo{author}{\bibfnamefont{N.}~\bibnamefont{Marzari}}, \bibnamefont{and}
  \bibinfo{author}{\bibfnamefont{A.~A.} \bibnamefont{Mostofi}},
  \bibinfo{journal}{npj Computational Materials} \textbf{\bibinfo{volume}{6}},
  \bibinfo{pages}{66} (\bibinfo{year}{2020}), ISSN \bibinfo{issn}{2057-3960},
  \urlprefix\url{https://doi.org/10.1038/s41524-020-0312-y}.

\bibitem[{\citenamefont{Aryasetiawan and Gunnarsson}(1998)}]{Aryasetiawan1998}
\bibinfo{author}{\bibfnamefont{F.}~\bibnamefont{Aryasetiawan}}
  \bibnamefont{and}
  \bibinfo{author}{\bibfnamefont{O.}~\bibnamefont{Gunnarsson}},
  \bibinfo{journal}{Reports Prog. Phys.} \textbf{\bibinfo{volume}{61}},
  \bibinfo{pages}{237} (\bibinfo{year}{1998}).

\bibitem[{\citenamefont{Shishkin and Kresse}(2006)}]{Shishkin2006}
\bibinfo{author}{\bibfnamefont{M.}~\bibnamefont{Shishkin}} \bibnamefont{and}
  \bibinfo{author}{\bibfnamefont{G.}~\bibnamefont{Kresse}},
  \bibinfo{journal}{Phys. Rev. B - Condens. Matter Mater. Phys.}
  \textbf{\bibinfo{volume}{74}}, \bibinfo{pages}{1} (\bibinfo{year}{2006}),
  ISSN \bibinfo{issn}{10980121}.

\bibitem[{\citenamefont{Harl and Kresse}(2008)}]{AsymConvergProof1}
\bibinfo{author}{\bibfnamefont{J.}~\bibnamefont{Harl}} \bibnamefont{and}
  \bibinfo{author}{\bibfnamefont{G.}~\bibnamefont{Kresse}},
  \bibinfo{journal}{Phys. Rev. B} \textbf{\bibinfo{volume}{77}},
  \bibinfo{pages}{045136} (\bibinfo{year}{2008}),
  \urlprefix\url{https://link.aps.org/doi/10.1103/PhysRevB.77.045136}.

\bibitem[{\citenamefont{Shepherd et~al.}(2012)\citenamefont{Shepherd,
  Gr\"uneis, Booth, Kresse, and Alavi}}]{AsymConvergProof2}
\bibinfo{author}{\bibfnamefont{J.~J.} \bibnamefont{Shepherd}},
  \bibinfo{author}{\bibfnamefont{A.}~\bibnamefont{Gr\"uneis}},
  \bibinfo{author}{\bibfnamefont{G.~H.} \bibnamefont{Booth}},
  \bibinfo{author}{\bibfnamefont{G.}~\bibnamefont{Kresse}}, \bibnamefont{and}
  \bibinfo{author}{\bibfnamefont{A.}~\bibnamefont{Alavi}},
  \bibinfo{journal}{Phys. Rev. B} \textbf{\bibinfo{volume}{86}},
  \bibinfo{pages}{035111} (\bibinfo{year}{2012}),
  \urlprefix\url{https://link.aps.org/doi/10.1103/PhysRevB.86.035111}.

\bibitem[{\citenamefont{Gulans}(2014)}]{AsymConvergProof3}
\bibinfo{author}{\bibfnamefont{A.}~\bibnamefont{Gulans}}, \bibinfo{journal}{The
  Journal of Chemical Physics} \textbf{\bibinfo{volume}{141}},
  \bibinfo{pages}{164127} (\bibinfo{year}{2014}), ISSN
  \bibinfo{issn}{0021-9606},
  \eprint{https://pubs.aip.org/aip/jcp/article-pdf/doi/10.1063/1.4900447/14113213/164127\_1\_online.pdf},
  \urlprefix\url{https://doi.org/10.1063/1.4900447}.

\bibitem[{\citenamefont{Schindlmayr}(2013)}]{AsymConvergProof4}
\bibinfo{author}{\bibfnamefont{A.}~\bibnamefont{Schindlmayr}},
  \bibinfo{journal}{Phys. Rev. B} \textbf{\bibinfo{volume}{87}},
  \bibinfo{pages}{075104} (\bibinfo{year}{2013}),
  \urlprefix\url{https://link.aps.org/doi/10.1103/PhysRevB.87.075104}.

\bibitem[{\citenamefont{Bj\"orkman et~al.}(2012)\citenamefont{Bj\"orkman,
  Gulans, Krasheninnikov, and Nieminen}}]{AsymConvergProof5}
\bibinfo{author}{\bibfnamefont{T.}~\bibnamefont{Bj\"orkman}},
  \bibinfo{author}{\bibfnamefont{A.}~\bibnamefont{Gulans}},
  \bibinfo{author}{\bibfnamefont{A.~V.} \bibnamefont{Krasheninnikov}},
  \bibnamefont{and} \bibinfo{author}{\bibfnamefont{R.~M.}
  \bibnamefont{Nieminen}}, \bibinfo{journal}{Phys. Rev. Lett.}
  \textbf{\bibinfo{volume}{108}}, \bibinfo{pages}{235502}
  (\bibinfo{year}{2012}),
  \urlprefix\url{https://link.aps.org/doi/10.1103/PhysRevLett.108.235502}.

\bibitem[{\citenamefont{Gonze et~al.}(2020)\citenamefont{Gonze, Amadon,
  Antonius, Arnardi, Baguet, Beuken, Bieder, Bottin, Bouchet, Bousquet
  et~al.}}]{Abinit_Gonze2020}
\bibinfo{author}{\bibfnamefont{X.}~\bibnamefont{Gonze}},
  \bibinfo{author}{\bibfnamefont{B.}~\bibnamefont{Amadon}},
  \bibinfo{author}{\bibfnamefont{G.}~\bibnamefont{Antonius}},
  \bibinfo{author}{\bibfnamefont{F.}~\bibnamefont{Arnardi}},
  \bibinfo{author}{\bibfnamefont{L.}~\bibnamefont{Baguet}},
  \bibinfo{author}{\bibfnamefont{J.-M.} \bibnamefont{Beuken}},
  \bibinfo{author}{\bibfnamefont{J.}~\bibnamefont{Bieder}},
  \bibinfo{author}{\bibfnamefont{F.}~\bibnamefont{Bottin}},
  \bibinfo{author}{\bibfnamefont{J.}~\bibnamefont{Bouchet}},
  \bibinfo{author}{\bibfnamefont{E.}~\bibnamefont{Bousquet}},
  \bibnamefont{et~al.}, \bibinfo{journal}{Comput. Phys. Commun.}
  \textbf{\bibinfo{volume}{248}}, \bibinfo{pages}{107042}
  (\bibinfo{year}{2020}),
  \urlprefix\url{https://doi.org/10.1016/j.cpc.2019.107042}.

\bibitem[{\citenamefont{Mortensen et~al.}(2024)\citenamefont{Mortensen, Larsen,
  Kuisma, Ivanov, Taghizadeh, Peterson, Haldar, Dohn, Schäfer, Jonsson
  et~al.}}]{GPAW}
\bibinfo{author}{\bibfnamefont{J.~J.} \bibnamefont{Mortensen}},
  \bibinfo{author}{\bibfnamefont{A.~H.} \bibnamefont{Larsen}},
  \bibinfo{author}{\bibfnamefont{M.}~\bibnamefont{Kuisma}},
  \bibinfo{author}{\bibfnamefont{A.~V.} \bibnamefont{Ivanov}},
  \bibinfo{author}{\bibfnamefont{A.}~\bibnamefont{Taghizadeh}},
  \bibinfo{author}{\bibfnamefont{A.}~\bibnamefont{Peterson}},
  \bibinfo{author}{\bibfnamefont{A.}~\bibnamefont{Haldar}},
  \bibinfo{author}{\bibfnamefont{A.~O.} \bibnamefont{Dohn}},
  \bibinfo{author}{\bibfnamefont{C.}~\bibnamefont{Schäfer}},
  \bibinfo{author}{\bibfnamefont{E.~O.} \bibnamefont{Jonsson}},
  \bibnamefont{et~al.}, \bibinfo{journal}{The Journal of Chemical Physics}
  \textbf{\bibinfo{volume}{160}}, \bibinfo{pages}{092503}
  (\bibinfo{year}{2024}), ISSN \bibinfo{issn}{0021-9606},
  \eprint{https://pubs.aip.org/aip/jcp/article-pdf/doi/10.1063/5.0182685/19717263/092503\_1\_5.0182685.pdf},
  \urlprefix\url{https://doi.org/10.1063/5.0182685}.

\bibitem[{\citenamefont{Gr\"uneis et~al.}(2014)\citenamefont{Gr\"uneis, Kresse,
  Hinuma, and Oba}}]{SOC_Gruneis}
\bibinfo{author}{\bibfnamefont{A.}~\bibnamefont{Gr\"uneis}},
  \bibinfo{author}{\bibfnamefont{G.}~\bibnamefont{Kresse}},
  \bibinfo{author}{\bibfnamefont{Y.}~\bibnamefont{Hinuma}}, \bibnamefont{and}
  \bibinfo{author}{\bibfnamefont{F.}~\bibnamefont{Oba}},
  \bibinfo{journal}{Phys. Rev. Lett.} \textbf{\bibinfo{volume}{112}},
  \bibinfo{pages}{096401} (\bibinfo{year}{2014}),
  \urlprefix\url{https://link.aps.org/doi/10.1103/PhysRevLett.112.096401}.

\bibitem[{Aii()}]{AiiDAVASP}
\emph{\bibinfo{title}{{AiiDA-VASP} plugin page}},
  \bibinfo{howpublished}{\url{https://github.com/aiida-vasp/aiida-vasp}},
  \bibinfo{note}{accessed: 2023-06-22}.

\bibitem[{\citenamefont{Damle et~al.}(2015)\citenamefont{Damle, Lin, and
  Ying}}]{SCDM_1}
\bibinfo{author}{\bibfnamefont{A.}~\bibnamefont{Damle}},
  \bibinfo{author}{\bibfnamefont{L.}~\bibnamefont{Lin}}, \bibnamefont{and}
  \bibinfo{author}{\bibfnamefont{L.}~\bibnamefont{Ying}},
  \bibinfo{journal}{Journal of Chemical Theory and Computation}
  \textbf{\bibinfo{volume}{11}}, \bibinfo{pages}{1463} (\bibinfo{year}{2015}),
  \bibinfo{note}{pMID: 26574357}, \eprint{https://doi.org/10.1021/ct500985f},
  \urlprefix\url{https://doi.org/10.1021/ct500985f}.

\bibitem[{\citenamefont{Damle and Lin}(2018)}]{SCDM_2}
\bibinfo{author}{\bibfnamefont{A.}~\bibnamefont{Damle}} \bibnamefont{and}
  \bibinfo{author}{\bibfnamefont{L.}~\bibnamefont{Lin}},
  \bibinfo{journal}{Multiscale Modeling \& Simulation}
  \textbf{\bibinfo{volume}{16}}, \bibinfo{pages}{1392} (\bibinfo{year}{2018}),
  \eprint{https://doi.org/10.1137/17M1129696},
  \urlprefix\url{https://doi.org/10.1137/17M1129696}.

\bibitem[{\citenamefont{Pizzi et~al.}(2020)\citenamefont{Pizzi, Vitale, Arita,
  Blügel, Freimuth, Géranton, Gibertini, Gresch, Johnson, Koretsune
  et~al.}}]{Wannier_Pizzi_2020}
\bibinfo{author}{\bibfnamefont{G.}~\bibnamefont{Pizzi}},
  \bibinfo{author}{\bibfnamefont{V.}~\bibnamefont{Vitale}},
  \bibinfo{author}{\bibfnamefont{R.}~\bibnamefont{Arita}},
  \bibinfo{author}{\bibfnamefont{S.}~\bibnamefont{Blügel}},
  \bibinfo{author}{\bibfnamefont{F.}~\bibnamefont{Freimuth}},
  \bibinfo{author}{\bibfnamefont{G.}~\bibnamefont{Géranton}},
  \bibinfo{author}{\bibfnamefont{M.}~\bibnamefont{Gibertini}},
  \bibinfo{author}{\bibfnamefont{D.}~\bibnamefont{Gresch}},
  \bibinfo{author}{\bibfnamefont{C.}~\bibnamefont{Johnson}},
  \bibinfo{author}{\bibfnamefont{T.}~\bibnamefont{Koretsune}},
  \bibnamefont{et~al.}, \bibinfo{journal}{Journal of Physics: Condensed Matter}
  \textbf{\bibinfo{volume}{32}}, \bibinfo{pages}{165902}
  (\bibinfo{year}{2020}),
  \urlprefix\url{https://dx.doi.org/10.1088/1361-648X/ab51ff}.

\bibitem[{\citenamefont{Franchini et~al.}(2012)\citenamefont{Franchini,
  Kováčik, Marsman, Murthy, He, Ederer, and Kresse}}]{Franchini2012}
\bibinfo{author}{\bibfnamefont{C.}~\bibnamefont{Franchini}},
  \bibinfo{author}{\bibfnamefont{R.}~\bibnamefont{Kováčik}},
  \bibinfo{author}{\bibfnamefont{M.}~\bibnamefont{Marsman}},
  \bibinfo{author}{\bibfnamefont{S.~S.} \bibnamefont{Murthy}},
  \bibinfo{author}{\bibfnamefont{J.}~\bibnamefont{He}},
  \bibinfo{author}{\bibfnamefont{C.}~\bibnamefont{Ederer}}, \bibnamefont{and}
  \bibinfo{author}{\bibfnamefont{G.}~\bibnamefont{Kresse}},
  \bibinfo{journal}{Journal of Physics: Condensed Matter}
  \textbf{\bibinfo{volume}{24}}, \bibinfo{pages}{235602}
  (\bibinfo{year}{2012}),
  \urlprefix\url{https://dx.doi.org/10.1088/0953-8984/24/23/235602}.

\bibitem[{\citenamefont{Shishkin et~al.}(2007)\citenamefont{Shishkin, Marsman,
  and Kresse}}]{Shishkin2007_Vertex}
\bibinfo{author}{\bibfnamefont{M.}~\bibnamefont{Shishkin}},
  \bibinfo{author}{\bibfnamefont{M.}~\bibnamefont{Marsman}}, \bibnamefont{and}
  \bibinfo{author}{\bibfnamefont{G.}~\bibnamefont{Kresse}},
  \bibinfo{journal}{Phys. Rev. Lett.} \textbf{\bibinfo{volume}{99}},
  \bibinfo{pages}{246403} (\bibinfo{year}{2007}),
  \urlprefix\url{https://link.aps.org/doi/10.1103/PhysRevLett.99.246403}.

\bibitem[{\citenamefont{Miglio et~al.}(2020)\citenamefont{Miglio,
  Brousseau-Couture, Godbout, Antonius, Chan, Louie, C{\^o}t{\'e}, Giantomassi,
  and Gonze}}]{ElPhonRen_5}
\bibinfo{author}{\bibfnamefont{A.}~\bibnamefont{Miglio}},
  \bibinfo{author}{\bibfnamefont{V.}~\bibnamefont{Brousseau-Couture}},
  \bibinfo{author}{\bibfnamefont{E.}~\bibnamefont{Godbout}},
  \bibinfo{author}{\bibfnamefont{G.}~\bibnamefont{Antonius}},
  \bibinfo{author}{\bibfnamefont{Y.-H.} \bibnamefont{Chan}},
  \bibinfo{author}{\bibfnamefont{S.~G.} \bibnamefont{Louie}},
  \bibinfo{author}{\bibfnamefont{M.}~\bibnamefont{C{\^o}t{\'e}}},
  \bibinfo{author}{\bibfnamefont{M.}~\bibnamefont{Giantomassi}},
  \bibnamefont{and} \bibinfo{author}{\bibfnamefont{X.}~\bibnamefont{Gonze}},
  \bibinfo{journal}{npj Computational Materials} \textbf{\bibinfo{volume}{6}},
  \bibinfo{pages}{167} (\bibinfo{year}{2020}), ISSN \bibinfo{issn}{2057-3960},
  \urlprefix\url{https://doi.org/10.1038/s41524-020-00434-z}.

\bibitem[{\citenamefont{Bhandari et~al.}(2018)\citenamefont{Bhandari, van
  Schilfgaarde, Kotani, and Lambrecht}}]{STO_2}
\bibinfo{author}{\bibfnamefont{C.}~\bibnamefont{Bhandari}},
  \bibinfo{author}{\bibfnamefont{M.}~\bibnamefont{van Schilfgaarde}},
  \bibinfo{author}{\bibfnamefont{T.}~\bibnamefont{Kotani}}, \bibnamefont{and}
  \bibinfo{author}{\bibfnamefont{W.~R.~L.} \bibnamefont{Lambrecht}},
  \bibinfo{journal}{Phys. Rev. Mater.} \textbf{\bibinfo{volume}{2}},
  \bibinfo{pages}{013807} (\bibinfo{year}{2018}),
  \urlprefix\url{https://link.aps.org/doi/10.1103/PhysRevMaterials.2.013807}.

\bibitem[{\citenamefont{Gant et~al.}(2022)\citenamefont{Gant, Haber, Filip,
  Sagredo, Wing, Ohad, Kronik, and Neaton}}]{Gant2022}
\bibinfo{author}{\bibfnamefont{S.~E.} \bibnamefont{Gant}},
  \bibinfo{author}{\bibfnamefont{J.~B.} \bibnamefont{Haber}},
  \bibinfo{author}{\bibfnamefont{M.~R.} \bibnamefont{Filip}},
  \bibinfo{author}{\bibfnamefont{F.}~\bibnamefont{Sagredo}},
  \bibinfo{author}{\bibfnamefont{D.}~\bibnamefont{Wing}},
  \bibinfo{author}{\bibfnamefont{G.}~\bibnamefont{Ohad}},
  \bibinfo{author}{\bibfnamefont{L.}~\bibnamefont{Kronik}}, \bibnamefont{and}
  \bibinfo{author}{\bibfnamefont{J.~B.} \bibnamefont{Neaton}},
  \bibinfo{journal}{Phys. Rev. Mater.} \textbf{\bibinfo{volume}{6}},
  \bibinfo{pages}{053802} (\bibinfo{year}{2022}),
  \urlprefix\url{https://link.aps.org/doi/10.1103/PhysRevMaterials.6.053802}.

\bibitem[{\citenamefont{Gao et~al.}(2018)\citenamefont{Gao, Xia, Wu, Ren, Gao,
  and Zhang}}]{Cu_Gao}
\bibinfo{author}{\bibfnamefont{W.}~\bibnamefont{Gao}},
  \bibinfo{author}{\bibfnamefont{W.}~\bibnamefont{Xia}},
  \bibinfo{author}{\bibfnamefont{Y.}~\bibnamefont{Wu}},
  \bibinfo{author}{\bibfnamefont{W.}~\bibnamefont{Ren}},
  \bibinfo{author}{\bibfnamefont{X.}~\bibnamefont{Gao}}, \bibnamefont{and}
  \bibinfo{author}{\bibfnamefont{P.}~\bibnamefont{Zhang}},
  \bibinfo{journal}{Phys. Rev. B} \textbf{\bibinfo{volume}{98}},
  \bibinfo{pages}{045108} (\bibinfo{year}{2018}),
  \urlprefix\url{https://link.aps.org/doi/10.1103/PhysRevB.98.045108}.

\bibitem[{\citenamefont{Varrassi et~al.}(2021)\citenamefont{Varrassi, Liu,
  Yavas, Bokdam, Kresse, and Franchini}}]{Varrassi2021}
\bibinfo{author}{\bibfnamefont{L.}~\bibnamefont{Varrassi}},
  \bibinfo{author}{\bibfnamefont{P.}~\bibnamefont{Liu}},
  \bibinfo{author}{\bibfnamefont{Z.~E.} \bibnamefont{Yavas}},
  \bibinfo{author}{\bibfnamefont{M.}~\bibnamefont{Bokdam}},
  \bibinfo{author}{\bibfnamefont{G.}~\bibnamefont{Kresse}}, \bibnamefont{and}
  \bibinfo{author}{\bibfnamefont{C.}~\bibnamefont{Franchini}},
  \bibinfo{journal}{Phys. Rev. Mater.} \textbf{\bibinfo{volume}{5}},
  \bibinfo{pages}{074601} (\bibinfo{year}{2021}),
  \urlprefix\url{https://link.aps.org/doi/10.1103/PhysRevMaterials.5.074601}.

\bibitem[{\citenamefont{Varrassi et~al.}(2024)\citenamefont{Varrassi, Liu, and
  Franchini}}]{Varrassi_2024}
\bibinfo{author}{\bibfnamefont{L.}~\bibnamefont{Varrassi}},
  \bibinfo{author}{\bibfnamefont{P.}~\bibnamefont{Liu}}, \bibnamefont{and}
  \bibinfo{author}{\bibfnamefont{C.}~\bibnamefont{Franchini}},
  \bibinfo{journal}{Phys. Rev. Mater.} \textbf{\bibinfo{volume}{8}},
  \bibinfo{pages}{024001} (\bibinfo{year}{2024}),
  \urlprefix\url{https://link.aps.org/doi/10.1103/PhysRevMaterials.8.024001}.

\bibitem[{\citenamefont{van Schilfgaarde
  et~al.}(2006{\natexlab{b}})\citenamefont{van Schilfgaarde, Kotani, and
  Faleev}}]{STO_AbInitio_test_1}
\bibinfo{author}{\bibfnamefont{M.}~\bibnamefont{van Schilfgaarde}},
  \bibinfo{author}{\bibfnamefont{T.}~\bibnamefont{Kotani}}, \bibnamefont{and}
  \bibinfo{author}{\bibfnamefont{S.~V.} \bibnamefont{Faleev}},
  \bibinfo{journal}{Phys. Rev. B} \textbf{\bibinfo{volume}{74}},
  \bibinfo{pages}{245125} (\bibinfo{year}{2006}{\natexlab{b}}),
  \urlprefix\url{https://link.aps.org/doi/10.1103/PhysRevB.74.245125}.

\bibitem[{\citenamefont{Tr\"oster et~al.}(2022)\citenamefont{Tr\"oster, Verdi,
  Dellago, Rychetsky, Kresse, and Schranz}}]{STO_AbInitio_test_2}
\bibinfo{author}{\bibfnamefont{A.}~\bibnamefont{Tr\"oster}},
  \bibinfo{author}{\bibfnamefont{C.}~\bibnamefont{Verdi}},
  \bibinfo{author}{\bibfnamefont{C.}~\bibnamefont{Dellago}},
  \bibinfo{author}{\bibfnamefont{I.}~\bibnamefont{Rychetsky}},
  \bibinfo{author}{\bibfnamefont{G.}~\bibnamefont{Kresse}}, \bibnamefont{and}
  \bibinfo{author}{\bibfnamefont{W.}~\bibnamefont{Schranz}},
  \bibinfo{journal}{Phys. Rev. Materials} \textbf{\bibinfo{volume}{6}},
  \bibinfo{pages}{094408} (\bibinfo{year}{2022}),
  \urlprefix\url{https://link.aps.org/doi/10.1103/PhysRevMaterials.6.094408}.

\bibitem[{\citenamefont{He and Franchini}(2012)}]{STO_AbInitio_test_3}
\bibinfo{author}{\bibfnamefont{J.}~\bibnamefont{He}} \bibnamefont{and}
  \bibinfo{author}{\bibfnamefont{C.}~\bibnamefont{Franchini}},
  \bibinfo{journal}{Phys. Rev. B} \textbf{\bibinfo{volume}{86}},
  \bibinfo{pages}{235117} (\bibinfo{year}{2012}),
  \urlprefix\url{https://link.aps.org/doi/10.1103/PhysRevB.86.235117}.

\bibitem[{\citenamefont{Friedrich et~al.}(2010)\citenamefont{Friedrich,
  Bl\"ugel, and Schindlmayr}}]{SeMa_STO_1}
\bibinfo{author}{\bibfnamefont{C.}~\bibnamefont{Friedrich}},
  \bibinfo{author}{\bibfnamefont{S.}~\bibnamefont{Bl\"ugel}}, \bibnamefont{and}
  \bibinfo{author}{\bibfnamefont{A.}~\bibnamefont{Schindlmayr}},
  \bibinfo{journal}{Phys. Rev. B} \textbf{\bibinfo{volume}{81}},
  \bibinfo{pages}{125102} (\bibinfo{year}{2010}),
  \urlprefix\url{https://link.aps.org/doi/10.1103/PhysRevB.81.125102}.

\bibitem[{\citenamefont{Sanna et~al.}(2011{\natexlab{a}})\citenamefont{Sanna,
  Thierfelder, Wippermann, Sinha, and Schmidt}}]{BaTiO3_1}
\bibinfo{author}{\bibfnamefont{S.}~\bibnamefont{Sanna}},
  \bibinfo{author}{\bibfnamefont{C.}~\bibnamefont{Thierfelder}},
  \bibinfo{author}{\bibfnamefont{S.}~\bibnamefont{Wippermann}},
  \bibinfo{author}{\bibfnamefont{T.~P.} \bibnamefont{Sinha}}, \bibnamefont{and}
  \bibinfo{author}{\bibfnamefont{W.~G.} \bibnamefont{Schmidt}},
  \bibinfo{journal}{Phys. Rev. B} \textbf{\bibinfo{volume}{83}},
  \bibinfo{pages}{054112} (\bibinfo{year}{2011}{\natexlab{a}}),
  \urlprefix\url{https://link.aps.org/doi/10.1103/PhysRevB.83.054112}.

\bibitem[{\citenamefont{Shih et~al.}(2010{\natexlab{b}})\citenamefont{Shih,
  Xue, Zhang, Cohen, and Louie}}]{ZnOwz_2}
\bibinfo{author}{\bibfnamefont{B.-C.} \bibnamefont{Shih}},
  \bibinfo{author}{\bibfnamefont{Y.}~\bibnamefont{Xue}},
  \bibinfo{author}{\bibfnamefont{P.}~\bibnamefont{Zhang}},
  \bibinfo{author}{\bibfnamefont{M.~L.} \bibnamefont{Cohen}}, \bibnamefont{and}
  \bibinfo{author}{\bibfnamefont{S.~G.} \bibnamefont{Louie}},
  \bibinfo{journal}{Phys. Rev. Lett.} \textbf{\bibinfo{volume}{105}},
  \bibinfo{pages}{146401} (\bibinfo{year}{2010}{\natexlab{b}}),
  \urlprefix\url{https://link.aps.org/doi/10.1103/PhysRevLett.105.146401}.

\bibitem[{\citenamefont{Friedrich
  et~al.}(2011{\natexlab{b}})\citenamefont{Friedrich, M\"uller, and
  Bl\"ugel}}]{ZnOwz_3}
\bibinfo{author}{\bibfnamefont{C.}~\bibnamefont{Friedrich}},
  \bibinfo{author}{\bibfnamefont{M.~C.} \bibnamefont{M\"uller}},
  \bibnamefont{and} \bibinfo{author}{\bibfnamefont{S.}~\bibnamefont{Bl\"ugel}},
  \bibinfo{journal}{Phys. Rev. B} \textbf{\bibinfo{volume}{83}},
  \bibinfo{pages}{081101} (\bibinfo{year}{2011}{\natexlab{b}}),
  \urlprefix\url{https://link.aps.org/doi/10.1103/PhysRevB.83.081101}.

\bibitem[{\citenamefont{Abedi et~al.}(2022)\citenamefont{Abedi, Ahmadpour,
  Baninajarian, Kahnouji, Hashemifar, Han, and
  Levchenko}}]{HTReference_abedi2022benchmark}
\bibinfo{author}{\bibfnamefont{S.}~\bibnamefont{Abedi}},
  \bibinfo{author}{\bibfnamefont{M.~T.} \bibnamefont{Ahmadpour}},
  \bibinfo{author}{\bibfnamefont{S.}~\bibnamefont{Baninajarian}},
  \bibinfo{author}{\bibfnamefont{H.}~\bibnamefont{Kahnouji}},
  \bibinfo{author}{\bibfnamefont{S.~J.} \bibnamefont{Hashemifar}},
  \bibinfo{author}{\bibfnamefont{Z.-K.} \bibnamefont{Han}}, \bibnamefont{and}
  \bibinfo{author}{\bibfnamefont{S.~V.} \bibnamefont{Levchenko}},
  \emph{\bibinfo{title}{A benchmark of first-principles methods for accurate
  prediction of semiconductor band gaps}} (\bibinfo{year}{2022}),
  \eprint{2211.14644}.

\bibitem[{\citenamefont{Zhang and Jiang}(2019)}]{AgCl_1}
\bibinfo{author}{\bibfnamefont{M.-Y.} \bibnamefont{Zhang}} \bibnamefont{and}
  \bibinfo{author}{\bibfnamefont{H.}~\bibnamefont{Jiang}},
  \bibinfo{journal}{Phys. Rev. B} \textbf{\bibinfo{volume}{100}},
  \bibinfo{pages}{205123} (\bibinfo{year}{2019}),
  \urlprefix\url{https://link.aps.org/doi/10.1103/PhysRevB.100.205123}.

\bibitem[{\citenamefont{Abedi et~al.}(2023)\citenamefont{Abedi,
  Tarighi~Ahmadpour, Baninajarian, Kahnouji, Hashemifar, Han, and
  Levchenko}}]{AgCl_2}
\bibinfo{author}{\bibfnamefont{S.}~\bibnamefont{Abedi}},
  \bibinfo{author}{\bibfnamefont{M.}~\bibnamefont{Tarighi~Ahmadpour}},
  \bibinfo{author}{\bibfnamefont{S.}~\bibnamefont{Baninajarian}},
  \bibinfo{author}{\bibfnamefont{H.}~\bibnamefont{Kahnouji}},
  \bibinfo{author}{\bibfnamefont{S.~J.} \bibnamefont{Hashemifar}},
  \bibinfo{author}{\bibfnamefont{Z.-K.} \bibnamefont{Han}}, \bibnamefont{and}
  \bibinfo{author}{\bibfnamefont{S.~V.} \bibnamefont{Levchenko}},
  \bibinfo{journal}{The Journal of Chemical Physics}
  \textbf{\bibinfo{volume}{158}}, \bibinfo{pages}{184109}
  (\bibinfo{year}{2023}), ISSN \bibinfo{issn}{0021-9606},
  \eprint{https://pubs.aip.org/aip/jcp/article-pdf/doi/10.1063/5.0138775/17415085/184109\_1\_5.0138775.pdf},
  \urlprefix\url{https://doi.org/10.1063/5.0138775}.

\bibitem[{\citenamefont{Yeh et~al.}(2022)\citenamefont{Yeh, Shee, Sun, Gull,
  and Zgid}}]{CuCl_1}
\bibinfo{author}{\bibfnamefont{C.-N.} \bibnamefont{Yeh}},
  \bibinfo{author}{\bibfnamefont{A.}~\bibnamefont{Shee}},
  \bibinfo{author}{\bibfnamefont{Q.}~\bibnamefont{Sun}},
  \bibinfo{author}{\bibfnamefont{E.}~\bibnamefont{Gull}}, \bibnamefont{and}
  \bibinfo{author}{\bibfnamefont{D.}~\bibnamefont{Zgid}},
  \bibinfo{journal}{Phys. Rev. B} \textbf{\bibinfo{volume}{106}},
  \bibinfo{pages}{085121} (\bibinfo{year}{2022}),
  \urlprefix\url{https://link.aps.org/doi/10.1103/PhysRevB.106.085121}.

\bibitem[{\citenamefont{Choudhary et~al.}(2019)\citenamefont{Choudhary, Bercx,
  Jiang, Pachter, Lamoen, and Tavazza}}]{CuBr_1}
\bibinfo{author}{\bibfnamefont{K.}~\bibnamefont{Choudhary}},
  \bibinfo{author}{\bibfnamefont{M.}~\bibnamefont{Bercx}},
  \bibinfo{author}{\bibfnamefont{J.}~\bibnamefont{Jiang}},
  \bibinfo{author}{\bibfnamefont{R.}~\bibnamefont{Pachter}},
  \bibinfo{author}{\bibfnamefont{D.}~\bibnamefont{Lamoen}}, \bibnamefont{and}
  \bibinfo{author}{\bibfnamefont{F.}~\bibnamefont{Tavazza}},
  \bibinfo{journal}{Chemistry of Materials} \textbf{\bibinfo{volume}{31}},
  \bibinfo{pages}{5900} (\bibinfo{year}{2019}),
  \eprint{https://doi.org/10.1021/acs.chemmater.9b02166},
  \urlprefix\url{https://doi.org/10.1021/acs.chemmater.9b02166}.

\bibitem[{\citenamefont{Pishtshev and Karazhanov}(2017)}]{CuI_1}
\bibinfo{author}{\bibfnamefont{A.}~\bibnamefont{Pishtshev}} \bibnamefont{and}
  \bibinfo{author}{\bibfnamefont{S.~Z.} \bibnamefont{Karazhanov}},
  \bibinfo{journal}{The Journal of Chemical Physics}
  \textbf{\bibinfo{volume}{146}}, \bibinfo{pages}{064706}
  (\bibinfo{year}{2017}), ISSN \bibinfo{issn}{0021-9606},
  \eprint{https://pubs.aip.org/aip/jcp/article-pdf/doi/10.1063/1.4975176/14778691/064706\_1\_online.pdf},
  \urlprefix\url{https://doi.org/10.1063/1.4975176}.

\bibitem[{\citenamefont{Riefer et~al.}(2011)\citenamefont{Riefer, Fuchs,
  R\"odl, Schleife, Bechstedt, and Goldhahn}}]{CaO_1}
\bibinfo{author}{\bibfnamefont{A.}~\bibnamefont{Riefer}},
  \bibinfo{author}{\bibfnamefont{F.}~\bibnamefont{Fuchs}},
  \bibinfo{author}{\bibfnamefont{C.}~\bibnamefont{R\"odl}},
  \bibinfo{author}{\bibfnamefont{A.}~\bibnamefont{Schleife}},
  \bibinfo{author}{\bibfnamefont{F.}~\bibnamefont{Bechstedt}},
  \bibnamefont{and} \bibinfo{author}{\bibfnamefont{R.}~\bibnamefont{Goldhahn}},
  \bibinfo{journal}{Phys. Rev. B} \textbf{\bibinfo{volume}{84}},
  \bibinfo{pages}{075218} (\bibinfo{year}{2011}),
  \urlprefix\url{https://link.aps.org/doi/10.1103/PhysRevB.84.075218}.

\bibitem[{\citenamefont{Poncé et~al.}(2013)\citenamefont{Poncé, Bertrand,
  Smet, Poelman, Mikami, and Gonze}}]{CaO_2}
\bibinfo{author}{\bibfnamefont{S.}~\bibnamefont{Poncé}},
  \bibinfo{author}{\bibfnamefont{B.}~\bibnamefont{Bertrand}},
  \bibinfo{author}{\bibfnamefont{P.}~\bibnamefont{Smet}},
  \bibinfo{author}{\bibfnamefont{D.}~\bibnamefont{Poelman}},
  \bibinfo{author}{\bibfnamefont{M.}~\bibnamefont{Mikami}}, \bibnamefont{and}
  \bibinfo{author}{\bibfnamefont{X.}~\bibnamefont{Gonze}},
  \bibinfo{journal}{Optical Materials} \textbf{\bibinfo{volume}{35}},
  \bibinfo{pages}{1477} (\bibinfo{year}{2013}), ISSN \bibinfo{issn}{0925-3467},
  \urlprefix\url{https://www.sciencedirect.com/science/article/pii/S0925346713001316}.

\bibitem[{\citenamefont{Nejatipour and Dadsetani}(2015)}]{CaO_3}
\bibinfo{author}{\bibfnamefont{H.}~\bibnamefont{Nejatipour}} \bibnamefont{and}
  \bibinfo{author}{\bibfnamefont{M.}~\bibnamefont{Dadsetani}},
  \bibinfo{journal}{Physica Scripta} \textbf{\bibinfo{volume}{90}},
  \bibinfo{pages}{085802} (\bibinfo{year}{2015}),
  \urlprefix\url{https://dx.doi.org/10.1088/0031-8949/90/8/085802}.

\bibitem[{\citenamefont{Berger et~al.}(2012)\citenamefont{Berger, Reining, and
  Sottile}}]{ZnO_GW_1}
\bibinfo{author}{\bibfnamefont{J.~A.} \bibnamefont{Berger}},
  \bibinfo{author}{\bibfnamefont{L.}~\bibnamefont{Reining}}, \bibnamefont{and}
  \bibinfo{author}{\bibfnamefont{F.}~\bibnamefont{Sottile}},
  \bibinfo{journal}{Phys. Rev. B} \textbf{\bibinfo{volume}{85}},
  \bibinfo{pages}{085126} (\bibinfo{year}{2012}),
  \urlprefix\url{https://link.aps.org/doi/10.1103/PhysRevB.85.085126}.

\bibitem[{\citenamefont{Usuda et~al.}(2002)\citenamefont{Usuda, Hamada, Kotani,
  and van Schilfgaarde}}]{ZnO_GW_2}
\bibinfo{author}{\bibfnamefont{M.}~\bibnamefont{Usuda}},
  \bibinfo{author}{\bibfnamefont{N.}~\bibnamefont{Hamada}},
  \bibinfo{author}{\bibfnamefont{T.}~\bibnamefont{Kotani}}, \bibnamefont{and}
  \bibinfo{author}{\bibfnamefont{M.}~\bibnamefont{van Schilfgaarde}},
  \bibinfo{journal}{Phys. Rev. B} \textbf{\bibinfo{volume}{66}},
  \bibinfo{pages}{125101} (\bibinfo{year}{2002}),
  \urlprefix\url{https://link.aps.org/doi/10.1103/PhysRevB.66.125101}.

\bibitem[{\citenamefont{Chen and Pasquarello}(2015{\natexlab{a}})}]{ZnO_GW_3}
\bibinfo{author}{\bibfnamefont{W.}~\bibnamefont{Chen}} \bibnamefont{and}
  \bibinfo{author}{\bibfnamefont{A.}~\bibnamefont{Pasquarello}},
  \bibinfo{journal}{Phys. Rev. B} \textbf{\bibinfo{volume}{92}},
  \bibinfo{pages}{041115} (\bibinfo{year}{2015}{\natexlab{a}}),
  \urlprefix\url{https://link.aps.org/doi/10.1103/PhysRevB.92.041115}.

\bibitem[{\citenamefont{Miglio et~al.}(2012)\citenamefont{Miglio, Waroquiers,
  Antonius, Giantomassi, Stankovski, C{\^o}t{\'e}, Gonze, and
  Rignanese}}]{SnO2_1}
\bibinfo{author}{\bibfnamefont{A.}~\bibnamefont{Miglio}},
  \bibinfo{author}{\bibfnamefont{D.}~\bibnamefont{Waroquiers}},
  \bibinfo{author}{\bibfnamefont{G.}~\bibnamefont{Antonius}},
  \bibinfo{author}{\bibfnamefont{M.}~\bibnamefont{Giantomassi}},
  \bibinfo{author}{\bibfnamefont{M.}~\bibnamefont{Stankovski}},
  \bibinfo{author}{\bibfnamefont{M.}~\bibnamefont{C{\^o}t{\'e}}},
  \bibinfo{author}{\bibfnamefont{X.}~\bibnamefont{Gonze}}, \bibnamefont{and}
  \bibinfo{author}{\bibfnamefont{G.-M.} \bibnamefont{Rignanese}},
  \bibinfo{journal}{The European Physical Journal B}
  \textbf{\bibinfo{volume}{85}}, \bibinfo{pages}{322} (\bibinfo{year}{2012}),
  ISSN \bibinfo{issn}{1434-6036},
  \urlprefix\url{https://doi.org/10.1140/epjb/e2012-30121-4}.

\bibitem[{\citenamefont{Li et~al.}(2015)\citenamefont{Li, Castelli, Thygesen,
  and Jacobsen}}]{SnO2_2}
\bibinfo{author}{\bibfnamefont{H.}~\bibnamefont{Li}},
  \bibinfo{author}{\bibfnamefont{I.~E.} \bibnamefont{Castelli}},
  \bibinfo{author}{\bibfnamefont{K.~S.} \bibnamefont{Thygesen}},
  \bibnamefont{and} \bibinfo{author}{\bibfnamefont{K.~W.}
  \bibnamefont{Jacobsen}}, \bibinfo{journal}{Phys. Rev. B}
  \textbf{\bibinfo{volume}{91}}, \bibinfo{pages}{045204}
  (\bibinfo{year}{2015}),
  \urlprefix\url{https://link.aps.org/doi/10.1103/PhysRevB.91.045204}.

\bibitem[{\citenamefont{Berger et~al.}(2010)\citenamefont{Berger, Reining, and
  Sottile}}]{SnO2_3}
\bibinfo{author}{\bibfnamefont{J.~A.} \bibnamefont{Berger}},
  \bibinfo{author}{\bibfnamefont{L.}~\bibnamefont{Reining}}, \bibnamefont{and}
  \bibinfo{author}{\bibfnamefont{F.}~\bibnamefont{Sottile}},
  \bibinfo{journal}{Phys. Rev. B} \textbf{\bibinfo{volume}{82}},
  \bibinfo{pages}{041103} (\bibinfo{year}{2010}),
  \urlprefix\url{https://link.aps.org/doi/10.1103/PhysRevB.82.041103}.

\bibitem[{\citenamefont{Chen et~al.}(2010)\citenamefont{Chen, Tegenkamp,
  Pfn\"ur, and Bredow}}]{NaCl_1}
\bibinfo{author}{\bibfnamefont{W.}~\bibnamefont{Chen}},
  \bibinfo{author}{\bibfnamefont{C.}~\bibnamefont{Tegenkamp}},
  \bibinfo{author}{\bibfnamefont{H.}~\bibnamefont{Pfn\"ur}}, \bibnamefont{and}
  \bibinfo{author}{\bibfnamefont{T.}~\bibnamefont{Bredow}},
  \bibinfo{journal}{Phys. Rev. B} \textbf{\bibinfo{volume}{82}},
  \bibinfo{pages}{104106} (\bibinfo{year}{2010}),
  \urlprefix\url{https://link.aps.org/doi/10.1103/PhysRevB.82.104106}.

\bibitem[{\citenamefont{Ren et~al.}(2021)\citenamefont{Ren, Merz, Jiang, Yao,
  Rampp, Lederer, Blum, and Scheffler}}]{NaCl_2}
\bibinfo{author}{\bibfnamefont{X.}~\bibnamefont{Ren}},
  \bibinfo{author}{\bibfnamefont{F.}~\bibnamefont{Merz}},
  \bibinfo{author}{\bibfnamefont{H.}~\bibnamefont{Jiang}},
  \bibinfo{author}{\bibfnamefont{Y.}~\bibnamefont{Yao}},
  \bibinfo{author}{\bibfnamefont{M.}~\bibnamefont{Rampp}},
  \bibinfo{author}{\bibfnamefont{H.}~\bibnamefont{Lederer}},
  \bibinfo{author}{\bibfnamefont{V.}~\bibnamefont{Blum}}, \bibnamefont{and}
  \bibinfo{author}{\bibfnamefont{M.}~\bibnamefont{Scheffler}},
  \bibinfo{journal}{Phys. Rev. Mater.} \textbf{\bibinfo{volume}{5}},
  \bibinfo{pages}{013807} (\bibinfo{year}{2021}),
  \urlprefix\url{https://link.aps.org/doi/10.1103/PhysRevMaterials.5.013807}.

\bibitem[{\citenamefont{Chen and Pasquarello}(2013)}]{NaCl_3}
\bibinfo{author}{\bibfnamefont{W.}~\bibnamefont{Chen}} \bibnamefont{and}
  \bibinfo{author}{\bibfnamefont{A.}~\bibnamefont{Pasquarello}},
  \bibinfo{journal}{Phys. Rev. B} \textbf{\bibinfo{volume}{88}},
  \bibinfo{pages}{119906} (\bibinfo{year}{2013}),
  \urlprefix\url{https://link.aps.org/doi/10.1103/PhysRevB.88.119906}.

\bibitem[{\citenamefont{Bechstedt et~al.}(2005)\citenamefont{Bechstedt, Seino,
  Hahn, and Schmidt}}]{NaCl_4}
\bibinfo{author}{\bibfnamefont{F.}~\bibnamefont{Bechstedt}},
  \bibinfo{author}{\bibfnamefont{K.}~\bibnamefont{Seino}},
  \bibinfo{author}{\bibfnamefont{P.~H.} \bibnamefont{Hahn}}, \bibnamefont{and}
  \bibinfo{author}{\bibfnamefont{W.~G.} \bibnamefont{Schmidt}},
  \bibinfo{journal}{Phys. Rev. B} \textbf{\bibinfo{volume}{72}},
  \bibinfo{pages}{245114} (\bibinfo{year}{2005}),
  \urlprefix\url{https://link.aps.org/doi/10.1103/PhysRevB.72.245114}.

\bibitem[{\citenamefont{Schena et~al.}(2015)\citenamefont{Schena, Wuttig, and
  Bl{\"u}gel}}]{RuS_1}
\bibinfo{author}{\bibfnamefont{T.}~\bibnamefont{Schena}},
  \bibinfo{author}{\bibfnamefont{M.}~\bibnamefont{Wuttig}}, \bibnamefont{and}
  \bibinfo{author}{\bibfnamefont{S.}~\bibnamefont{Bl{\"u}gel}}
  (\bibinfo{year}{2015}),
  \urlprefix\url{https://api.semanticscholar.org/CorpusID:113461905}.

\bibitem[{\citenamefont{Gorelov et~al.}(2023)\citenamefont{Gorelov, Reining,
  Lambrecht, and Gatti}}]{V2O5_1}
\bibinfo{author}{\bibfnamefont{V.}~\bibnamefont{Gorelov}},
  \bibinfo{author}{\bibfnamefont{L.}~\bibnamefont{Reining}},
  \bibinfo{author}{\bibfnamefont{W.~R.~L.} \bibnamefont{Lambrecht}},
  \bibnamefont{and} \bibinfo{author}{\bibfnamefont{M.}~\bibnamefont{Gatti}},
  \bibinfo{journal}{Phys. Rev. B} \textbf{\bibinfo{volume}{107}},
  \bibinfo{pages}{075101} (\bibinfo{year}{2023}),
  \urlprefix\url{https://link.aps.org/doi/10.1103/PhysRevB.107.075101}.

\bibitem[{\citenamefont{Rasmussen and Thygesen}(2015{\natexlab{b}})}]{SnSe2_1}
\bibinfo{author}{\bibfnamefont{F.~A.} \bibnamefont{Rasmussen}}
  \bibnamefont{and} \bibinfo{author}{\bibfnamefont{K.~S.}
  \bibnamefont{Thygesen}}, \bibinfo{journal}{The Journal of Physical Chemistry
  C} \textbf{\bibinfo{volume}{119}}, \bibinfo{pages}{13169}
  (\bibinfo{year}{2015}{\natexlab{b}}),
  \eprint{https://doi.org/10.1021/acs.jpcc.5b02950},
  \urlprefix\url{https://doi.org/10.1021/acs.jpcc.5b02950}.

\bibitem[{\citenamefont{Zhang et~al.}(2016)\citenamefont{Zhang, Gong, Nie, Min,
  Liang, Oh, Zhang, Wang, Hong, Colombo et~al.}}]{SnSe2_2}
\bibinfo{author}{\bibfnamefont{C.}~\bibnamefont{Zhang}},
  \bibinfo{author}{\bibfnamefont{C.}~\bibnamefont{Gong}},
  \bibinfo{author}{\bibfnamefont{Y.}~\bibnamefont{Nie}},
  \bibinfo{author}{\bibfnamefont{K.-A.} \bibnamefont{Min}},
  \bibinfo{author}{\bibfnamefont{C.}~\bibnamefont{Liang}},
  \bibinfo{author}{\bibfnamefont{Y.~J.} \bibnamefont{Oh}},
  \bibinfo{author}{\bibfnamefont{H.}~\bibnamefont{Zhang}},
  \bibinfo{author}{\bibfnamefont{W.}~\bibnamefont{Wang}},
  \bibinfo{author}{\bibfnamefont{S.}~\bibnamefont{Hong}},
  \bibinfo{author}{\bibfnamefont{L.}~\bibnamefont{Colombo}},
  \bibnamefont{et~al.}, \bibinfo{journal}{2D Materials}
  \textbf{\bibinfo{volume}{4}}, \bibinfo{pages}{015026} (\bibinfo{year}{2016}),
  \urlprefix\url{https://dx.doi.org/10.1088/2053-1583/4/1/015026}.

\bibitem[{\citenamefont{Kotani and {van Schilfgaarde}}(2002)}]{GaN_1}
\bibinfo{author}{\bibfnamefont{T.}~\bibnamefont{Kotani}} \bibnamefont{and}
  \bibinfo{author}{\bibfnamefont{M.}~\bibnamefont{{van Schilfgaarde}}},
  \bibinfo{journal}{Solid State Communications} \textbf{\bibinfo{volume}{121}},
  \bibinfo{pages}{461} (\bibinfo{year}{2002}), ISSN \bibinfo{issn}{0038-1098},
  \urlprefix\url{https://www.sciencedirect.com/science/article/pii/S0038109802000285}.

\bibitem[{\citenamefont{Rohlfing et~al.}(1998)\citenamefont{Rohlfing, Kr\"uger,
  and Pollmann}}]{GaN_2}
\bibinfo{author}{\bibfnamefont{M.}~\bibnamefont{Rohlfing}},
  \bibinfo{author}{\bibfnamefont{P.}~\bibnamefont{Kr\"uger}}, \bibnamefont{and}
  \bibinfo{author}{\bibfnamefont{J.}~\bibnamefont{Pollmann}},
  \bibinfo{journal}{Phys. Rev. B} \textbf{\bibinfo{volume}{57}},
  \bibinfo{pages}{6485} (\bibinfo{year}{1998}),
  \urlprefix\url{https://link.aps.org/doi/10.1103/PhysRevB.57.6485}.

\bibitem[{\citenamefont{Rinke et~al.}(2006)\citenamefont{Rinke, Scheffler,
  Qteish, Winkelnkemper, Bimberg, and Neugebauer}}]{GaN_4}
\bibinfo{author}{\bibfnamefont{P.}~\bibnamefont{Rinke}},
  \bibinfo{author}{\bibfnamefont{M.}~\bibnamefont{Scheffler}},
  \bibinfo{author}{\bibfnamefont{A.}~\bibnamefont{Qteish}},
  \bibinfo{author}{\bibfnamefont{M.}~\bibnamefont{Winkelnkemper}},
  \bibinfo{author}{\bibfnamefont{D.}~\bibnamefont{Bimberg}}, \bibnamefont{and}
  \bibinfo{author}{\bibfnamefont{J.}~\bibnamefont{Neugebauer}},
  \bibinfo{journal}{Applied Physics Letters} \textbf{\bibinfo{volume}{89}},
  \bibinfo{pages}{161919} (\bibinfo{year}{2006}), ISSN
  \bibinfo{issn}{0003-6951},
  \eprint{https://pubs.aip.org/aip/apl/article-pdf/doi/10.1063/1.2364469/14661288/161919\_1\_online.pdf},
  \urlprefix\url{https://doi.org/10.1063/1.2364469}.

\bibitem[{\citenamefont{Rinke et~al.}(2008)\citenamefont{Rinke, Winkelnkemper,
  Qteish, Bimberg, Neugebauer, and Scheffler}}]{GaN_5}
\bibinfo{author}{\bibfnamefont{P.}~\bibnamefont{Rinke}},
  \bibinfo{author}{\bibfnamefont{M.}~\bibnamefont{Winkelnkemper}},
  \bibinfo{author}{\bibfnamefont{A.}~\bibnamefont{Qteish}},
  \bibinfo{author}{\bibfnamefont{D.}~\bibnamefont{Bimberg}},
  \bibinfo{author}{\bibfnamefont{J.}~\bibnamefont{Neugebauer}},
  \bibnamefont{and}
  \bibinfo{author}{\bibfnamefont{M.}~\bibnamefont{Scheffler}},
  \bibinfo{journal}{Phys. Rev. B} \textbf{\bibinfo{volume}{77}},
  \bibinfo{pages}{075202} (\bibinfo{year}{2008}),
  \urlprefix\url{https://link.aps.org/doi/10.1103/PhysRevB.77.075202}.

\bibitem[{\citenamefont{Lopez-Candales
  et~al.}(2021)\citenamefont{Lopez-Candales, Tang, Xia, Jia, and
  Zhang}}]{STO_BTO_PhysRevB.103.035128}
\bibinfo{author}{\bibfnamefont{G.}~\bibnamefont{Lopez-Candales}},
  \bibinfo{author}{\bibfnamefont{Z.}~\bibnamefont{Tang}},
  \bibinfo{author}{\bibfnamefont{W.}~\bibnamefont{Xia}},
  \bibinfo{author}{\bibfnamefont{F.}~\bibnamefont{Jia}}, \bibnamefont{and}
  \bibinfo{author}{\bibfnamefont{P.}~\bibnamefont{Zhang}},
  \bibinfo{journal}{Phys. Rev. B} \textbf{\bibinfo{volume}{103}},
  \bibinfo{pages}{035128} (\bibinfo{year}{2021}),
  \urlprefix\url{https://link.aps.org/doi/10.1103/PhysRevB.103.035128}.

\bibitem[{\citenamefont{Sanna et~al.}(2011{\natexlab{b}})\citenamefont{Sanna,
  Thierfelder, Wippermann, Sinha, and Schmidt}}]{BTO_1}
\bibinfo{author}{\bibfnamefont{S.}~\bibnamefont{Sanna}},
  \bibinfo{author}{\bibfnamefont{C.}~\bibnamefont{Thierfelder}},
  \bibinfo{author}{\bibfnamefont{S.}~\bibnamefont{Wippermann}},
  \bibinfo{author}{\bibfnamefont{T.~P.} \bibnamefont{Sinha}}, \bibnamefont{and}
  \bibinfo{author}{\bibfnamefont{W.~G.} \bibnamefont{Schmidt}},
  \bibinfo{journal}{Phys. Rev. B} \textbf{\bibinfo{volume}{83}},
  \bibinfo{pages}{054112} (\bibinfo{year}{2011}{\natexlab{b}}),
  \urlprefix\url{https://link.aps.org/doi/10.1103/PhysRevB.83.054112}.

\bibitem[{\citenamefont{Bilc et~al.}(2008)\citenamefont{Bilc, Orlando, Shaltaf,
  Rignanese, \'I\~niguez, and Ghosez}}]{BTO_2}
\bibinfo{author}{\bibfnamefont{D.~I.} \bibnamefont{Bilc}},
  \bibinfo{author}{\bibfnamefont{R.}~\bibnamefont{Orlando}},
  \bibinfo{author}{\bibfnamefont{R.}~\bibnamefont{Shaltaf}},
  \bibinfo{author}{\bibfnamefont{G.-M.} \bibnamefont{Rignanese}},
  \bibinfo{author}{\bibfnamefont{J.}~\bibnamefont{\'I\~niguez}},
  \bibnamefont{and} \bibinfo{author}{\bibfnamefont{P.}~\bibnamefont{Ghosez}},
  \bibinfo{journal}{Phys. Rev. B} \textbf{\bibinfo{volume}{77}},
  \bibinfo{pages}{165107} (\bibinfo{year}{2008}),
  \urlprefix\url{https://link.aps.org/doi/10.1103/PhysRevB.77.165107}.

\bibitem[{\citenamefont{Gierlich and Bl{\"u}gel}(2010)}]{PbTiO3_1}
\bibinfo{author}{\bibfnamefont{A.}~\bibnamefont{Gierlich}} \bibnamefont{and}
  \bibinfo{author}{\bibfnamefont{S.}~\bibnamefont{Bl{\"u}gel}},
  \bibinfo{journal}{Bulletin of the American Physical Society}
  \textbf{\bibinfo{volume}{2010}} (\bibinfo{year}{2010}),
  \urlprefix\url{https://api.semanticscholar.org/CorpusID:94225183}.

\bibitem[{\citenamefont{Brehm et~al.}(2014)\citenamefont{Brehm, Takenaka, Lee,
  Grinberg, Bennett, Schoenberg, and Rappe}}]{PbTiO3_2}
\bibinfo{author}{\bibfnamefont{J.~A.} \bibnamefont{Brehm}},
  \bibinfo{author}{\bibfnamefont{H.}~\bibnamefont{Takenaka}},
  \bibinfo{author}{\bibfnamefont{C.-W.} \bibnamefont{Lee}},
  \bibinfo{author}{\bibfnamefont{I.}~\bibnamefont{Grinberg}},
  \bibinfo{author}{\bibfnamefont{J.~W.} \bibnamefont{Bennett}},
  \bibinfo{author}{\bibfnamefont{M.~R.} \bibnamefont{Schoenberg}},
  \bibnamefont{and} \bibinfo{author}{\bibfnamefont{A.~M.} \bibnamefont{Rappe}},
  \bibinfo{journal}{Phys. Rev. B} \textbf{\bibinfo{volume}{89}},
  \bibinfo{pages}{195202} (\bibinfo{year}{2014}),
  \urlprefix\url{https://link.aps.org/doi/10.1103/PhysRevB.89.195202}.

\bibitem[{\citenamefont{Bendaoudi et~al.}(2023)\citenamefont{Bendaoudi,
  Ouahrani, Daouli, Rerbal, Boufatah, Morales-García, Franco, Bedrane, Badawi,
  and Errandonea}}]{PbTiO_3}
\bibinfo{author}{\bibfnamefont{L.}~\bibnamefont{Bendaoudi}},
  \bibinfo{author}{\bibfnamefont{T.}~\bibnamefont{Ouahrani}},
  \bibinfo{author}{\bibfnamefont{A.}~\bibnamefont{Daouli}},
  \bibinfo{author}{\bibfnamefont{B.}~\bibnamefont{Rerbal}},
  \bibinfo{author}{\bibfnamefont{R.~M.} \bibnamefont{Boufatah}},
  \bibinfo{author}{\bibfnamefont{A.}~\bibnamefont{Morales-García}},
  \bibinfo{author}{\bibfnamefont{R.}~\bibnamefont{Franco}},
  \bibinfo{author}{\bibfnamefont{Z.}~\bibnamefont{Bedrane}},
  \bibinfo{author}{\bibfnamefont{M.}~\bibnamefont{Badawi}}, \bibnamefont{and}
  \bibinfo{author}{\bibfnamefont{D.}~\bibnamefont{Errandonea}},
  \bibinfo{journal}{Dalton Trans.} \textbf{\bibinfo{volume}{52}},
  \bibinfo{pages}{11965} (\bibinfo{year}{2023}),
  \urlprefix\url{http://dx.doi.org/10.1039/D3DT01478A}.

\bibitem[{\citenamefont{Begum et~al.}(2019)\citenamefont{Begum, Gruner, and
  Pentcheva}}]{STO_1}
\bibinfo{author}{\bibfnamefont{V.}~\bibnamefont{Begum}},
  \bibinfo{author}{\bibfnamefont{M.~E.} \bibnamefont{Gruner}},
  \bibnamefont{and}
  \bibinfo{author}{\bibfnamefont{R.}~\bibnamefont{Pentcheva}},
  \bibinfo{journal}{Phys. Rev. Mater.} \textbf{\bibinfo{volume}{3}},
  \bibinfo{pages}{065004} (\bibinfo{year}{2019}),
  \urlprefix\url{https://link.aps.org/doi/10.1103/PhysRevMaterials.3.065004}.

\bibitem[{\citenamefont{Berger et~al.}(2011)\citenamefont{Berger, Fennie, and
  Neaton}}]{STO_3}
\bibinfo{author}{\bibfnamefont{R.~F.} \bibnamefont{Berger}},
  \bibinfo{author}{\bibfnamefont{C.~J.} \bibnamefont{Fennie}},
  \bibnamefont{and} \bibinfo{author}{\bibfnamefont{J.~B.}
  \bibnamefont{Neaton}}, \bibinfo{journal}{Phys. Rev. Lett.}
  \textbf{\bibinfo{volume}{107}}, \bibinfo{pages}{146804}
  (\bibinfo{year}{2011}),
  \urlprefix\url{https://link.aps.org/doi/10.1103/PhysRevLett.107.146804}.

\bibitem[{\citenamefont{Sponza et~al.}(2013)\citenamefont{Sponza, V\'eniard,
  Sottile, Giorgetti, and Reining}}]{STO_4}
\bibinfo{author}{\bibfnamefont{L.}~\bibnamefont{Sponza}},
  \bibinfo{author}{\bibfnamefont{V.}~\bibnamefont{V\'eniard}},
  \bibinfo{author}{\bibfnamefont{F.}~\bibnamefont{Sottile}},
  \bibinfo{author}{\bibfnamefont{C.}~\bibnamefont{Giorgetti}},
  \bibnamefont{and} \bibinfo{author}{\bibfnamefont{L.}~\bibnamefont{Reining}},
  \bibinfo{journal}{Phys. Rev. B} \textbf{\bibinfo{volume}{87}},
  \bibinfo{pages}{235102} (\bibinfo{year}{2013}),
  \urlprefix\url{https://link.aps.org/doi/10.1103/PhysRevB.87.235102}.

\bibitem[{\citenamefont{Liu et~al.}(2019)\citenamefont{Liu, Franchini, Marsman,
  and Kresse}}]{MnO_1}
\bibinfo{author}{\bibfnamefont{P.}~\bibnamefont{Liu}},
  \bibinfo{author}{\bibfnamefont{C.}~\bibnamefont{Franchini}},
  \bibinfo{author}{\bibfnamefont{M.}~\bibnamefont{Marsman}}, \bibnamefont{and}
  \bibinfo{author}{\bibfnamefont{G.}~\bibnamefont{Kresse}},
  \bibinfo{journal}{Journal of Physics: Condensed Matter}
  \textbf{\bibinfo{volume}{32}}, \bibinfo{pages}{015502}
  (\bibinfo{year}{2019}),
  \urlprefix\url{https://dx.doi.org/10.1088/1361-648X/ab4150}.

\bibitem[{\citenamefont{Jiang et~al.}(2010)\citenamefont{Jiang, Gomez-Abal,
  Rinke, and Scheffler}}]{MnO_2}
\bibinfo{author}{\bibfnamefont{H.}~\bibnamefont{Jiang}},
  \bibinfo{author}{\bibfnamefont{R.~I.} \bibnamefont{Gomez-Abal}},
  \bibinfo{author}{\bibfnamefont{P.}~\bibnamefont{Rinke}}, \bibnamefont{and}
  \bibinfo{author}{\bibfnamefont{M.}~\bibnamefont{Scheffler}},
  \bibinfo{journal}{Phys. Rev. B} \textbf{\bibinfo{volume}{82}},
  \bibinfo{pages}{045108} (\bibinfo{year}{2010}),
  \urlprefix\url{https://link.aps.org/doi/10.1103/PhysRevB.82.045108}.

\bibitem[{\citenamefont{Das et~al.}(2015)\citenamefont{Das, Coulter, and
  Manousakis}}]{MnO_3}
\bibinfo{author}{\bibfnamefont{S.}~\bibnamefont{Das}},
  \bibinfo{author}{\bibfnamefont{J.~E.} \bibnamefont{Coulter}},
  \bibnamefont{and}
  \bibinfo{author}{\bibfnamefont{E.}~\bibnamefont{Manousakis}},
  \bibinfo{journal}{Phys. Rev. B} \textbf{\bibinfo{volume}{91}},
  \bibinfo{pages}{115105} (\bibinfo{year}{2015}),
  \urlprefix\url{https://link.aps.org/doi/10.1103/PhysRevB.91.115105}.

\bibitem[{\citenamefont{Chen and Pasquarello}(2015{\natexlab{b}})}]{NiO_1}
\bibinfo{author}{\bibfnamefont{W.}~\bibnamefont{Chen}} \bibnamefont{and}
  \bibinfo{author}{\bibfnamefont{A.}~\bibnamefont{Pasquarello}},
  \bibinfo{journal}{Phys. Rev. B} \textbf{\bibinfo{volume}{92}},
  \bibinfo{pages}{041115} (\bibinfo{year}{2015}{\natexlab{b}}),
  \urlprefix\url{https://link.aps.org/doi/10.1103/PhysRevB.92.041115}.

\bibitem[{\citenamefont{Riley et~al.}(2014)\citenamefont{Riley, Mazzola,
  Dendzik, Michiardi, Takayama, Bawden, Graner{\o}d, Leandersson,
  Balasubramanian, Hoesch et~al.}}]{SOC_WSe2}
\bibinfo{author}{\bibfnamefont{J.~M.} \bibnamefont{Riley}},
  \bibinfo{author}{\bibfnamefont{F.}~\bibnamefont{Mazzola}},
  \bibinfo{author}{\bibfnamefont{M.}~\bibnamefont{Dendzik}},
  \bibinfo{author}{\bibfnamefont{M.}~\bibnamefont{Michiardi}},
  \bibinfo{author}{\bibfnamefont{T.}~\bibnamefont{Takayama}},
  \bibinfo{author}{\bibfnamefont{L.}~\bibnamefont{Bawden}},
  \bibinfo{author}{\bibfnamefont{C.}~\bibnamefont{Graner{\o}d}},
  \bibinfo{author}{\bibfnamefont{M.}~\bibnamefont{Leandersson}},
  \bibinfo{author}{\bibfnamefont{T.}~\bibnamefont{Balasubramanian}},
  \bibinfo{author}{\bibfnamefont{M.}~\bibnamefont{Hoesch}},
  \bibnamefont{et~al.}, \bibinfo{journal}{Nature Physics}
  \textbf{\bibinfo{volume}{10}}, \bibinfo{pages}{835} (\bibinfo{year}{2014}),
  ISSN \bibinfo{issn}{1745-2481},
  \urlprefix\url{https://doi.org/10.1038/nphys3105}.

\bibitem[{\citenamefont{Ganose et~al.}(2016)\citenamefont{Ganose, Cuff, Butler,
  Walsh, and Scanlon}}]{SOC_BiOCl}
\bibinfo{author}{\bibfnamefont{A.~M.} \bibnamefont{Ganose}},
  \bibinfo{author}{\bibfnamefont{M.}~\bibnamefont{Cuff}},
  \bibinfo{author}{\bibfnamefont{K.~T.} \bibnamefont{Butler}},
  \bibinfo{author}{\bibfnamefont{A.}~\bibnamefont{Walsh}}, \bibnamefont{and}
  \bibinfo{author}{\bibfnamefont{D.~O.} \bibnamefont{Scanlon}},
  \bibinfo{journal}{Chemistry of Materials} \textbf{\bibinfo{volume}{28}},
  \bibinfo{pages}{1980} (\bibinfo{year}{2016}), ISSN \bibinfo{issn}{0897-4756},
  \urlprefix\url{https://doi.org/10.1021/acs.chemmater.6b00349}.

\bibitem[{\citenamefont{Garcia et~al.}(2004)\citenamefont{Garcia, Scolfaro,
  Leite, Lino, Freire, Farias, and da~Silva}}]{SOC_HfO2}
\bibinfo{author}{\bibfnamefont{J.~C.} \bibnamefont{Garcia}},
  \bibinfo{author}{\bibfnamefont{L.~M.~R.} \bibnamefont{Scolfaro}},
  \bibinfo{author}{\bibfnamefont{J.~R.} \bibnamefont{Leite}},
  \bibinfo{author}{\bibfnamefont{A.~T.} \bibnamefont{Lino}},
  \bibinfo{author}{\bibfnamefont{V.~N.} \bibnamefont{Freire}},
  \bibinfo{author}{\bibfnamefont{G.~A.} \bibnamefont{Farias}},
  \bibnamefont{and} \bibinfo{author}{\bibfnamefont{J.}~\bibnamefont{da~Silva},
  \bibfnamefont{E.~F.}}, \bibinfo{journal}{Applied Physics Letters}
  \textbf{\bibinfo{volume}{85}}, \bibinfo{pages}{5022} (\bibinfo{year}{2004}),
  ISSN \bibinfo{issn}{0003-6951},
  \eprint{https://pubs.aip.org/aip/apl/article-pdf/85/21/5022/18601074/5022\_1\_online.pdf},
  \urlprefix\url{https://doi.org/10.1063/1.1823584}.

\bibitem[{\citenamefont{K{\"u}hn and Weigend}(2015)}]{SOC_Kuhn2015}
\bibinfo{author}{\bibfnamefont{M.}~\bibnamefont{K{\"u}hn}} \bibnamefont{and}
  \bibinfo{author}{\bibfnamefont{F.}~\bibnamefont{Weigend}},
  \bibinfo{journal}{Journal of Chemical Theory and Computation}
  \textbf{\bibinfo{volume}{11}}, \bibinfo{pages}{969} (\bibinfo{year}{2015}),
  ISSN \bibinfo{issn}{1549-9618},
  \urlprefix\url{https://doi.org/10.1021/ct501069b}.

\bibitem[{\citenamefont{Scherpelz et~al.}(2016)\citenamefont{Scherpelz, Govoni,
  Hamada, and Galli}}]{SOC_Scherpelz2016}
\bibinfo{author}{\bibfnamefont{P.}~\bibnamefont{Scherpelz}},
  \bibinfo{author}{\bibfnamefont{M.}~\bibnamefont{Govoni}},
  \bibinfo{author}{\bibfnamefont{I.}~\bibnamefont{Hamada}}, \bibnamefont{and}
  \bibinfo{author}{\bibfnamefont{G.}~\bibnamefont{Galli}},
  \bibinfo{journal}{Journal of Chemical Theory and Computation}
  \textbf{\bibinfo{volume}{12}}, \bibinfo{pages}{3523} (\bibinfo{year}{2016}),
  ISSN \bibinfo{issn}{1549-9618},
  \urlprefix\url{https://doi.org/10.1021/acs.jctc.6b00114}.

\bibitem[{\citenamefont{Aggoune et~al.}(2022)\citenamefont{Aggoune, Eljarrat,
  Nabok, Irmscher, Zupancic, Galazka, Albrecht, Koch, and
  Draxl}}]{BaSnO3_Aggoune2022}
\bibinfo{author}{\bibfnamefont{W.}~\bibnamefont{Aggoune}},
  \bibinfo{author}{\bibfnamefont{A.}~\bibnamefont{Eljarrat}},
  \bibinfo{author}{\bibfnamefont{D.}~\bibnamefont{Nabok}},
  \bibinfo{author}{\bibfnamefont{K.}~\bibnamefont{Irmscher}},
  \bibinfo{author}{\bibfnamefont{M.}~\bibnamefont{Zupancic}},
  \bibinfo{author}{\bibfnamefont{Z.}~\bibnamefont{Galazka}},
  \bibinfo{author}{\bibfnamefont{M.}~\bibnamefont{Albrecht}},
  \bibinfo{author}{\bibfnamefont{C.}~\bibnamefont{Koch}}, \bibnamefont{and}
  \bibinfo{author}{\bibfnamefont{C.}~\bibnamefont{Draxl}},
  \bibinfo{journal}{Communications Materials} \textbf{\bibinfo{volume}{3}},
  \bibinfo{pages}{12} (\bibinfo{year}{2022}), ISSN \bibinfo{issn}{2662-4443},
  \urlprefix\url{https://doi.org/10.1038/s43246-022-00234-6}.

\bibitem[{\citenamefont{Engel et~al.}(2022)\citenamefont{Engel, Miranda,
  Chaput, Togo, Verdi, Marsman, and Kresse}}]{ElPhonRen_4}
\bibinfo{author}{\bibfnamefont{M.}~\bibnamefont{Engel}},
  \bibinfo{author}{\bibfnamefont{H.}~\bibnamefont{Miranda}},
  \bibinfo{author}{\bibfnamefont{L.}~\bibnamefont{Chaput}},
  \bibinfo{author}{\bibfnamefont{A.}~\bibnamefont{Togo}},
  \bibinfo{author}{\bibfnamefont{C.}~\bibnamefont{Verdi}},
  \bibinfo{author}{\bibfnamefont{M.}~\bibnamefont{Marsman}}, \bibnamefont{and}
  \bibinfo{author}{\bibfnamefont{G.}~\bibnamefont{Kresse}},
  \bibinfo{journal}{Phys. Rev. B} \textbf{\bibinfo{volume}{106}},
  \bibinfo{pages}{094316} (\bibinfo{year}{2022}),
  \urlprefix\url{https://link.aps.org/doi/10.1103/PhysRevB.106.094316}.

\bibitem[{\citenamefont{He and Franchini}(2014)}]{SRPdO3_Franchini}
\bibinfo{author}{\bibfnamefont{J.}~\bibnamefont{He}} \bibnamefont{and}
  \bibinfo{author}{\bibfnamefont{C.}~\bibnamefont{Franchini}},
  \bibinfo{journal}{Phys. Rev. B} \textbf{\bibinfo{volume}{89}},
  \bibinfo{pages}{045104} (\bibinfo{year}{2014}),
  \urlprefix\url{https://link.aps.org/doi/10.1103/PhysRevB.89.045104}.

\bibitem[{\citenamefont{Maggio et~al.}(2017)\citenamefont{Maggio, Liu, van
  Setten, and Kresse}}]{Maggio2017}
\bibinfo{author}{\bibfnamefont{E.}~\bibnamefont{Maggio}},
  \bibinfo{author}{\bibfnamefont{P.}~\bibnamefont{Liu}},
  \bibinfo{author}{\bibfnamefont{M.~J.} \bibnamefont{van Setten}},
  \bibnamefont{and} \bibinfo{author}{\bibfnamefont{G.}~\bibnamefont{Kresse}},
  \bibinfo{journal}{Journal of Chemical Theory and Computation}
  \textbf{\bibinfo{volume}{13}}, \bibinfo{pages}{635} (\bibinfo{year}{2017}),
  ISSN \bibinfo{issn}{1549-9618},
  \urlprefix\url{https://doi.org/10.1021/acs.jctc.6b01150}.

\bibitem[{\citenamefont{Gorelov et~al.}(2022)\citenamefont{Gorelov, Reining,
  Feneberg, Goldhahn, Schleife, Lambrecht, and Gatti}}]{V2O5_2}
\bibinfo{author}{\bibfnamefont{V.}~\bibnamefont{Gorelov}},
  \bibinfo{author}{\bibfnamefont{L.}~\bibnamefont{Reining}},
  \bibinfo{author}{\bibfnamefont{M.}~\bibnamefont{Feneberg}},
  \bibinfo{author}{\bibfnamefont{R.}~\bibnamefont{Goldhahn}},
  \bibinfo{author}{\bibfnamefont{A.}~\bibnamefont{Schleife}},
  \bibinfo{author}{\bibfnamefont{W.~R.~L.} \bibnamefont{Lambrecht}},
  \bibnamefont{and} \bibinfo{author}{\bibfnamefont{M.}~\bibnamefont{Gatti}},
  \bibinfo{journal}{npj Computational Materials} \textbf{\bibinfo{volume}{8}},
  \bibinfo{pages}{94} (\bibinfo{year}{2022}), ISSN \bibinfo{issn}{2057-3960},
  \urlprefix\url{https://doi.org/10.1038/s41524-022-00754-2}.

\bibitem[{\citenamefont{Roginskii et~al.}(2021)\citenamefont{Roginskii,
  Smirnov, Smirnov, Baddour-Hadjean, Pereira-Ramos, Smirnov, and
  Davydov}}]{V2O5_3}
\bibinfo{author}{\bibfnamefont{E.~M.} \bibnamefont{Roginskii}},
  \bibinfo{author}{\bibfnamefont{M.~B.} \bibnamefont{Smirnov}},
  \bibinfo{author}{\bibfnamefont{K.~S.} \bibnamefont{Smirnov}},
  \bibinfo{author}{\bibfnamefont{R.}~\bibnamefont{Baddour-Hadjean}},
  \bibinfo{author}{\bibfnamefont{J.-P.} \bibnamefont{Pereira-Ramos}},
  \bibinfo{author}{\bibfnamefont{A.~N.} \bibnamefont{Smirnov}},
  \bibnamefont{and} \bibinfo{author}{\bibfnamefont{V.~Y.}
  \bibnamefont{Davydov}}, \bibinfo{journal}{The Journal of Physical Chemistry
  C} \textbf{\bibinfo{volume}{125}}, \bibinfo{pages}{5848}
  (\bibinfo{year}{2021}), \eprint{https://doi.org/10.1021/acs.jpcc.0c11285},
  \urlprefix\url{https://doi.org/10.1021/acs.jpcc.0c11285}.

\bibitem[{\citenamefont{Perdew et~al.}(1996)\citenamefont{Perdew, Burke, and
  Ernzerhof}}]{Perdew1996}
\bibinfo{author}{\bibfnamefont{J.~P.} \bibnamefont{Perdew}},
  \bibinfo{author}{\bibfnamefont{K.}~\bibnamefont{Burke}}, \bibnamefont{and}
  \bibinfo{author}{\bibfnamefont{M.}~\bibnamefont{Ernzerhof}},
  \bibinfo{journal}{Phys. Rev. Lett.} \textbf{\bibinfo{volume}{77}},
  \bibinfo{pages}{3865} (\bibinfo{year}{1996}), ISSN \bibinfo{issn}{10797114},
  \urlprefix\url{https://journals.aps.org/prl/abstract/10.1103/PhysRevLett.77.3865}.

\bibitem[{\citenamefont{Kresse and Joubert}(1999)}]{KresseG;Joubert1999}
\bibinfo{author}{\bibfnamefont{G.}~\bibnamefont{Kresse}} \bibnamefont{and}
  \bibinfo{author}{\bibfnamefont{D.}~\bibnamefont{Joubert}},
  \bibinfo{journal}{Phys. Rev. B - Condens. Matter Mater. Phys.}
  \textbf{\bibinfo{volume}{59}}, \bibinfo{pages}{1758} (\bibinfo{year}{1999}),
  ISSN \bibinfo{issn}{1550235X},
  \urlprefix\url{https://journals.aps.org/prb/abstract/10.1103/PhysRevB.59.1758}.

\end{thebibliography}

\end{document}